\documentclass[linenumbers,fleqn,usenatbib]{mnras}
\usepackage{tikz}

\newcommand{\tikzcmark}{%
\tikz[scale=0.23] {
    \draw[line width=0.7,line cap=round] (0.25,0) to [bend left=10] (1,1);
    \draw[line width=0.8,line cap=round] (0,0.35) to [bend right=1] (0.23,0);
}}

\usepackage{newtxtext,newtxmath}
\usepackage[T1]{fontenc}

\DeclareRobustCommand{\VAN}[3]{#2}
\let\VANthebibliography\thebibliography
\def\thebibliography{\DeclareRobustCommand{\VAN}[3]{##3}\VANthebibliography}

\usepackage{graphicx}	% Including figure files
\usepackage{amsmath}	% Advanced maths commands

\title[Nuclear dust in AGN with JWST]{JWST reveals the diversity of nuclear obscuring dust in nearby AGN:
\\ nuclear isolation of MIRI/MRS datacubes and continuum spectral fitting}

\author[González-Martín et al.]{
Omaira Gonz\'alez-Mart\'in,$^{1}$\thanks{E-mail: o.gonzalez@irya.unam.mx}
Daniel J. D\'iaz-Gonz\'alez,$^{1}$
Mariela Martínez-Paredes,$^{2}$
Almudena Alonso-Herrero,$^{3}$ 
\newauthor
Enrique López-Rodríguez,$^{4,5}$
Begoña García-Lorenzo,$^{6,7}$
Cristina Ramos Almeida,$^{6,7}$
Ismael García-Bernete,$^{3}$ 
\newauthor
Donaji Esparza-Arredondo,$^{6,7}$
Sebastian F. Hoenig,$^{8}$ 
Santiago Garc\'ia-Burillo,$^{9}$
Chris Packham,$^{10}$
\newauthor
Nancy A. Levenson,$^{11}$
Alvaro Labiano,$^{12}$
Miguel Pereira-Santaella,$^{13}$
Francoise Combes,$^{14}$
Anelise Audibert,$^{6,7}$
\newauthor
Erin K. S. Hicks,$^{15,10,16}$
Lulu Zhang,$^{10}$
Enrica Bellocchi,$^{17,18}$
Richard I. Davies,$^{19}$
Laura Hermosa Muñoz,$^{3}$ 
\newauthor
Masatoshi Imanishi,$^{20}$
Claudio Ricci,$^{21,22}$ and 
Marko Stalevski,$^{23,24}$
\newauthor
\\
$^{1}$Instituto de Radioastrononomía y Astrofísica (IRyA), Universidad Nacional Autónoma de México, Antigua Carretera a Pátzcuaro \\ \#8701, 1087 ExHda. San José de la Huerta, Morelia, Michoacán, México C.P. 1088 58089 \\
$^{2}$1142 Sunset Point Rd., Clearwater, Florida, USA, 33755 \\
$^{3}$Centro de Astrobiología (CAB), CSIC-INTA, Camino Bajo del Castillo s/n, 28692 Villanueva de la Cañada, Madrid, Spain \\
$^{4}$Department of Physics \& Astronomy, University of South Carolina, Columbia, SC 29208, USA \\
$^{5}$Kavli Institute for Particle Astrophysics \& Cosmology (KIPAC), Stanford University, Stanford, CA 94305, USA \\
$^{6}$Instituto de Astrofśica de Canarias, Calle Vía Láctea, s/n, E-38205, La Laguna, Tenerife, Spain \\
$^{7}$Departamento de Astrofísica, Universidad de La Laguna, E-38206 La Laguna, Tenerife, Spain \\
$^{8}$School of Physics \& Astronomy, University of Southampton, Southampton SO17 1BJ, UK \\
$^{9}$Observatorio Astronómico Nacional (OAN-IGN)-Observatorio de Madrid, Alfonso XII, 3, 28014 Madrid, Spain \\
$^{10}$The University of Texas at San Antonio, One UTSA Circle, San Antonio, TX 78249, USA \\
$^{11}$Space Telescope Science Institute, 3700 San Martin Drive Baltimore, MD 21218, USA \\
$^{12}$Telespazio UK for ESA, ESAC, Camino Bajo del Castillo s/n, 28692 Villanueva de la Cañada, Madrid, Spain \\
$^{13}$Instituto de Física Fundamental, CSIC, Calle Serrano 123, 28006 Madrid, Spain \\
$^{14}$Observatoire de Paris, LUX, Collège de France, PSL University, CNRS, Sorbonne University, Paris, France \\
$^{15}$Department of Physics and Astronomy, University of Alaska Anchorage, Anchorage, AK 99508-4664, USA \\
$^{16}$Department of Physics, University of Alaska, Fairbanks, Alaska 99775-5920, USA \\
$^{17}$Departmento de F\'isica de la Tierra y Astrof\'isica, Fac. de CC F\'isicas, Universidad Complutense de Madrid, E28040 Madrid, Spain \\
$^{18}$Instituto de F\'isica de Part\'iculas y del Cosmos IPARCOS, Fac. CC F\'isicas, Universidad Complutense de Madrid, E-28040 Madrid, Spain \\
$^{19}$Max Planck Institute for extraterrestrial Physics, Giessenbachstrasse 1, 85748, Garching, Germany \\
$^{20}$National Astronomical Observatory of Japan, 2-21-1 Osawa, Mitaka, Tokyo 181-8588, Japan \\
$^{21}$Instituto de Estudios Astrof\'isicos, Facultad de Ingenier\'ia y Ciencias, Universidad Diego Portales, Av. Ej\'ercito Libertador 441, Santiago, Chile \\
$^{22}$Kavli Institute for Astronomy and Astrophysics, Peking University, Beijing 100871, China \\
$^{23}$Astronomical Observatory, Volgina 7, 11060 Belgrade, Serbia  \\
$^{24}$Sterrenkundig Observatorium, Universiteit Gent, Krijgslaan 281-S9, Gent B-9000, Belgium 
}

\date{Accepted XXX. Received YYY; in original form ZZZ}

\pubyear{\the\year{}}

\begin{document}
\label{firstpage}
\pagerange{\pageref{firstpage}--\pageref{lastpage}}
\maketitle

\begin{abstract}
We investigate the capabilities of the mid-infrared instrument (MIRI) of James Webb Space Telescope (\emph{JWST}) to advance our knowledge of AGN dust using the spectral fitting technique on an AGN collection of 21 nearby ($z<0.05$) AGN (7 type-1 and 14 type-2) observations obtained with the medium resolution spectroscopy (MRS) mode. This collection includes publicly available AGN and data from the collaboration of Galactic Activity, Torus, and Outflow Survey (GATOS). We developed a tool named MRSPSFisol that decomposes MRS cubes into point-like and extended contributions. We found statistically good fits for 12 targets with current AGN dust models. The model that provides good fits ($\rm{\chi^2/dof<2}$) for {these 12 targets} assumes a combination of clumpy and smooth distribution of dust in a flare-disk geometry where the dust grain size is a free parameter. Still, two and one AGN statistically prefer the disk$+$wind and the classical clumpy torus model, respectively. However, the currently available models fail to reproduce 40\% of the targets, likely due to the extreme silicate features not well reproduced by the models and signatures of water-ice and aliphatic hydrocarbon absorption features in most targets. New models exploring, for instance, new chemistry, are needed to explain the complexity of AGN dust continuum emission observed by \emph{JWST}. 
\end{abstract}


%%%%%%%%%%%%%%%%%%%%%%%%%%%%%%%%%%%%%%%%%%%%%%%%%%

%%%%%%%%%%%%%%%%% BODY OF PvailPER %%%%%%%%%%%%%%%%%%

\section{Introduction} \label{sec:intro}

Despite the enormous observational and theoretical effort, the active galactic nucleus (AGN) fuel origin and the triggering mechanism nature are two of the main unresolved questions in the field \citep{Netzer15}. Feeding should rely on gas falling into the super-massive black hole (SMBH), driven by magnetic fields and loss of angular momentum \citep[][]{Blandford77,Emmering92,Lopez-Rodriguez20}. Due to nuclear dust being coupled with gas \citep{Lauer05, Esparza-Arredondo21}, understanding the physical properties of the dust is essential to understanding AGN feeding \citep{Tran01}. 

Infrared wavelengths are valuable for studying AGN dust, as the dust that hides many AGN at optical and UV wavelengths is emitted in the infrared. Indeed, thanks to infrared ground- and space-based observations, our perception of the AGN dusty central regions has significantly changed in recent years \citep{Ramos-Almeida17}. It had been proposed that a thick torus, {present in most AGN}, surrounded the accretion disk and broad line region \citep{Antonucci93,Urry95}. However, high-angular-resolution interferometric observations of local Seyfert galaxies in the infrared showed that the dust at pc-scale is not distributed in a single, toroidal structure for some sources {\citep[][and references therein]{Cameron93,Bock00,Radomski03,Hoenig13, Lopez-Gonzaga14, Leftley18, Hoenig19, Gamez-Rosas22,Isbell25}}. Instead, these observations suggest a combination of an equatorial, thin disk and a polar-extended component, which would have originated from a dusty wind \citep{Hoenig17}. The dust motion is then an infall throughout a thin disk (warm disk traced by infrared and cold and larger disk traced by sub-mm) and an outflow in the perpendicular direction \citep{Combes23,Garcia-Burillo24}. The thin disk has a well-known counterpart at sub-mm wavelengths \citep[][]{Garcia-Burillo16, Imanishi18,Alonso-Herrero18,Alonso-Herrero19,Alonso-Herrero23, Combes19, Impellizzeri19, Imanishi20, Garcia-Burillo19, Lopez-Rodriguez18, Garcia-Burillo21, Nikutta21A}. The AGN obscuring structure is part of a complex and dynamic accretion flow \citep{Norman88}, and therefore, it must evolve with the AGN phases, as suggested by multiwavelength observations that reveal a multi-phase and multi-component nature \citep{Ramos-Almeida17, Hoenig19}. 

The nuclear dust analysis in statistically significant AGN samples is needed to understand the geometry and distribution of this material and how it depends on the AGN evolutionary state. The spectral energy distribution (SED) fitting technique can give insights into the best geometry and distribution \citep{Gonzalez-Martin19A,Gonzalez-Martin23}. This is achieved by giving definitive probes of the AGN dust diversity of properties (that might be linked to the evolution) for extensive AGN collections, which cannot be achieved with any other technique (e.g., infrared interferometry) due to the faintness of {most AGN} at mid-infrared wavelengths and limits of the instrument sensitivity. The SED fitting technique has been applied to collections of AGN spectra obtained from space-based \citep{Gonzalez-Martin17, Gonzalez-Martin19B, Gonzalez-Martin23} and also from ground-based \citep{Ramos-Almeida09,Ramos-Almeida11, Alonso-Herrero11, Audibert17, Martinez-Paredes17, Martinez-Paredes20, Martinez-Paredes21, Garcia-Bernete19, Garcia-Bernete22B} telescopes. The emission of the AGN dust peaks in the mid-infrared wavelengths, where many parts of the spectrum were virtually unexplored from the ground due to strong atmospheric absorption. Ground-based observations rely on photometric data points combined with 7-13$\rm{\mu m}$ spectra, except for the brightest nearby AGN \citep[see, for instance, the work by][in NGC\,1068]{Victoria-Ceballos22}. However, photometric points alone cannot constrain the models \citep{Gonzalez-Martin19A}, and high angular resolution spectra are available for a few bright sources. Space-based observations (e.g., those provided by \emph{Spitzer}) had a relatively poor spatial resolution. Even for intermediate and luminous nearby AGN, their central region consists of a combination of AGN and a non-negligible contribution of star formation (SF) activity component \citep[][]{Martinez-Paredes15,Herrero-Illana17,Efstathiou22}. This might imply a bias in the AGN samples to those objects where the inner hundred parsecs are free of host galaxy contamination. 

The James Webb Space Telescope (\emph{JWST}) has already been revolutionary in the field of nearby AGN \citep[see][]{Dasyra24,Pereira-Santaella24,Donnan24,Garcia-Bernete24A,Garcia-Bernete24B,Gonzalez-Alfonso24,Davies24,Haidar24,Hermosa-Muñoz24,Goold24}. For this work, with the 6.5m \emph{JWST} mirror and advanced instrument suite, we can now resolve the dust at an angular resolution of $\rm{\sim}$0.3-0.8 arcsec (50-120\,pc at D=20\,Mpc) in the mid-infrared wavelengths \citep[5-28$\rm{\mu m}$][]{Gardner23, Wright24}, being able to detect AGN as faint as 0.1 mJy at 12$\rm{\mu m}$\footnote{Detection limit for SNR=10 in 10\,ksec.} \citep{Glasse15}. This is a decisive step forward compared to ground-based mid-infrared facilities where the N-band ($\rm{\sim 7-13\mu m}$) spectrum can be obtained only for AGN brighter than 50$-$100\,mJy at 12$\rm{\mu m}$ \citep{Gonzalez-Martin13,Alonso-Herrero16}. There have been only a few attempts to model the AGN dust continuum obtained with \emph{JWST} observations. \citet{Garcia-Bernete24B} presented the case of II Zw 096-D1, the brightest infrared sources within II\,Zw\,096, a complex interacting system and luminous infrared galaxy. They investigated the adequacy of several torus models, finding the presence of a dust-obscured AGN, which has an exceptionally high covering factor. However, they also found that models do not reproduce the silicate absorption feature and the 12-15$\rm{\mu m}$ continuum simultaneously. \citet{Garcia-Bernete24A} explored the effectiveness of clumpy and smooth torus models to explain the silicate absorption features in the Galaxy Activity, Torus, and Outflow Survey \citep[GATOS][]{Alonso-Herrero21,Garcia-Burillo21}, finding that both models can explain these features. They also attempted to perform spectral fitting using clumpy and smooth torus models for NGC\,5728. They saw the need for developing torus models that include the expected water-ice and aliphatic hydrocarbon molecular content. 
\begin{table*}
	\centering
    \footnotesize
    \renewcommand{\tabcolsep}{0.07cm}
	\caption{General properties of our AGN collection; from col. 1 to 8: object name, right ascension, declination, redshift, optical classification obtained from the NASA Extragalactic Database (NED), spatial scale in parsecs for 1 arcsec, and X-ray luminosity. References for 2-10\,keV intrinsic X-ray luminosities are: (1) \citet[][]{Ricci17}; (2) \citet[][]{Fernandez-Ontiveros20}; (3) \citet{Ohyama15}; (4) \citet{Goold24}; (5) \citet{Gonzalez-Martin09}; and (6) \citet{Iwasawa18}. Columns 9-12 give the proposal identification, observation numbers for the target and background (separated by `/'), principal investigator of the proposal, and observation date. Col.\,13 shows whether \emph{Spitzer} spectra are available, and Col.\,14 shows the 12$\rm{\mu m}$ PSF flux (in mJy) obtained from ground-based observations when available. Objects marked with an asterisk belong to the GATOS sample. Objects marked as $^{p}$ were in the proprietary period when we selected the AGN collection.}
	\label{tab:collection}
	\begin{tabular}{lrrl ccc ccc cccc} % four columns, alignment for each
		\hline \hline
  &  &  &  &  &  &  &  & \multicolumn{4}{c}{JWST data} &  &  \\ \cline{9-12}
 Name & {R.A.} & {Dec.} & {redshift} & {Type}  & {pc/$''$}  & {$\rm{log(L_{X})}$} & {Ref.} & {Prop.ID} & {Obs.ID} & {PI}  & {Date} & {\emph{Spitzer}}  &  {$\rm{F_{12\mu m}}$ (ground)} \\
  &  & $^{\circ}$ & $^{\circ}$ &   &  & {(2-10 keV)} &  &  & {(t/b)} &  &  &  & {mJy } \\ 
 (1)  & (2) & (3) & (4) & (5) & (6) & (7) & (8) & (9) & (10) & (11) & (12) & (13) & (14)  \\  \hline 
NGC\,1052       &  40.2700252 & -8.2557428  &   0.00482 &  L1.9  & 133  &  41.5    &  1  & 02016  & 11/12  &  Seth       & 2022-08-11 & \checkmark & 137.6 \\ % &  8.82   &   41.5    &   -3.97  
ESO\,420-G13    &  63.4570420 & -32.0069720 &   0.01191 &  S2    & 330  &  43.6    &  2  & 01875  & 01/02  &  Fernandez  & 2022-11-30 & \checkmark & 192.3 \\ % &  6.90   &   43.6    &   -1.43  
UGC\,05101      & 143.9649642 & 61.3532726  &   0.03937 &  L1.5  &1090  &  43.7    &  1  & 01717  & 05/06  &  U          & 2023-04-06 & \checkmark & 227.0 \\ % &  8.35   &   43.7    &   -1.78  
MCG\,-05-23-016*& 146.9172878 & -30.9489000 &   0.00849 &  S2    & 235  &  43.5    &  1   & 01670  & 06/12  &  Shimizu    & 2023-04-03 & \checkmark & 714.3 \\ % &  7.43   &   43.5    &   -1.08  
NGC\,3031       & 148.8882210  &  69.0652950  &   0.00077 &  S1.8  &  21  &  40.3   &  1  & 02016  & 17/18  &  Seth       & 2023-04-03 &            & 136.9 \\ % &  7.80   &   40.3   &   -4.71  
NGC\,3081*$^{p}$ & 149.8730800  & -22.8262770  &   0.00591 &  S2    & 164  &  43.1    &  1   & 01670  & 11/05  &  Shimizu    & 2023-12-09 & \checkmark & 162.1 \\ % &  7.72   &   43.1    &   -1.82  
NGC\,3256NUC1   & 156.9634188 & -43.9052237 &   0.00935 &  S2    & 259  & 40.8    &  3  & 01328  & 09/10  &  Armus      & 2022-12-27 & \checkmark & $\dots$ \\ % & $\dots$ &   40.8    &  $\dots$ 
% NGC\,3256NUC2   & 156.9634753 & -43.9037901 &   0.00935 &  HII   & 259  & 37.4    &     & 01328  & 08/10  &  Armus      & 2022-12-27 & \checkmark & $\dots$ \\ % & $\dots$ &   37.4    &  $\dots$ 
% CGCG\,012-070   & 176.119984  & -3.570855   &   0.04808 &  S2    &1331  & $\dots$   &     & 01928  & 05/06  &  Riffel     & 2023-06-10 & \checkmark \\ % & $\dots$ & $\dots$   &  $\dots$ 
% NGC\,3884       & 176.550767  & 20.391647   &   0.02312 &  S1    & 640  & 41.92   &     & 01928  & 02/03  &  Riffel     & 2023-06-09 & \checkmark$N$  \\ % & $\dots$ &   41.92   &  $\dots$
NGC\,4395       & 186.4535920  & 33.5469280   &   0.00106 &  S1.8  &  29  &  40.5    &  1  & 02016  & 02/03  &  Seth       & 2023-06-11 & \checkmark & 9.7 \\ % &  4.99   &   40.5    &  -1.72   
NGC\,4594       & 189.9976051 & -11.6230287 &   0.00364 &  L2    & 101  &  40.0    &  4  & 02016  & 23/24  &  Seth       & 2022-07-04 & \checkmark & 4.4 \\ % &  8.83   &   40.0    &  -5.66   
NGC\,4736       & 192.7210880  & 41.1204580   &   0.00103 &  L2    &  29  &  38.6    &  5   & 02016  & 14/15  &  Seth       & 2023-06-12 & \checkmark & 12.9 \\ % &  6.98   &   39.3    &  -4.72   
MRK\,231        & 194.0593080  & 56.8736770   &   0.04217 &  S1    &1168  &  45.6    &  1   & 01268  & 01/02  &  Maiolino   & 2023-04-05 & \checkmark & $\dots$ \\ % &  8.40   &   45.6    &  -0.84   
Mrk\,273SW      & 206.1752182 & 55.8868658  &   0.03734 &  S2    &1034  &  43.1    &  6  & 01717  & 01/06  &  U          & 2022-06-11 & \checkmark & $\dots$ \\ % &  7.96   &   43.1    &  -2.05   
Mrk\,273        & 206.1755048 & 55.8871075  &   0.03734 &  S2    &1034  &  43.1    &  6  & 01717  & 01/06  &  U          & 2022-06-11 & \checkmark & $\dots$ \\ % &  7.96   &   43.1    &  -2.05   
NGC\,5506*$^{p}$ & 213.3120500 & -3.2075769  &   0.00561 &  S1.9   & 155  &  43.3    &  1  & 01670  & 07/01  &  Shimizu    & 2023-07-06 & \checkmark & 870.8 \\ % &  7.51   &   43.3    &  -1.39   
NGC\,5728*      & 220.5995381 & -17.2530630 &   0.00932 &  S2    & 258  &  43.0    &  1  & 01670  & 04/10  &  Shimizu    & 2023-05-06 & \checkmark & 49.1 \\ % & 7.85   &   43.0    &  -2.06   
ESO\,137-G034*  & 248.8087920  & -58.0800280  &   0.00769 &  S2    & 213  &  42.2    &  1  & 01670  & 03/08  &  Shimizu    & 2023-05-06 & \checkmark & $\dots$ \\ % &  8.02   &   42.2    &  -3.07   
NGC\,6552       & 270.0301638 & 66.6151606  &   0.02649 &  S2    & 733  &  42.8    &  1  & 01039  & 05/09  &  Dicken     & 2022-05-07 &            & $\dots$ \\ % &  7.63   &   42.8    &  -2.06   
IC\,5063        & 313.0097500  & -57.0687780  &   0.01135 &  S2    & 314  &  43.1   &  1  & 02004  & 03/04  &  Dasyra     & 2023-05-13 & \checkmark & 820.6 \\ % &  7.63   &   43.1   &  -1.75   
NGC\,7172*      & 330.5078800 & -31.8696658 &   0.00791 &  S2    & 219  &  43.4    &  1  & 01670  & 09/02  &  Shimizu    & 2023-06-20 & \checkmark & 185.0 \\ % &  8.09   &   43.4    &  -1.86  
NGC\,7319       & 339.0149629 & 33.9758636  &   0.02251 &  S2    & 623  &  42.2    &  1  & 02732  & 04/05  &  Pontoppidan& 2022-06-20 &            & $\dots$ \\ % &  8.10   &   42.2    &  -1.69   
NGC\,7469       & 345.8150769 &  8.8739358  &   0.01627 &  S1.5   & 450  &  43.0    &  1  & 01328  & 15/16  &  Armus      & 2022-06-22 & \checkmark & 485.2 \\ % &  6.96   &   43.0    &  -1.17   
		\hline \hline
	\end{tabular}
\end{table*}

The present work is the first attempt to use the spectral fitting technique in an AGN collection observed with the mid-infrared instrument (MIRI) of \emph{JWST} in Medium-resolution spectrometer (MRS) mode, confronting up to seven available libraries of models. The main objectives of the current work are: (1) to understand if the currently available models can reproduce the continuum features observed with \emph{JWST} and (2) to report spectral features that should be included in future developments of the torus models for a better match of \emph{JWST} observations. While the data analysis was performed, we saw the need to further decompose MIRI/MRS data cubes to isolate the point-like nuclear component from the extended emission. This manuscript reports the details of the MRSPSFisol routine, which performs this decomposition of MIRI/MRS cubes, thanks to the synthetic point spread function (PSF).  

This manuscript is organized as follows. Section\,\ref{sec:data} describes the data collection, and section\,\ref{sec:processing} reports the data acquisition and general processing. Section\,\ref{sec:decomposition} reports the details of MRSPSFisol routine to decompose MRS data cubes into point-like and extended data cubes, section\,\ref{sec:extraction} describes the spectral extractions performed in this analysis, and section\,\ref{sec:spectralfitting} {describes} the general procedure applied for the spectral fitting. Sections\,\ref{sec:results} and \ref{sec:discussion} present and discuss the results, respectively. The main findings of this work are summarized in section\,\ref{sec:summary}. We assume in this work a cosmology with $\rm{H_{0} = 70\, kms^{-1}\,Mpc^{-1}}$, $\Omega_{M} = 0.27$, and $\Omega_{\Lambda} = 0.73$.

\section{Data collection} \label{sec:data}

We include the six AGN (see Table\,\ref{tab:collection}) from \emph{JWST} proposal ID 1670 (PI Shimizu) in this AGN collection as part of the GATOS\footnote{https://gatos.myportfolio.com/.} collaboration effort. We also searched for all AGN observed with the MRS mode \citep[][]{Argyriou23} at mid-infrared with \emph{JWST}/MIRI \citep[][]{Wright23} and publicly available by June 15, 2024, which includes observations performed before June 15, 2023. To accomplish this, we used the \emph{JWST} observing schedule\footnote{https://jwstfeed.com/Schedule/} to select all MRS observations within the `Galaxies' category with keywords related to AGN, such as active galaxies, quasars, galaxy nuclei, radio galaxies, active galactic nuclei, and Seyfert galaxies. This process resulted in a parent sample of 40 AGN. We initially excluded two galaxies (CGCG\,012-070 and NGC\,3884) because the observational setup does not include the full MIRI wavelength coverage. Although the study of dust in high-redshift AGN is intriguing and deserves attention, we decided to focus this work on nearby AGN as an initial step to explore the range of conditions the nuclear environment can exhibit within the first few kpcs from the SMBH with \emph{JWST}/MIRI. We then sought the redshifts of the sources to narrow down our analysis to nearby ($\rm{z<0.1}$) AGN. This limit imposes a minimum spatial resolution of $\rm{\sim 1\,kpc}$ at 20$\rm{\mu m}$. We excluded 18 AGN at redshifts above $\rm{z=0.1}$, with the nearest exclusion being PDS\,456 at $\rm{z=0.184}$, while the others are at $\rm{z>0.4}$.  

Our sample includes 21 targets from 20 \emph{JWST}/MIRI observations. One of them, Mrk\,273, shows two nuclei separated at 0.75\,kpc. These two nuclei are observed in a single observation because they fall within the same field of view (FOV) with the MRS mode. We study both nuclei individually, named Mrk\,273 and Mrk\,273SW, confirmed by X-rays as AGN \citep{Liu19}. Particular attention is also needed for NGC\,3256, which is a nearby (D$\rm{\sim}$40Mpc) gas-rich advanced merger that contains two nuclei observed at two different \emph{JWST} points \citep{Bohn24,Donnan24}. Various authors have claimed that the southern nucleus, referred to as NGC\,3256NUC1 in this work, may harbor a heavily obscured AGN \citep[][]{Neff03,Ohyama15,Pereira-Santaella24}. However, X-ray studies have not provided conclusive evidence of this claim \citep[][]{Lira02, Pereira-Santaella11}, although they do suggest the existence of a low-luminosity AGN ($\rm{L(2-10\,keV)\sim 10^{40}erg\,s^{-1}}$). The northern component is not included in our data collections because it is fueled by a bright starburst with no indication of nuclear activity \citep{Lira08}.

\begin{figure}
\includegraphics[width=1.\columnwidth]{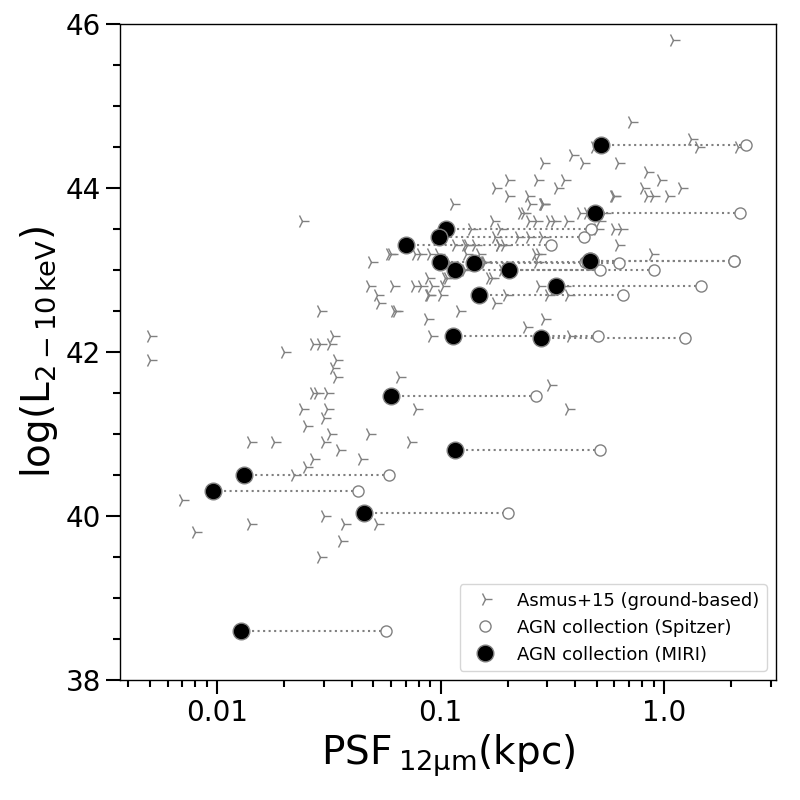}
\caption{Logarithm of the 2-10\,keV intrinsic luminosity versus PSF spatial scale at $\rm{12\,\mu m}$ (in units of kpc) of our AGN collection achieved by MIRI/MRS ($\rm{\sim 0.4 arcsec}$, black dots) and \emph{Spitzer} ($\rm{\sim 2 arcsec}$, small empty circles). The X-ray luminosities and spatial scales obtained using ground-based mid-infrared observation of nearby AGN reported by \citet{Asmus14,Asmus15} are also shown with gray blades.  \label{fig:LxSpatialScale}}
\end{figure}

Altogether, our AGN collection of MIRI observations includes 21 AGN. Table\,\ref{tab:collection} lists the main properties of our AGN collection and the observational details for MIRI/MRS observations. Seven are type-1, and 14 are type-2 AGN. The distance of the sources yields spatial scales as high as $\rm{\sim20\,pc/arcsec}$ for NGC\,3031 down to 1.1\,kpc/arcsec for the furthest AGN in our sample, Mrk\,231. The sample contains sources with X-ray luminosities in the range {$\rm{log(L_{X} (erg/s))=40-45.6}$}.

Fig.\,\ref{fig:LxSpatialScale} illustrates the gain in spatial resolution by \emph{JWST}/MIRI (black-filled dots) when compared with \emph{Spitzer}/IRS spectra (small empty circles)\footnote{The apparent correlation in Fig.\,\ref{fig:LxSpatialScale} {is a sensitivity bias} because both PSF and X-ray luminosities are proportional to the distance.}. It also shows that the spatial scales achieved by MIRI/MRS at 12$\rm{\mu m}$ are similar to those obtained in imaging mode from the ground for a sample of 150 nearby AGN compiled by \citet{Asmus14,Asmus15}. However, the PSF stability of \emph{JWST} is better than that from the ground, allowing to better recover the extended emission.  

Note that this sample collection is heterogeneous by construction due to the different goals of the individual proposals. The proposals with the largest number of objects within our sample are: propID. 1670 (six objects) GATOS survey containing hard X-ray selected Seyfert galaxies in the nearby universe \citep[][]{Shimizu21}; propID. 2016 (five objects) to characterize the mid-infrared properties of low-luminosity AGN \citep[][]{Seth21}; propID. 1717 (three objects) to study AGN activity in luminous infrared galaxies \citep[][]{U21}; and propID. 1328 (two objects) to study the starburst-AGN connection in luminous infrared galaxies \citep[][]{Armus17}. The remaining five objects belong to proposals aiming at studying individual targets for various reasons \citep[][]{Fernandez-Ontiveros21, Alvarez-Marquez23, Dasyra21, Maiolino17, Pontoppidan22, Alonso-Herrero24}. Despite that, this AGN collection well represents the range of X-ray luminosities of the $\rm{\sim}$150 nearby ($z<0.4$) AGN compiled by \citet{Asmus14, Asmus15}, including a volume-limited sample selected from the 9-months \emph{Swift}/BAT catalog \citep[see Fig.\,\ref{fig:LxSpatialScale}][]{Tueller05}.

\begin{figure*}
\begin{center}
\includegraphics[width=0.5\columnwidth]{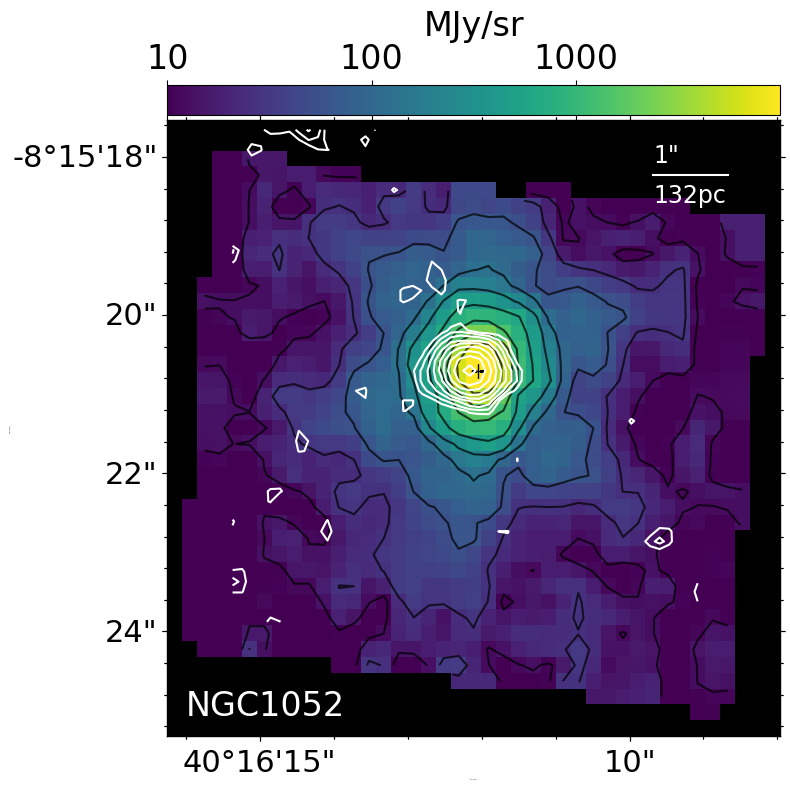}
\includegraphics[width=0.5\columnwidth]{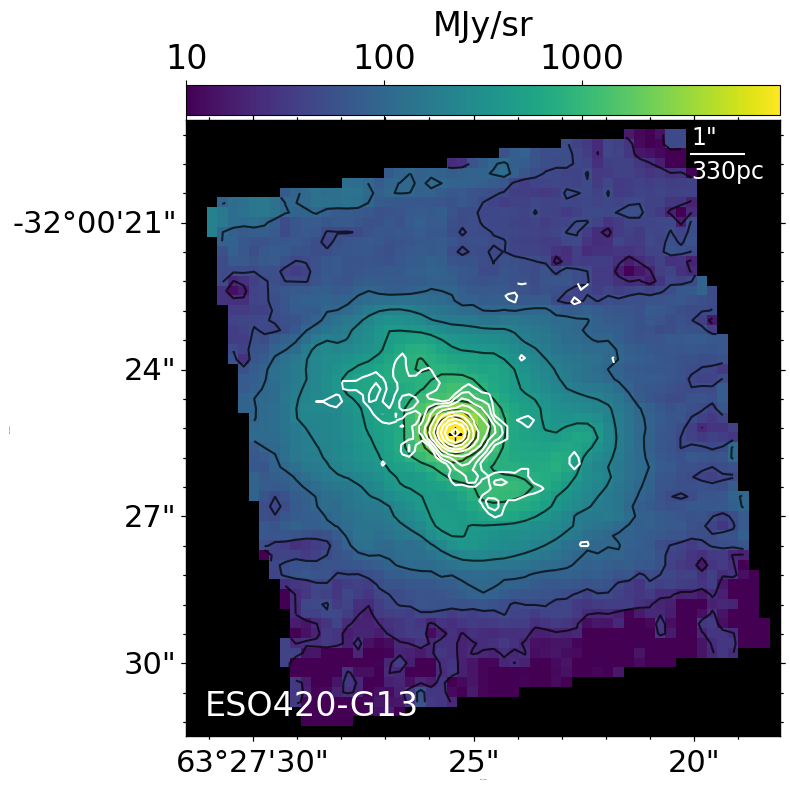}
\includegraphics[width=0.5\columnwidth]{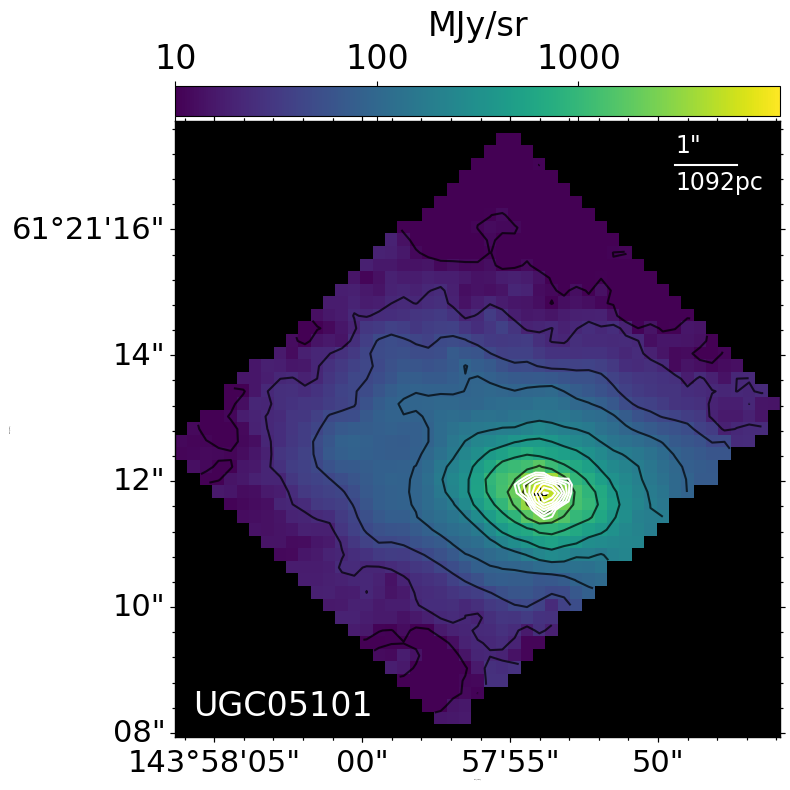}
\includegraphics[width=0.5\columnwidth]{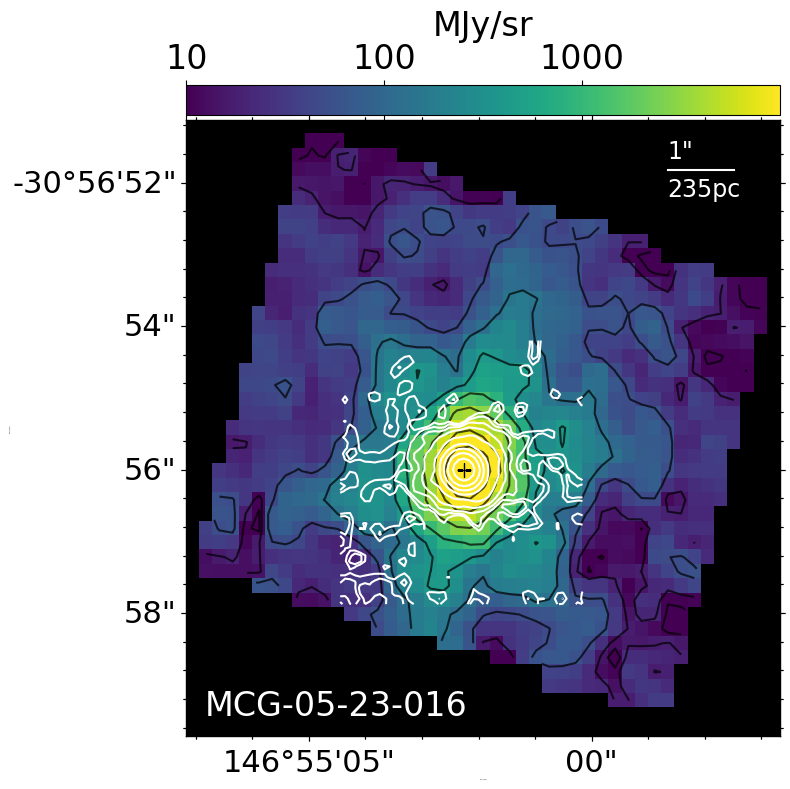}
\includegraphics[width=0.5\columnwidth]{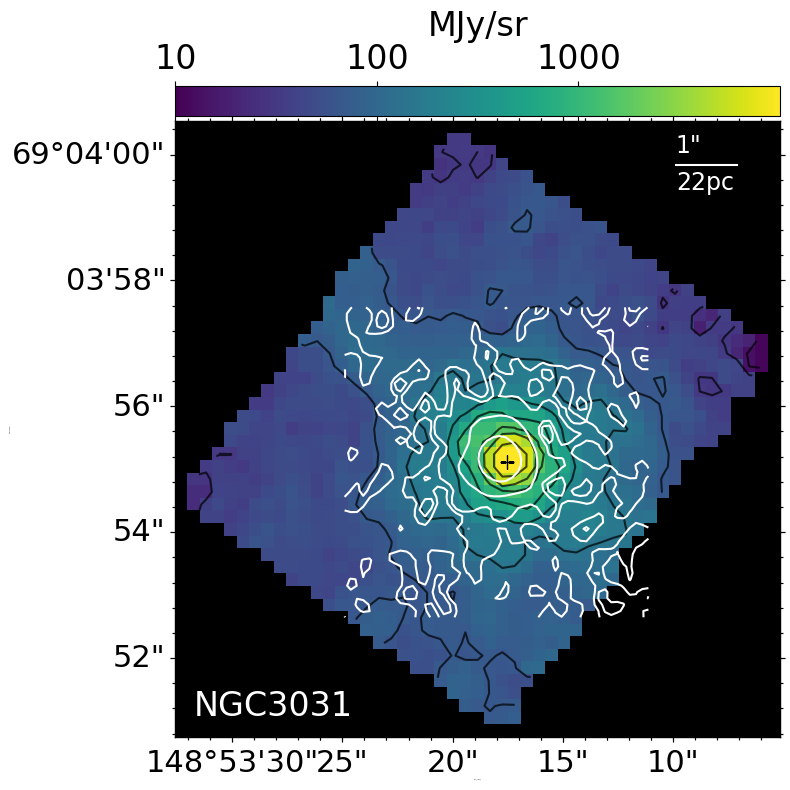}
\includegraphics[width=0.5\columnwidth]{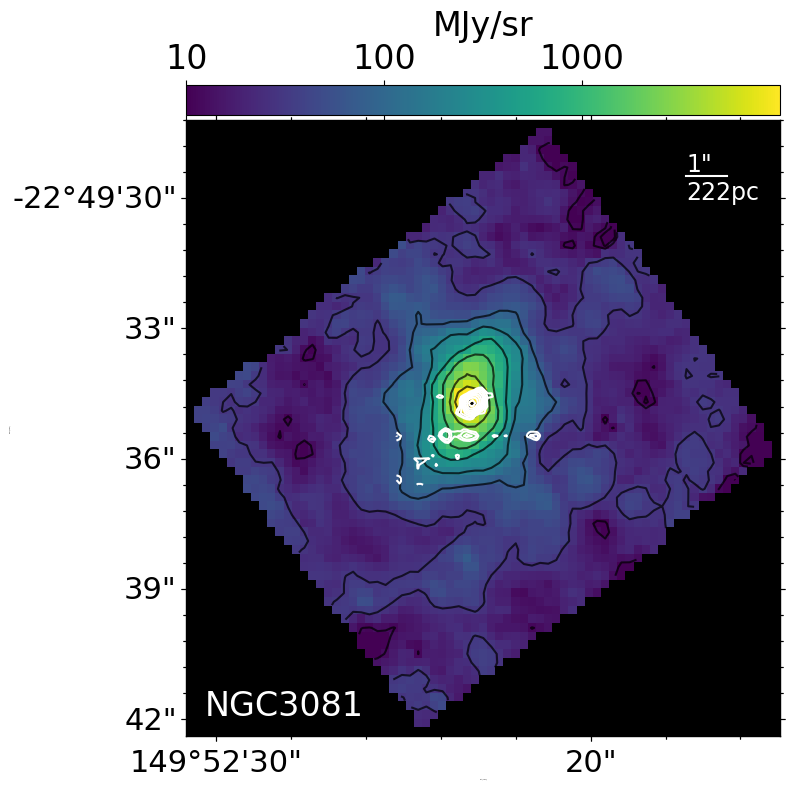}
\includegraphics[width=0.5\columnwidth]{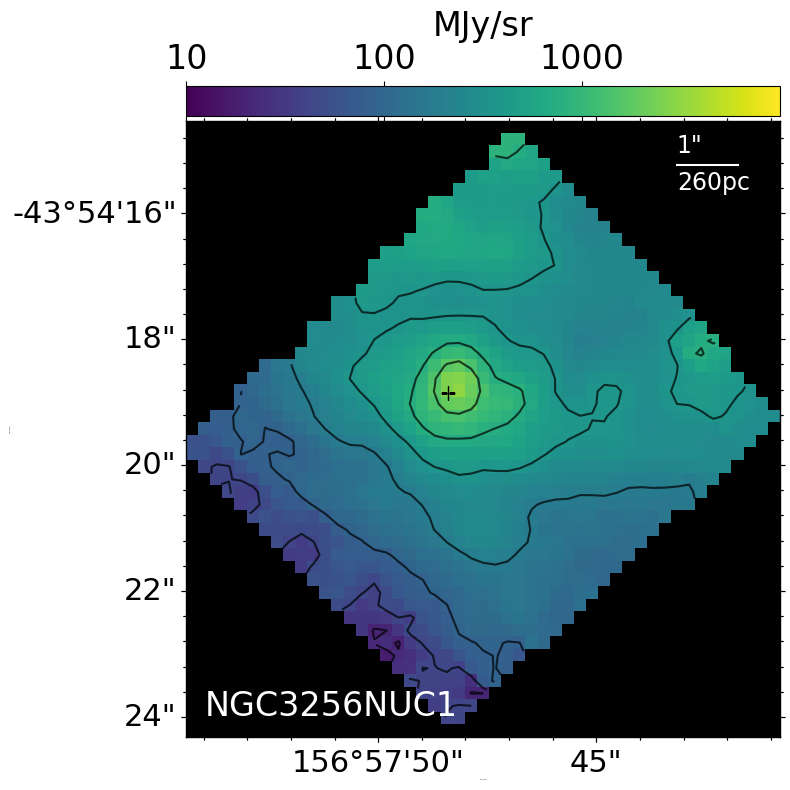}
\includegraphics[width=0.5\columnwidth]{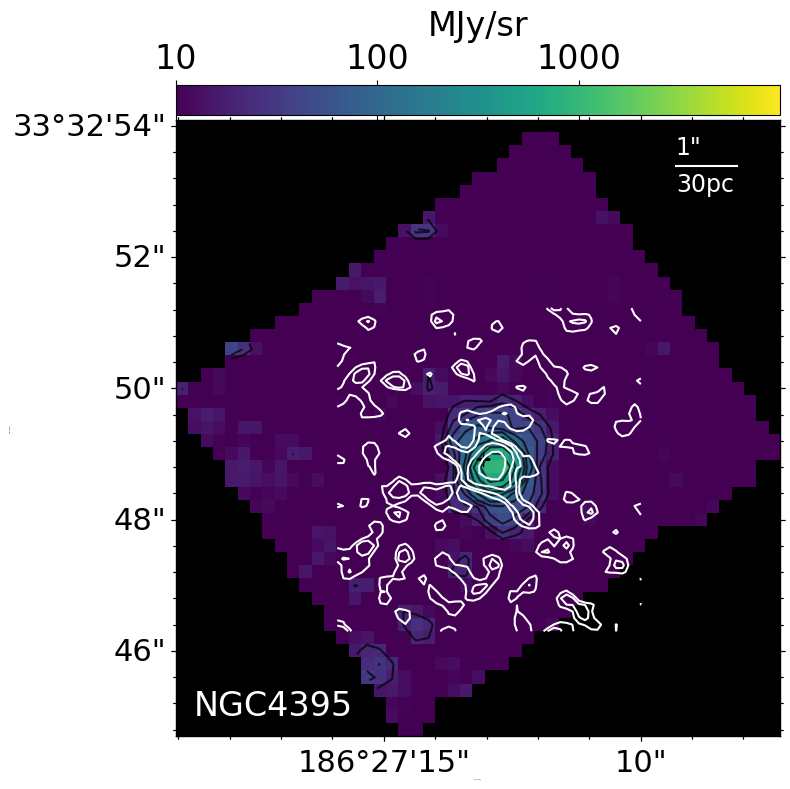}
\includegraphics[width=0.5\columnwidth]{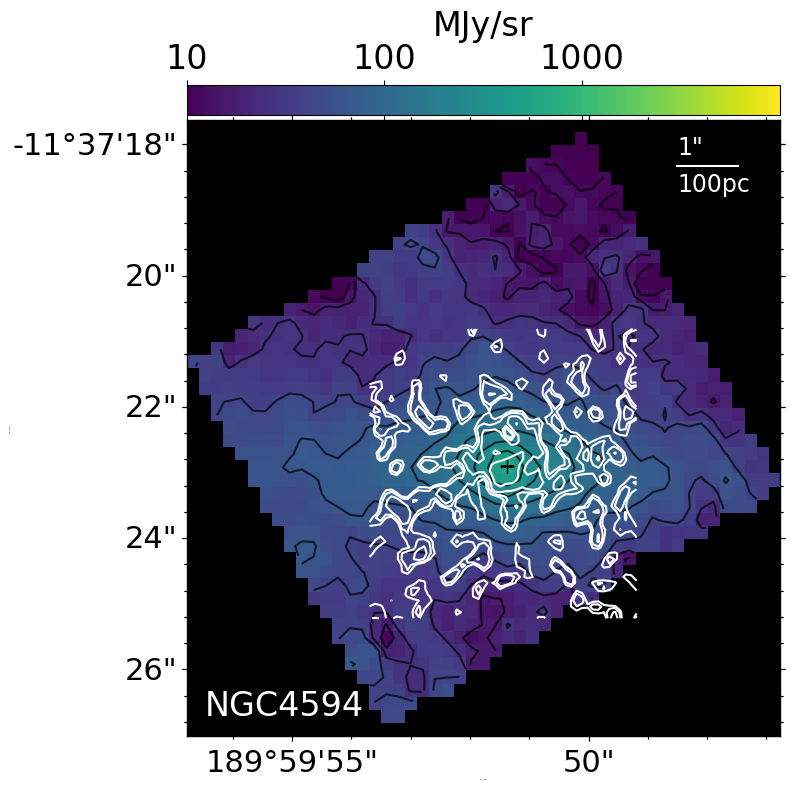}
\includegraphics[width=0.5\columnwidth]{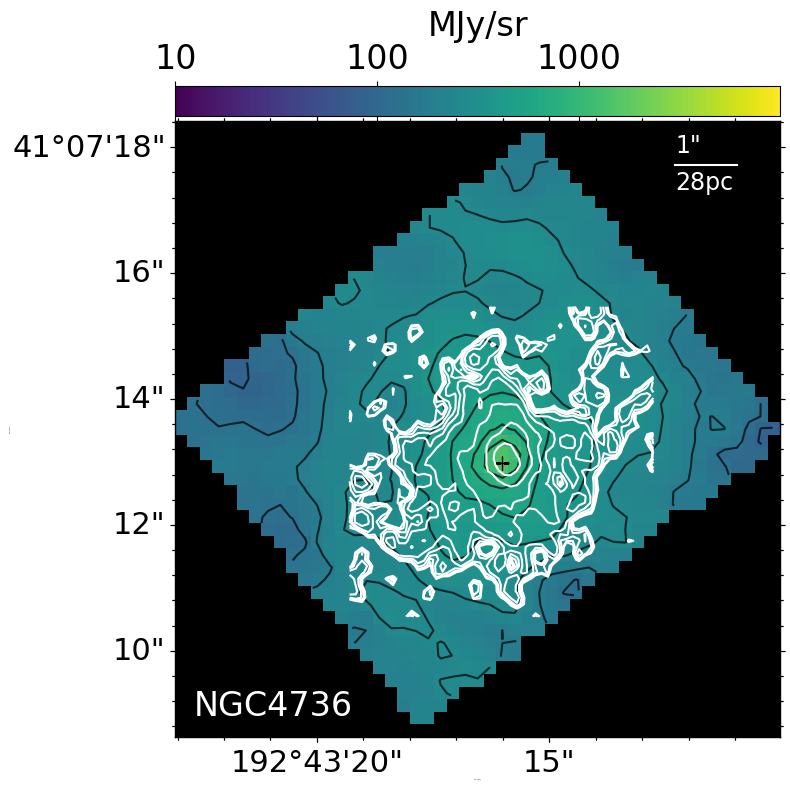}
\includegraphics[width=0.5\columnwidth]{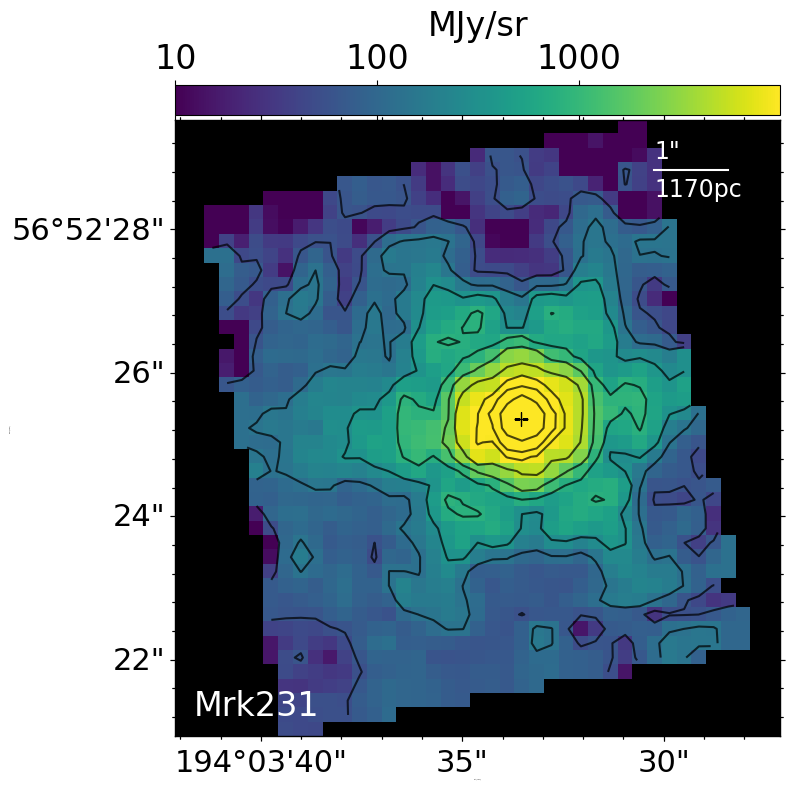}
\includegraphics[width=0.5\columnwidth]{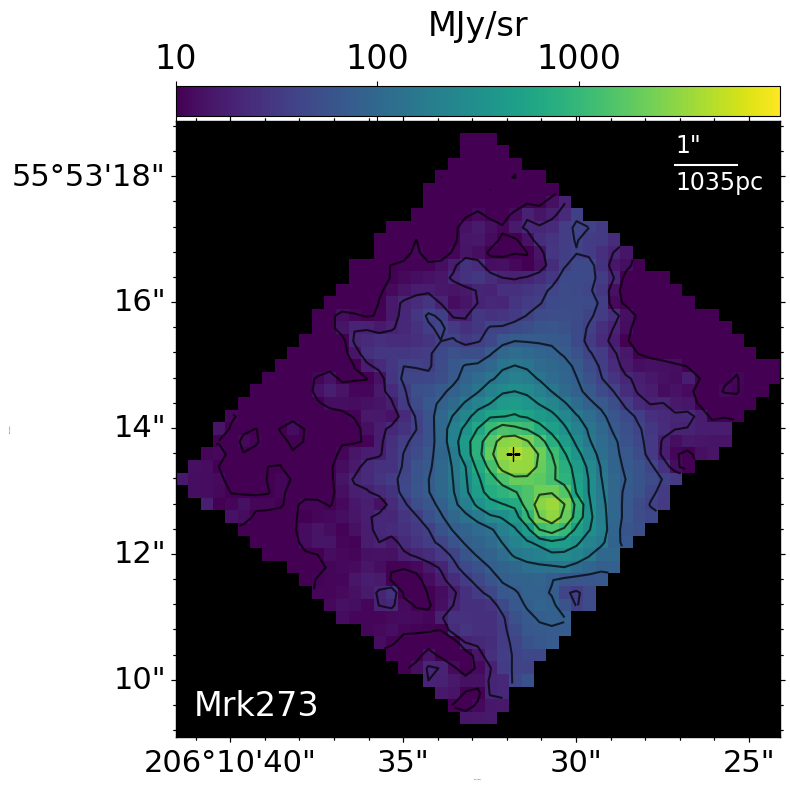}
\includegraphics[width=0.5\columnwidth]{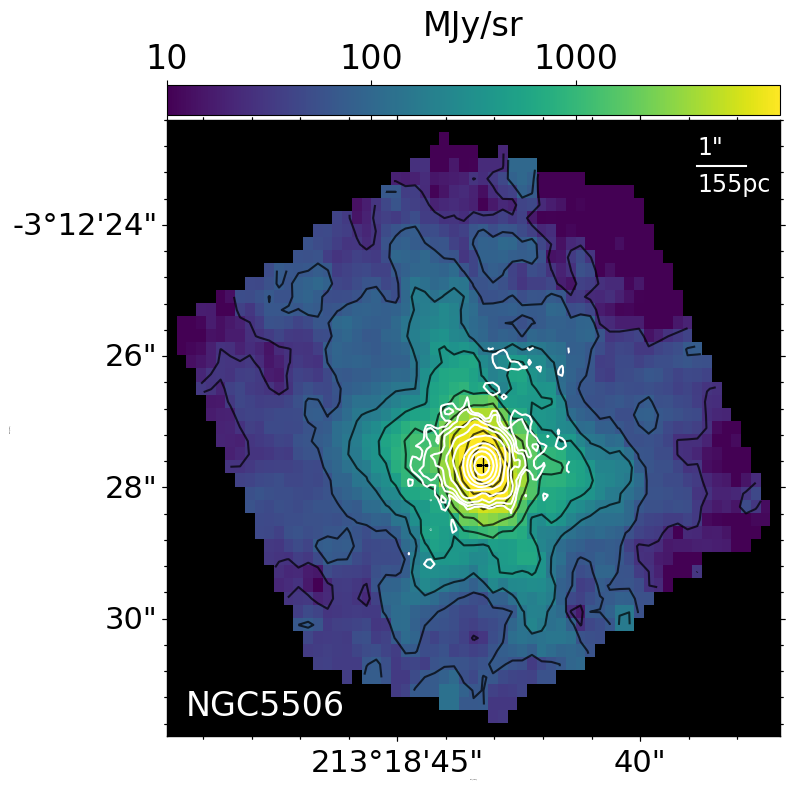}
\includegraphics[width=0.5\columnwidth]{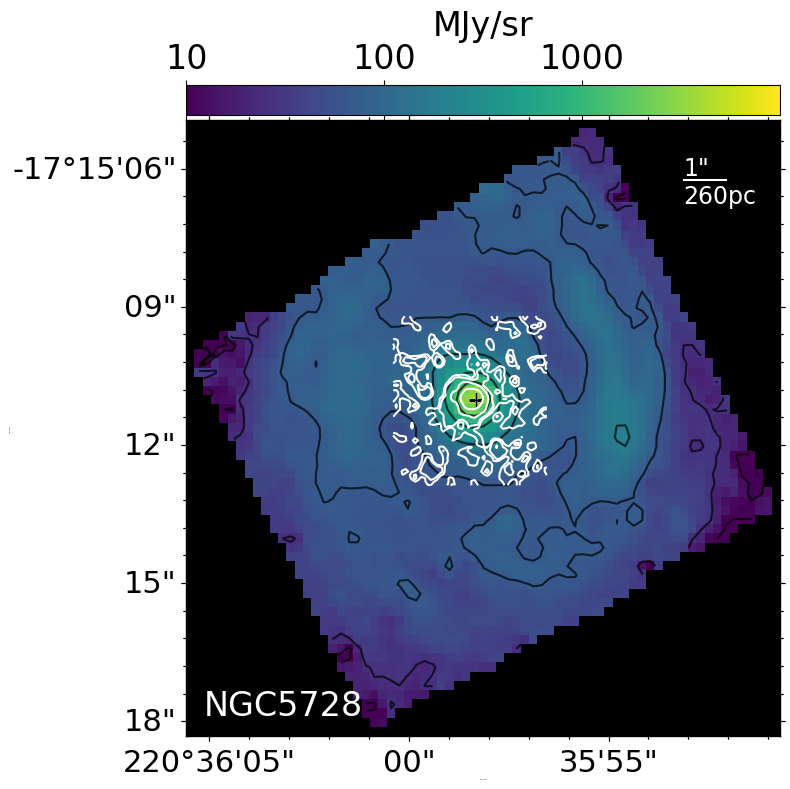}
\includegraphics[width=0.5\columnwidth]{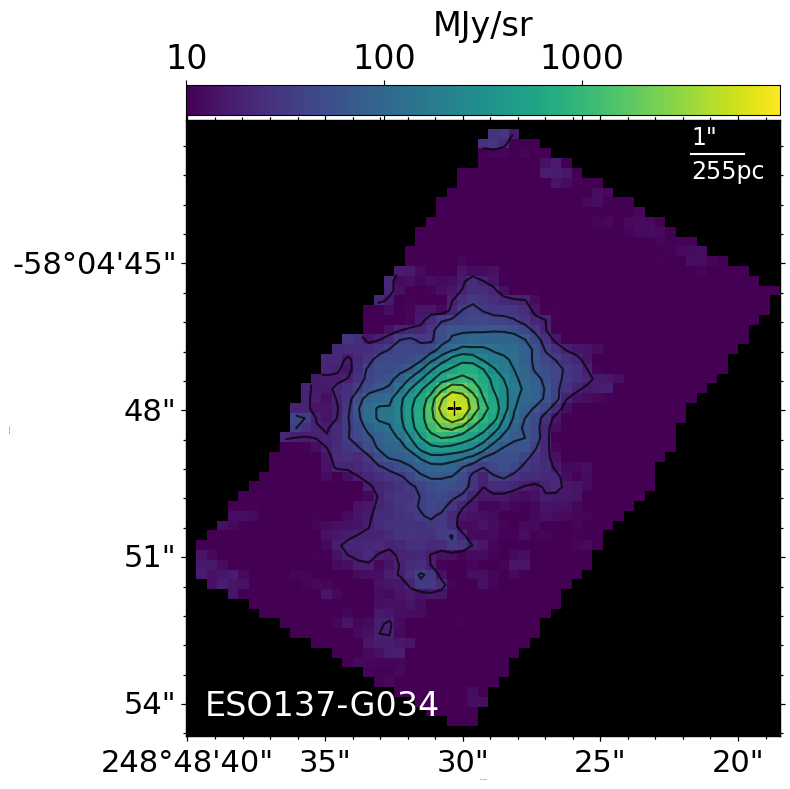}
\includegraphics[width=0.5\columnwidth]{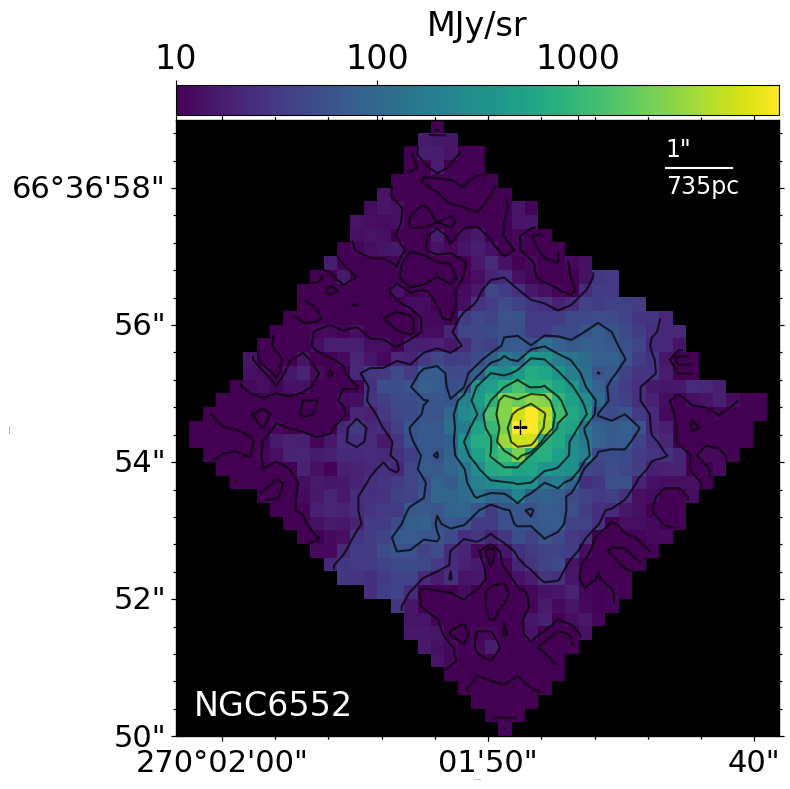}
\includegraphics[width=0.5\columnwidth]{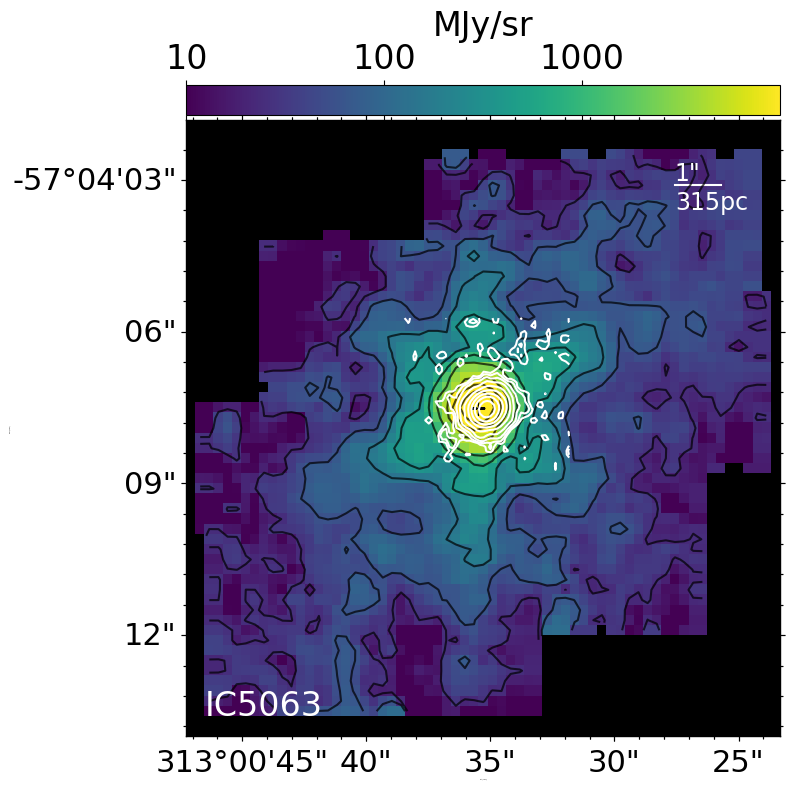}
\includegraphics[width=0.5\columnwidth]{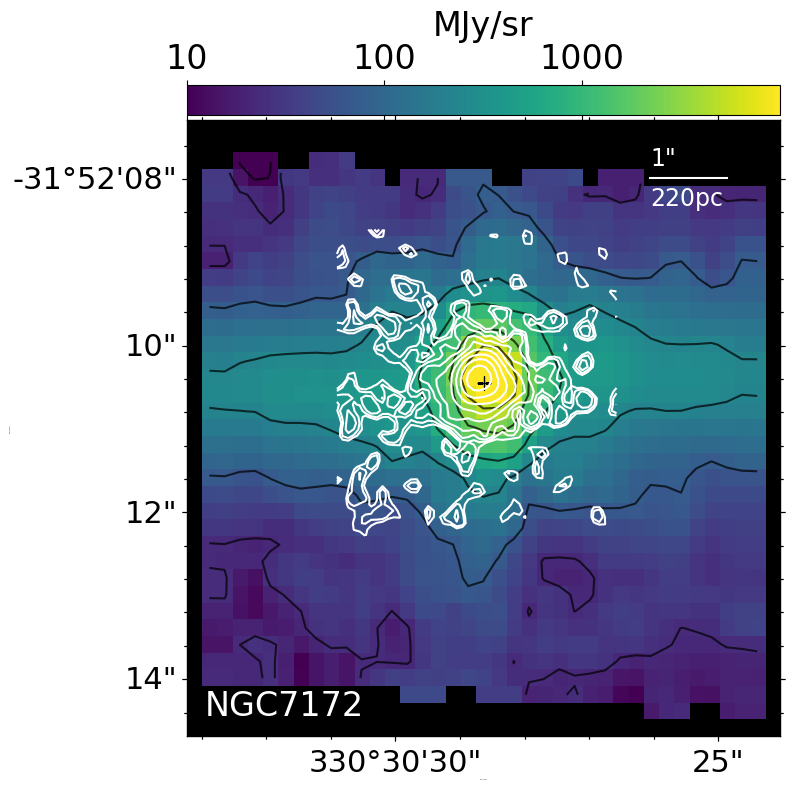}
\includegraphics[width=0.5\columnwidth]{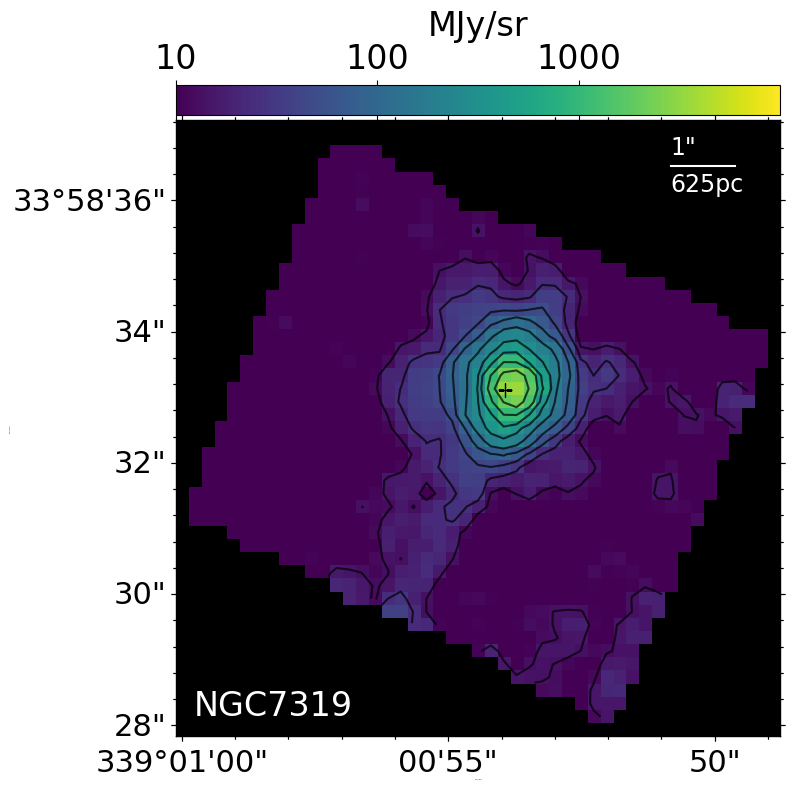}
\includegraphics[width=0.5\columnwidth]{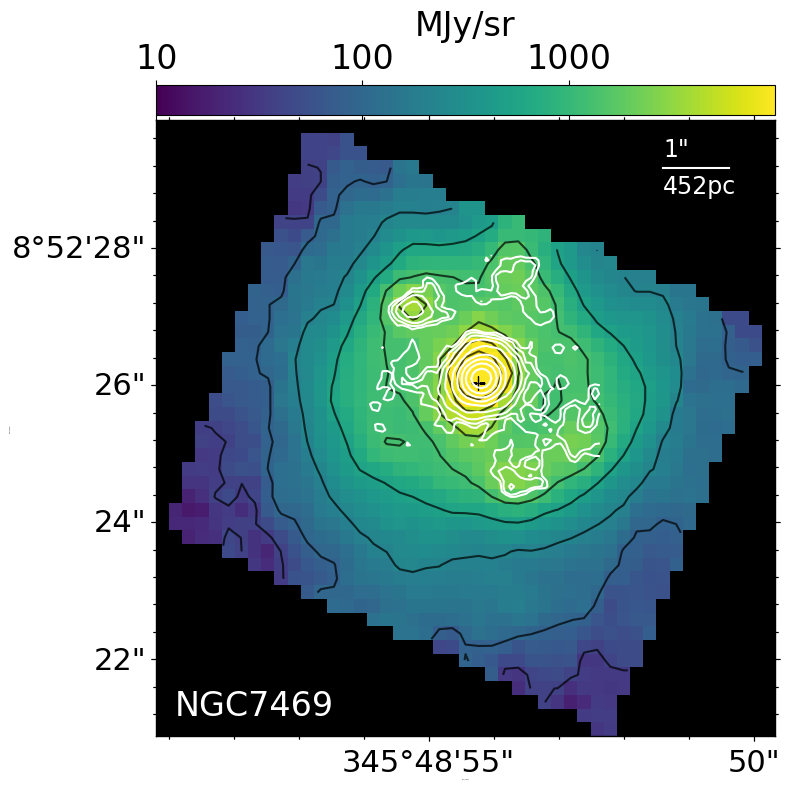}
\caption{MIRI/MRS images at 12$\rm{\mu m}$. Object names are written in the bottom-left corner of each panel. Black contours show any detection above the 3$\sigma$ level (10 levels up to the maximum of the image). White contours show the comparison with ground-based N-band mid-infrared observations when available. The top-right corner shows the 1-arcsec spatial scale and its physical correspondence for each target in parsec units. The black cross identifies the location of the center. \label{fig:12umsample}}
\end{center}
\end{figure*}

Fig.\,\ref{fig:12umsample} shows the slice at 12$\rm{\mu m}$ from the MIRI/MRS data cubes for each of our targets. Note that all images presented here are shown in a single slice. We found publicly available ground-based mid-infrared observations for 14 of the 21 objects in the SASMIRALA database\footnote{http://dc.zah.uni-heidelberg.de/sasmirala/q/cone/form}. We chose the observation at a filter closer to 12$\rm{\mu m}$ when more than one observation is available. Table\,\ref{tab:collection} (Col.\,14) shows the 12\,$\rm{\mu m}$ flux (in mJy) from the SASMIRALA database \citep{Asmus14} and, when available, these images are overlaid as white contours in Fig.\,\ref{fig:12umsample}. Note that the Mrk\,273/Mrk\,273SW system is shown in a single panel. Extended emission is recovered in most of the targets. For instance, a spiral structure is visible in ESO\,420-G13, which is also visible with CO observations \citep[shown in][]{Fernandez-Ontiveros20} and NGC\,5728 \citep[shown in][]{Davies24} and a circumnuclear ring is visible for NGC\,7469 \citep[shown in][]{Garcia-Bernete22A,Armus23,Zhang23}. This visual inspection of the data cubes already shows strong spikes associated with the MIRI PSF, which sometimes hinders any possible detection of the extended emission. This is the case for MCG\,05-23-16 reported by \citet{Esparza-Arredondo24} and Mrk\,231 by \citet{Alonso-Herrero24}. %\citep[e.g., MCG\,05-23-16 reported by][]{Esparza-Arredondo24}. 
It is also worth noticing how some objects show a weak central source embedded in a strong diffuse component. Examples of that are NGC\,3256 and NGC\,4594 \citep[shown in][]{Goold24} and NGC\,4736. 

\section{Data acquisition and general processing} \label{sec:processing}

All observations were taken using all four channels (ch1-4) and three sub-channels (short, medium, and long) to cover the full mid-infrared wavelength range; for ch1 (4.9–7.65\,$\rm{\mu m}$) and ch2 (7.51–11.71\,$\rm{\mu m}$) the MIRIFU SHORT Detector is used, while for ch3 (11.55–17.98\,$\rm{\mu m}$) and ch4 (17.71–27.9\,$\rm{\mu m}$) the MIRIFU LONG Detector is used. Each channel has an increasing field of view (FoV): ch1 (3.2$\rm{\times}$3.7 arcsec), ch2 (4.0$\rm{\times}$4.8 arcsec), ch3 (5.2$\rm{\times}$6.2 arcsec), and ch4 (6.6$\rm{\times}$7.7 arcsec) and pixel size: ch1 and ch2 with 0.196 arcsec/pixel, ch3 with 0.245 arcsec/pixel, and ch4 with 0.273 arcsec/pixel \citep[][]{Argyriou23}. Note that the ‘multi-channel’ drizzling option within the \emph{JWST} pipeline allows the linear spatial sampling to be set to the ideal value for the shortest wavelengths and provides roughly the Nyquist sampling of the spectral resolving power at all wavelengths \citep[see Fig.\,6 in][]{Law23}. Background exposures were taken using offset blank fields except for NGC\,6552 because this object was observed as part of the commissioning, and, therefore, the goal was not to obtain a final optimized datacube. For the same reason, MRS observations of NGC\,6552 were taken with no dithering, resulting in a small number of spaxels affected by the lack of data due to bad or hot pixels in the detector \citep[][]{Garcia-Bernete22A,Alvarez-Marquez23}. Our postprocessing analysis mitigates this issue (see Section\,\ref{sec:decomposition}).

After the search through the MAST archive website, publicly available uncalibrated science and background observations (i.e., raw data) were downloaded using the MAST Portal through the {\sc jwst\_mast\_query} package, which uses the MAST API to query the MAST archive. Observations in the proprietary period were provided by the GATOS collaboration.

Although the MAST archive provides reprocessed data and GATOS collaboration has reprocessed their observations, we self-consistently reprocessed all raw data using the same procedure and version of the \emph{JWST} pipeline. This guarantees a uniform quality of the resultant data cubes for all the observations analyzed here. We process the raw observations through version 1.14.1 of the \emph{JWST} pipeline \citep{Bushouse24} released on 2024-03-29 (calibration context 1210) and run within Python version 3.9.7.

The calibration of the data is divided within the pipeline into three main stages of processing: {\sc detector1}, {\sc spec2}, and {\sc spec3}. The {\sc detector1} pipeline applies detector-level corrections and ramp fitting to the individual exposures. The output rate images were subsequently processed outside the \emph{JWST} pipeline to flag newly-acquired bad pixels and additional cosmic-ray artifacts and remove vertical stripes and zero-point residuals after the pipeline dark subtraction. These additional corrections broadly follow the steps taken for \emph{JWST} observations described by \citet{Pontoppidan22} \citep[see also][]{Labiano16}. The resulting rate files are then processed with the \emph{JWST} {\sc spec2} pipeline for distortion and wavelength calibration, flux calibration, and other 2D detector-level steps. The cube-building step in {\sc spec3} pipeline assembles a single 3D data cube from all the individual 2D calibrated detector images, combining data across individual wavelength bands and channels. Residual fringe corrections using prototype pipeline code have been applied to the Stage 2 products and the 3D data cube resulting from the {\sc spec3} pipeline, incorporated in the most recent version of \emph{JWST} pipeline. The resulting data products generated by the pipeline are 12 fringe-corrected, wavelength, and flux-calibrated data cubes, one for every combination of channel and grating settings. 

We then applied background subtraction using the blank field observed before or after each target. Note that we did not find background observation for NGC\,6552 \citep[][]{Alvarez-Marquez23}. We used the background taken for NGC\,7469 in this case. Note that other backgrounds from our target collection were also tested, and no noticeable differences were found for the nuclear spectrum (there might be differences in the extended emission that do not affect the analysis performed here). The master background is a 1D median sigma-clipped spectrum calculated over the FOV of the background observations. This value is subtracted from the entire 2D detector array at each spaxel. 

\begin{figure*}
\includegraphics[width=2.\columnwidth,trim={0cm 0cm 0cm 0cm},clip]{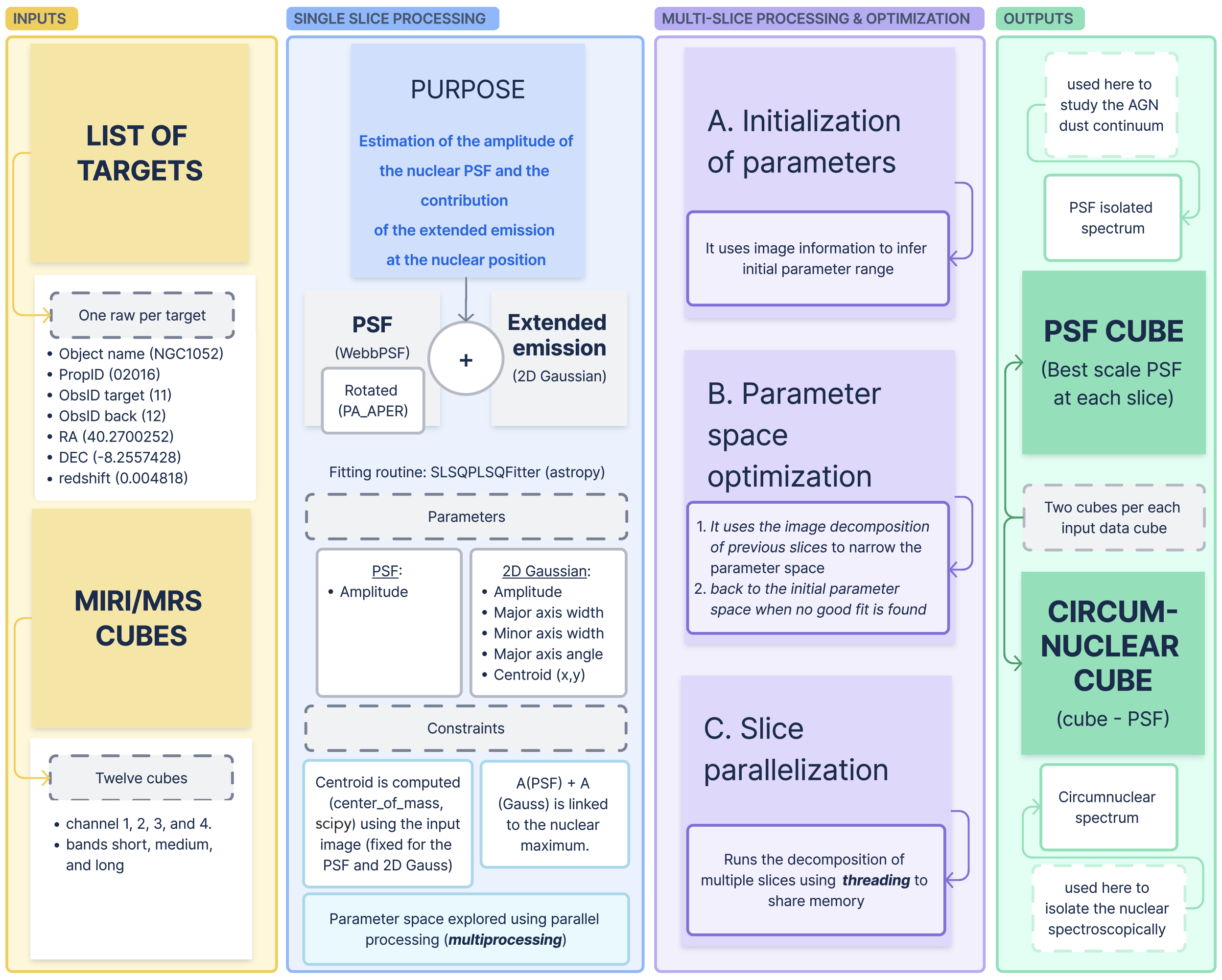}
\label{fig:pipelinescketch}
\caption{Sketch of the MIRIMRSisol tool. From left to right, inputs (yellow), single-slice decomposition (blue), multi-slice processing and optimization (violet), and outputs (green) details are given (see Section\,\ref{sec:decomposition} for more information).}
\end{figure*}

\section{Decomposition technique} \label{sec:decomposition}

Previous works have used deconvolution techniques to the MIRI images when a bright point source outshines the MIRI FOV to recover the extended emission \citep{Leist24, Haidar24}. Complementary to this, the main goal of the present analysis is to subtract the nuclear-unresolved component from each image plane (i.e., at each wavelength) of the data cubes so we can obtain the isolated, unresolved nuclear spectrum. The byproduct of this procedure is a data cube where the nuclear flux is subtracted (henceforth named circumnuclear data cube) and a data cube of the contribution of the PSF to the original data cube (henceforth named PSF data cube). As explained in Section\,\ref{sec:extraction}, the final isolation of AGN dust is performed in two steps. First, we use this technique to produce the PSF spectrum from the PSF data cube. Second, we use the circumnuclear data cube to create a spectral template of the circumnuclear emission to decontaminate the AGN dust continuum from other contributors. Therefore, the PSF data cube is used in the subsequent analysis to perform spectral fitting to the isolated nuclear component, and the circumnuclear data cube is used to produce a template of the circumnuclear emission to isolate the nuclear spectrum further when performing the spectral fittings. These circumnuclear data cubes could be used for other purposes. For instance, they can help to study the kinematics of the host/circumnuclear environment after removing the contribution of nuclear flux for the emission lines used to derive these kinematic properties. This might be particularly useful when the AGN is so bright that it outshines the feeble signatures of the host galaxy. 

Our approach relies on previous ground-based treatments of mid-infrared images. Nuclear isolation has been developed in the AGN field to quantify the unresolved nuclear flux and to allow the study of the extended emission for bright sources \citep{Radomski08, Ramos-Almeida14, Garcia-Bernete15, Garcia-Bernete16, Asmus16, Gonzalez-Martin17}. Most of these studies use a 2D Gaussian profile to mimic the instrumental PSF. Furthermore, the amplitude of this 2D Gaussian component is scaled in most cases to an arbitrary fraction of the maximum flux (typically 90-95\% of the maximum flux) of the image at the nuclear position. The technique applied here is based on the treatment used by \citet{Gonzalez-Martin17} for ground-based mid-infrared observations of faint AGN. \citet{Gonzalez-Martin17} models the image as two components, one for the nuclear and another for the extended emission, to estimate the fractional contribution of the nuclear component by minimizing the residuals of the fit. The nuclear and circumnuclear components are mimicked with 2D Gaussian profiles. Although simplistic, this method has been proven to work fine for mid-infrared images, X-ray images obtained with \emph{Chandra} \citep{Osorio-Clavijo23}, optical imaging \citep[e.g.][]{Garcia-Lorenzo05, Husemann14}, and at sub-mm wavelengths with ALMA observations \citep[][]{Garcia-Burillo21}.

Several peculiarities of applying this technique to MIRI/MRS data cubes are discussed below, including the complexity of the PSF shape, minimization algorithm, determination of the centroid of the nuclear component, initialization of the range parameters of the fitting, and optimization to speed up the computational time. We created a tool called MRSPSFisol, written in Python, to be distributed upon request. MRSPSFisol runs through a list of targets included in an input file the user should provide in the command line and other options. See Appendix\,\ref{app:running} for the details on how to run the code. Fig.\,\ref{fig:pipelinescketch} gives a view of all the critical aspects of the MRSPSFisol tool.

\subsection{\emph{JWST} PSF}

{Using a 2D Gaussian profile for the MIRI PSF in MIRI/MRS data is far from a good approximation}. The actual MIRI PSF shows diffraction spikes that might extend the entire FOV of the MRS mode \citep[see {Mrk\,231 in} Fig.\,\ref{fig:12umsample} and also][]{Leist24}. In this work, we use the MIRI PSF simulations taken from the WebbPSF tool\footnote{https://github.com/spacetelescope/webbpsf} (version v1.2.1) to produce 12 data cubes of the expected PSF. {This simulated PSF clearly shows the actual spikes seen in the MIRI/MRS data cubes (see, for instance, the PSF shown in Fig.\,\ref{fig:decompspec} for NGC\,6552 at 11.6$\rm{\mu m}$}). This follows previous works using the observed PSF to decompose the images \citep[e.g.][]{Radomski03,Lopez-Rodriguez18}. These cubes are included within the tool. 

Furthermore, the spikes of \emph{JWST} PSF are artifacts due to the hexagonal mirrors of \emph{JWST}. These spikes are rotated into different angles at the MIRI/MRS FOV, depending on the rotation of the observatory. This angle is reported in the $PA\_APER$ keyword in the observation headers, which is the angle compared to the north when projected on the sky. Simulated PSF, on the other hand, are obtained for non-rotated setups. To overcome this issue, the MRSPSFisol tool uses the $PA\_APER$ keyword to rotate the simulated PSF before fitting. This procedure includes subsampling the pixel size to avoid losing information when rotating the PSF image. The MRSPSFisol tool returns to the original pixel size before fitting. {We confirmed that any other rotation of the PSF produces worse results for all the objects in our sample.}  

We tried using empirical PSF data cubes from observed stars (in particular HD\,163466, A-type, PropIDs 1050, 1536, and 4499, observed seven times with MIRI/MRS). However, the SNR of the star at long wavelengths is not enough to use it as a PSF template. Furthermore, differences in the background of the star and the target complicate the procedure, resulting in systematic oversubtraction of the source. We acknowledge here that the simulated PSF from WebbPSF shows departures from the real one. In particular, at short wavelengths, the observed PSF has a preferential axis along the slit axis not seen in the simulated PSF. We expect the simulated PSF to change when tested against observations. Although these tests are out of the scope of this research, the ancillary data cubes of the simulated PSF will be updated within the pipeline when a significant update of the WebbPSF simulations tool is made.

\subsection{General fitting procedure} 

The PSF is fitted together with the extended emission:
\begin{equation}
    F_{s_i}(x,y) = A_{PSF}\,PSF_{s_i}(x,y) + A_{Gauss} Gaussian_{s_i}(x,y)
\end{equation}

\noindent where $\rm{s_i}$ denotes the decomposition applied to the slice $i$. $\rm{A_{PSF}}$ and $\rm{A_{Gauss}}$ are the amplitured of the PSF and 2D Gaussian profiles, respectively. 

This method can subtract multiple point-like sources by subtracting the source with the highest flux and using the subtracted data cubes as inputs for the subsequent subtraction (see Section\,\ref{sec:results}). To do so, it subtracts the PSF fixed at the center of mass/brightness of the target and scales to a certain amplitude (this amplitude is computed to minimize residuals; see below). The subtracted image is fitted to a 2D Gaussian profile using the SLSQPLSQFitter task within {\sc astropy}. This function minimizes the residuals to obtain the best fit. This minimization process is optimized in a small region of the image covering 60\% of the expected width of nuclear PSF and in a large area up to the maximum radius allowed by the data from the target position. 

The code is optimized to accurately estimate the amplitude of the PSF ($\rm{A_{PSF}}$) because it is the one needed to produce the PSF and circumnuclear map (i.e., PSF subtracted) by minimizing the residuals in the inner regions to compute the fractional contribution of the PSF to the central source. Therefore, the technique works well irrespective of the assumed shape for the extended emission as long as it allows the determination of this fractional contribution. The code iterates along a grid of parameter values to ensure the global minimum is found (see Section\,\ref{subsec:parameterrange}). The following subsections give the details of this general procedure. Note that all these considerations happen in the MRSPSFisol tool at the same time.

\subsection{Nuclear centroid}

The center position of the PSF and 2D Gaussian (used to model the extended emission) profiles are fixed to the center of mass/brightness of the system ($center\_of\_mass$ within {\sc scipy}), computed within the maximum radius from the input position of the target and the border of the MRS FOV. This process is iterated several times, and the new position is selected as the input position so that the final center of mass is accurate. {Once the position of the centroid is determined for each data cube, it is fixed for all the slices in the data cube.} This step could be avoided (by choosing \emph{recent=False}, it is set to true by default). If avoided, the central position is fixed to the position of the target chosen by the user and written in the input file (see Appendix\,\ref{app:running}). Note that using the pixel with the maximum flux is not recommended because it maximizes distortion patterns (such as fringing) when decomposition is applied. We realize that the extended emission could have complexity that could benefit from allowing the central position of the 2D Gaussian to vary. However, the method is optimized to estimate the PSF contribution at the nuclear position to produce accurate PSF and circumnuclear data cubes. 

Within our sample, we avoided the use of the automatic center of mass for the low-luminosity AGN NGC\,4594, NGC\,4736, and NGC\,5728 and the merger system Mrk\,273/Mrk\,273SW. The nuclei of NGC\,4594 and NGC\,4736 are embedded in a strong circumnuclear environment that dominates the total flux. In these cases, the center of mass tends to reflect the extended emission instead of the nuclear position (see Fig.\,\ref{fig:12umsample}). A slight shift of the center of mass from the actual center is found in NGC\,5728, maybe due to an asymmetric extended structure or due to the fact the nucleus is not at the center of the MRS FOV (slight distortions of the PSF are expected when the source is in the corner of the detector). NGC\,5728 shows a complex structure within the inner 1\,kpc, with a large stellar bar and prominent dust lanes settling gas into a circumnuclear ring \citep[][]{Shimizu19, Garcia-Bernete24B}. Furthermore, \citet{Davies24} noticed a shift in the nuclear position due to the broad silicate absorption feature at the nucleus. Mrk\,273/Mrk\,273SW system shows two nuclei where the center of mass reflects none of the two nuclei. 

\subsection{Free parameters and initial values}\label{subsec:freepar}

The PSF has one free parameter left, which is the amplitude of the PSF ($\rm{A_{PSF}}$). The 2D Gaussian (i.e., elliptical Gaussian) component has the following free parameters: amplitude ($\rm{A_{Gauss}}$) widths ($\rm{[\sigma_x,\sigma_y]_{Gauss}}$), and long-axis position angle ($\rm{\Theta_{Gauss}}$). The observations are used to determine the initial ranges of these parameters. Both amplitudes are set to zero up to the maximum of each image slice. To optimize the parameter space, the code only tests combinations of amplitudes where the sum of the PSF and 2D Gaussian amplitudes is between 80-120\% of the maximum of the {source}. Percentages below and above this range will underpredict and overpredict the actual maximum of the image. Although, in principle, this could be fixed to 100\%, allowing this range helps find the best PSF amplitude due to the inherent difference between the simulated and the actual PSF. The range of 2D Gaussian widths is set between 1.2 times the expected width of the PSF at each wavelength (to avoid confusion between the wings of the PSF and the extended component) and the maximum radius allowed by the observation from the target. We used the MRS PSF models as a first guess of the minimum width of the 2D Gaussian profile. This information is enclosed in the CRDS calibration file of the aperture calibration \footnote{Aperute calibration file: jwst\_miri\_apcorr\_0008.asdf} obtained using in-flight data based on observations of multiple standard stars observed throughout Cycle 1 \citep{Argyriou23}. Only values where ${\sigma_x > \sigma_y}$ are tested with long-axis angles $\rm{\Theta_{Gauss}<180^{\circ}}$ to optimize the parameter space (i.e., remove redundant models). Each parameter grid runs entirely in parallel thanks to the {\sc multiprocessing} package within Python.  

\begin{figure*}
\includegraphics[width=2.1\columnwidth,trim={0.cm 0cm 0.cm 0cm},clip]{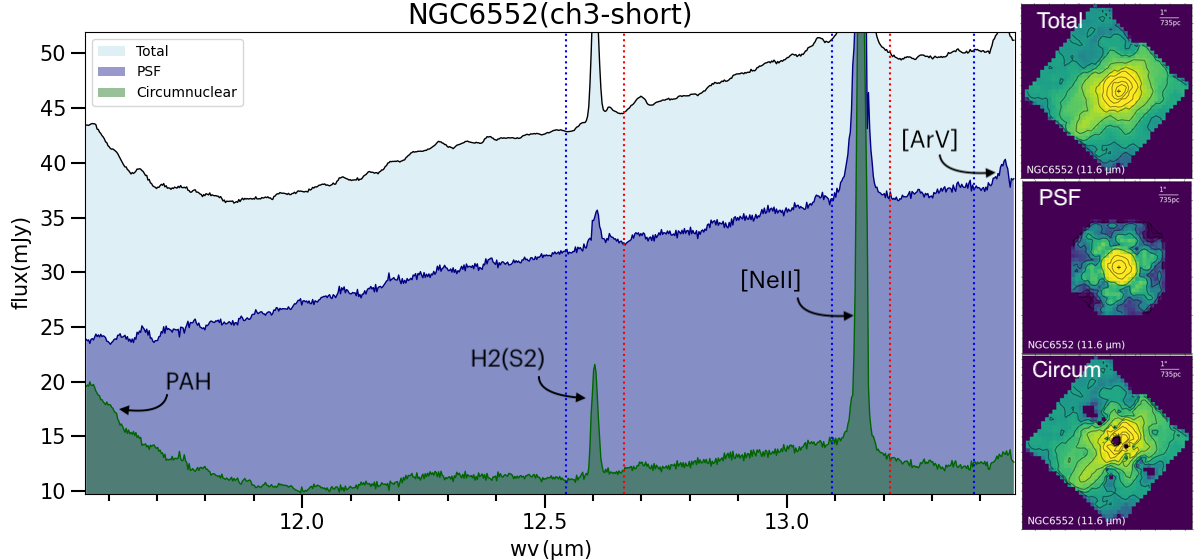}
\caption{Resulting spectra for the total (black line filled in light-blue color), PSF (blue line and filled in blue color), and extended (green and filled in green) for the channel 3 (short band) data cube of NGC\,6552. All of them are extracted from an aperture of eight pixels (see text). Emission lines expected in the wavelength range are enclosed between vertical blue and red dashed lines. The right panels show the total (top), PSF (middle), and Circumnuclear (bottom) image centered at 11.6$\rm{\mu m}$. \label{fig:decompspec}} 
\end{figure*}

\subsection{Parameter optimization and speed-up procedures} \label{subsec:parameterrange}

These initial ranges of parameters are particularly time-consuming because of the large number of iterations needed to cover the entire grid. For this reason, when the MRSPSFisol tool has enough slices already fitted, the range of parameters is set to the mean value and two times the standard deviation from previous fits to reduce the number of iterations performed for each 2D image slice of the data cube. In practice, it performs this new range based on the first four slices when available, changes to the first 15 slices, and finally estimates the parameter ranges based on the last 75 slices for the rest of the data cube. At the same time, it also starts to parallel the process, computing 3, 5, and 25 slices in parallel as soon as it reaches 4, 15, and 75 slices already performed. Given the nature of the algorithm, we implemented two levels of parallelization. At the first level, we prioritized I/O performance and needed the threads to share memory, so we used \textit{threading}. For the second level, where memory sharing between processes was not required, we chose process-level parallelization using the \textit{multiprocessing} package. As Section\,\ref{subsec:freepar} mentions, the code also uses parallel processing to test the range of parameters for each spectral fit performed. Therefore, it performs nested parallel processing, i.e., several slices simultaneously and multiple fittings for each of these slices. 

This narrow range of parameters speeds up the process, but it is not accurate for abrupt changes in the morphological structure. An example is when the MRSPSFisol tool approaches emission lines where the amplitude of any of the two components might abruptly change (and perhaps the morphology of the extended emission also abruptly changes). The parameters are reset to the initial parameter ranges when no fit is found to overcome this issue. The MRSPSFisol tool also automatically redefines any parameter range when the best-fit parameter is found at the border of the range tested. In practice, steep spectral slopes require more time because they usually require the resetting of parameters (mainly the amplitude), and all emission lines are performed with the initial conditions to account for the abrupt changes in amplitude and morphology. 

\subsection{Products and outputs}

The MRSPSFisol tool produces a figure showing the resulting spectral decomposition of the PSF and extended emission in a standard aperture of 8 pixels {centred} at the center of mass/brightness of the target for each data cube. Note that this aperture is performed only for a preliminary view of the spectral decomposition. Fig.\,\ref{fig:decompspec} shows the resulting decomposition of NGC\,6552 for channel 3 (short band) as an example of this by-product. This output is updated along the MRSPSFisol tool procedure to visualize how the decomposition is performed. Note that this plot shows several expected emission lines (marked with red- and blue-dashed vertical lines) for visual inspection if the code can recover the decomposition in such abrupt changes of space parameters. This plot already shows how the tail of the PAH feature at 11.3$\rm{\mu m}$ is associated with the extended emission for this particular source, and $\rm{H_{2}S(2)}$, [NeII], and [ArV] (see Table\,\ref{tab:emissionlines}) emission lines show a contribution from both nuclear and extended emission (although [ArV] is mainly associated to the nuclear spectrum). The final scientific outputs of the MRSPSFisol tool are three data cubes: the PSF map (simulated PSF scaled to the best amplitude for each slice), the circumnuclear map (original map where the PSF is subtracted), and the model (best-fit model consisting of the sum of the PSF and the 2D Gaussian best fitted at each slice).

\begin{table}
	\centering
    \footnotesize
    \renewcommand{\tabcolsep}{0.07cm}
	\caption{Spatial scales (radius) for spectral extractions}
	\label{tab:spatialscales}
	\begin{tabular}{lccc} 
		\hline \hline
{Name}       &   {Total}     &  {Nuclear}    &  {Circumnuclear} \\ %&    \colhead{\# SEDs}    
{}       &   {("/parsec)}     &  {("/parsec)}    &  {("/parsec)} \\ %&    \colhead{\# SEDs}    
(1) & (2) & (3) & (4) \\
\hline
NGC1052 & 0.8-3.2 / 107-421 & 0.4-1.6 / 53-210  & 3.6 / 514 \\
ESO420-G13 & 0.8-3.2 / 264-1040 & 0.4-1.6 / 132-520  & 6.2 / 1616 \\
UGC05101 & 0.7-2.6 / 728-2865 & 0.4-1.6 / 437-1719  & 2.4 / 3815 \\
MCG-05-23-016 & 0.8-3.2 / 188-741 & 0.4-1.6 / 94-371  & 3.8 / 823 \\
NGC3031 & 0.8-3.2 / 17-67 & 0.4-1.6 / 9-34  & 3.8 / 75 \\
NGC3081 & 0.8-3.2 / 177-697 & 0.4-1.6 / 89-348  & 6.8 / 1082 \\
NGC3256NUC1 & 0.8-3.2 / 208-817 & 0.4-1.6 / 104-408  & 3.6 / 906 \\
% NGC3256NUC2 & 0.8-3.2 / 208-817 & 0.4-1.6 / 104-408  & 3.8 / 906 \\
NGC4395 & 0.5-2.1 / 16-62 & 0.4-1.6 / 12-46  & 2.4 / 103 \\
NGC4594 & 0.8-3.2 / 81-318 & 0.4-1.6 / 40-159  & 3.6 / 353 \\
NGC4736 & 0.8-3.2 / 23-90 & 0.4-1.6 / 11-45  & 3.8 / 100 \\
Mrk231 & 0.8-3.2 / 936-3683 & 0.4-1.6 / 468-1841  & 3.6 / 4086 \\
Mrk273SW & 0.5-2.1 / 553-2174 & 0.4-1.6 / 414-1630  & 1.8 / 3618 \\
Mrk273 & 0.8-3.2 / 829-3261 & 0.4-1.6 / 414-1630  & 3.8 / 3618 \\
NGC5506 & 0.8-3.2 / 125-490 & 0.4-1.6 / 62-245  & 4.6 / 653 \\
NGC5728 & 0.8-3.2 / 208-818 & 0.4-1.6 / 104-409  & 6.8 / 1271 \\
ESO137-G034 & 0.8-3.2 / 203-799 & 0.4-1.6 / 102-399  & 3.8 / 886 \\
NGC6552 & 0.7-2.6 / 490-1928 & 0.4-1.6 / 294-1157  & 2.8 / 2054 \\
IC5063 & 0.8-3.2 / 252-991 & 0.4-1.6 / 126-496  & 5.2 / 1430 \\
NGC7172 & 0.8-3.2 / 176-691 & 0.4-1.6 / 88-346  & 3.8 / 767 \\
NGC7319 & 0.8-3.2 / 500-1966 & 0.4-1.6 / 250-983  & 3.6 / 2181 \\
NGC7469 & 0.8-3.2 / 361-1421 & 0.4-1.6 / 181-710  & 3.6 / 1576 \\
		\hline \hline
	\end{tabular}
\end{table}

\section{Spectral extraction} \label{sec:extraction}

The \emph{JWST} pipeline can perform spectral extractions. The MIRI MRS documentation shows that optimizing the center where the nuclear spectrum is extracted for each data cube in point-like sources is critical. This is done with high accuracy when the nuclear point-like source dominates. However, these capabilities of the \emph{JWST} pipeline are not optimized for sources highly embedded in diffuse emission or for the circumnuclear map obtained throughout our decomposition technique where the nuclear PSF is removed from the original data cube (see Section\,\ref{sec:decomposition}). Therefore, due to the peculiarities of our analysis, we performed our spectral extractions and stitched the individual segments of the spectra into a single spectrum covering the full 5-27$\rm{\mu m}$ spectral range using general Python routines. This way, we perform the same extraction (described below) centered on the nuclear position obtained from the decomposition technique for all our data products (original, PSF, and circumnuclear data cubes). The nuclear positions found with the \emph{JWST} pipeline and MRSPSFisol agree for bright point-like AGN. As expected, this is not the case for the circumnuclear spectrum, faint sources with strong diffuse emission, or double sources where the true center of mass of each object is not easily computed.

We performed four different 1D spectral extractions using these data cubes, called below total, nuclear, PSF, and circumnuclear spectra. Table\,\ref{tab:spatialscales} gives each spectral extraction and the corresponding physical scales in parsecs. We compared the results of both the \emph{JWST} pipeline and our routines for the total spectrum, showing that the final spectra are the same within the 3\% of flux (this is not the case for the circumnuclear maps due to the inability of the \emph{JWST} pipeline to find that centroid, see above). We use this percentage to estimate the flux error for spectral fitting. 

\subsection{Total spectrum}

We perform the extraction of the spectrum from the original data cubes (before the decomposition) at the position of the target obtained from our analysis with an aperture that depends on wavelength to account for the changing width of the PSF. This aperture is generally 0.8 arcsec in radius at 5$\rm{\mu m}$ (growing up to 3.2 arcsec at 27$\rm{\mu m}$). This corresponds to 2.7$\rm{\times FWHM(\lambda)}$\footnote{$\rm{FWHM(\lambda)=0.30\,arcsec}$ for $\rm{\lambda < 8 \mu m}$ and $\rm{FWHM(\lambda)=0.31 \times \lambda[\mu m]/8\,arcsec}$ for $\rm{\lambda > 8 \mu m}$\citep[][]{Rigby23}.}. It encircles well above 90\% of the PSF at all wavelengths. Since the central region contains a point source, we apply a wavelength-dependent aperture correction to this spectrum. We used the MRS PSF models to correct the aperture losses in the 1D spectra \citep{Argyriou23}\footnote{This information is enclosed in the CRDS calibration file jwst\_miri\_apcorr\_0008.asdf.}. There are three objects where we performed a slightly smaller aperture, namely UGC\,05101 (0.7$-$2.6 arcsec), Mrk\,273SW (0.5$-$2.1 arcsec), and NGC\,6552 (0.7$-$2.6 arcsec) because the location of the target within the MRS FOV prevents us from a larger aperture. These apertures translate into a wide range of physical scales, from 20$-$70\,pc for our nearest source (NGC\,3031) up to 0.9$-$3.7\,kpc for our farthest source (Mrk\,231). 

\subsection{Nuclear spectrum}

We performed a second extraction from the original data cubes, again at the position of the target, but now we are trying to minimize the contribution of the circumnuclear components. For that purpose, we extracted an aperture that starts at 0.4 arcsec in radius at 5$\rm{\mu m}$ and grows up to 1.6 arcsec at 27$\rm{\mu m}$. This corresponds to 1.33 times the FWHM at that wavelength. This aperture corresponds to spatial regions of 10-35\,pc for NGC 3031 up to 0.5-1.8\,kpc for Mrk\,231. Aperture corrections are also applied to the nuclear spectrum to correct for flux losses. Note that although there are aperture corrections for extractions as small as 0.3 arcsecs at 5$\rm{\mu m}$ (i.e., one time the FWHM at that wavelength), our tests comparing this aperture with several others for point-like sources show that this aperture correction is not as reliable as those using apertures beyond the one selected here (it over-predicts the flux at long wavelengths). Indeed, the percentage of flux lost for this aperture is 17\% for ch1 and increases up to 30\% in ch4. Therefore, an aperture of 1.33 times the FWHM at that wavelength is the smallest reliable aperture to optimize the isolation of the nucleus. 

\subsection{PSF spectrum}

We then performed the same nuclear aperture above in the PSF data cube obtained in our decomposition technique, i.e., we extracted an aperture that starts at 0.4 arcsecs at 5$\rm{\mu m}$ and grows up to 1.6 arcsecs at 27$\rm{\mu m}$. Aperture corrections are also applied to the {PSF spectrum} to correct for flux losses. 

\subsection{Circumnuclear spectrum}

Using the circumnuclear data cube, we extracted a spectrum in a circular aperture with a constant radius for all MRS channels. This aperture radius is delimited by the largest aperture allowed by the MRS FOV of channel 1 (short band), where the FOV is the smallest. Note that we also computed the circumnuclear spectrum with several smaller apertures, and the spectral shape does not depend on the maximum radius used. This aperture depends not only on the distance of the target but also on the position of the source within the FOV and the strategy used to perform the observations. In our sample, the smallest corresponds to NGC\,3031 (75\,pc), and the largest is Mrk\,231 (4\,kpc). This circumnuclear spectrum is used in our analysis to decontaminate the AGN spectrum from other contributors (see Section\,\ref{sec:spectralfitting}). 

\begin{table}
	\centering
    \footnotesize
    \renewcommand{\tabcolsep}{0.07cm}
	\caption{Summary AGN dust models}
	\label{tab:models}
	\begin{tabular}{lccc} 
		\hline \hline
{Name}       &   {Geometry}     &  {Distrib.}    &  {Reference} \\ %&    \colhead{\# SEDs}    
(1) & (2) & (3) & (4) \\ \hline
Fritz06    &   flared disk  & smooth       & \citet{Fritz06}           \\ %&    24,000     
Nenkova08  &   torus        & clumpy       & \citet{Nenkova08}         \\  %& 1,247,000     
Hoenig10   &   flared disk  & clumpy       & \citet{Hoenig10}          \\ %&     1,680     
Hoenig17   &   disk+wind    & clumpy       & \citet{Hoenig17}          \\ %&   132,300     
Hoenig17D  &   disk         & clumpy       & \citet{Hoenig17}          \\  %&               
Stalev16   &   flared disk  & two-phase    & \citet{Stalevski16}       \\ %&    19,200     
GoMar23    &   flared disk  & two-phase    & \citet{Gonzalez-Martin23} \\  %&   691,200     
		\hline \hline
	\end{tabular}
\end{table}

\section{Spectral fitting} \label{sec:spectralfitting}

%Once the nuclear spectrum is isolated using MRSPSFisol, we fit the PSF spectra to a set of available SED libraries of AGN dust models reported in the literature. 

Using the isolated PSF spectrum, we attempted to constrain the parameters of the seven AGN dust models listed in Table\,\ref{tab:models}. We assume that the unresolved component in the mid-infrared is dominated by the AGN dust emission (in the form of a disk, torus, or wind) \footnote{This is not the case for sub-mm observations where the colder dust is distributed in an extended disk \citep{Garcia-Burillo21}.}. Following the general procedure explained by \citet{Gonzalez-Martin19B}, we used the software XSPEC \citep{Arnaud96} to perform the spectral fitting. XSPEC is a command-driven, interactive, spectral-fitting program within the HEASOFT package and provides a wide range of tools to perform spectral fittings. Additionally, it can work in parallel processes to speed them up, which is ideal for our spectral fitting requirements. To do so, we converted the PSF spectra into XSPEC format using the {\sc flx2xsp} task within HEASOFT. XSPEC quickly reads these files to perform statistical tests when they fit the models. Note that errors need to be included along the spectral fitting procedure. Unfortunately, the error estimate for MIRI/MRS data cubes is an open issue that needs to be fixed in the \emph{JWST} pipeline, with the current derived quantity yielding errors much smaller than expected. {Instead, we used the PSF data cube to extract several spectra with different apertures and compute their standard deviation. We also compared our extracted spectrum from the original data cube with that of the JWST pipeline to quantify the errors associated with the extraction method. Although this needs to be addressed in future releases of the \emph{JWST} pipeline, we estimate an error corresponding to 3\% of the spectral flux after running these tests.} %, which seems to work well for our AGN collection. % Note that this error does not include any flux calibration error because this source of error should not be included when performing spectral fits. 

We confront this data set against seven models available in the literature, including morphologies (torus, disk, and wind) and dust distributions (smooth, clumpy, and two-phase) for the AGN dust. Table\,\ref{tab:models} summarizes the libraries of models used in this analysis. We refer to \citet{Gonzalez-Martin19A,Gonzalez-Martin19B,Martinez-Paredes21,Garcia-Bernete22B} for works where these models have been used before. They are already converted into XSPEC format by \citet{Gonzalez-Martin19A}, \citet{Garcia-Bernete22B}, and \citet{Gonzalez-Martin23}. Note that we do not attempt to go into the details of the resulting parameters of the models because the main scope of this analysis is to investigate if any of them satisfactorily fit the AGN collection in hand. Any derived property obtained from the best fits should be taken with caution because the sample is heterogeneous, and these models do not easily describe the complexity of the data, as we will show below. 

\begin{table}
	\centering
    \footnotesize
    \renewcommand{\tabcolsep}{0.07cm}
	\caption{Percentage of PSF contribution}
	\label{tab:percentage}
	\begin{tabular}{lc cccc c cccc} 
		\hline \hline
{Name} & {Type} &\multicolumn{4}{c}{PSF vs total \%} & & \multicolumn{4}{c}{PSF vs nuclear \%} \\ \cline{3-6}  \cline{8-11}
                &  & min   & max  & mean & $\rm{12\mu m}$ & & min  & max  & mean & $\rm{12\mu m}$ \\ \hline
NGC\,1052         & S1.9 & 54.5  & 100  & 88.4 & 87.5 & & 68.4 & 95.2 & 89.6 & 89.3 \\
UGC\,05101        & S1.5 & 28.7  & 100  & 82.5 & 69.5 & & 42.5 & 100  & 88.7 & 81.0 \\
NGC\,3081         & S1.8 & 47.6  & 81.7 & 75.8 & 77.7 & & 66.3 & 100  & 86.5 & 90.9 \\
NGC\,4395         & S1.8 & 49.5  & 100  & 94.7 & 92.4 & & 52.6 & 100  & 92.0 & 89.5 \\
Mrk\,231          & S1   & 78.7  & 99.2 & 95.4 & 95.8 & & 79.8 & 100  & 96.8 & 97.4 \\
NGC\,5506         & S1.9 & 43.1  & 85.8 & 81.1 & 80.7 & & 55.9 & 88.9 & 84.1 & 84.2 \\ 
NGC\,7469         & S1.5 & 13.1  & 79.7 & 58.3 & 63.3 & & 37.9 & 92.0 & 79.2 & 84.3 \\ \hline
ESO\,420-G13      & S2   & 16.0  & 60.9 & 46.1 & 49.3 & & 37.2 & 83.5 & 69.9 & 76.1 \\
MCG\,-05-23-016   & S2   & 66.5  & 95.4 & 88.4 & 86.5 & & 69.8 & 96.1 & 88.1 & 86.4 \\
NGC\,3031         & S2   & 42.9  & 81.8 & 69.5 & 70.0 & & 59.1 & 88.8 & 79.5 & 79.6 \\
NGC\,3256NUC1     & S2   & 24.8  & 74.1 & 42.3 & 35.6 & & 46.6 & 100  & 63.2 & 65.3 \\
NGC\,4594         & S2   &  4.6  & 48.8 & 24.0 & 17.3 & & 11.1 & 69.7 & 37.6 & 31.9 \\
NGC\,4736         & S2   &  3.0  & 23.9 & 15.7 & 8.0  & & 8.3 & 37.0 & 28.4 & 19.2 \\
Mrk\,273SW        & S2   &  8.3  & 70.6 & 29.9 & 38.4 & & 10.0 & 78.0 & 40.2 & 56.4 \\
Mrk\,273          & S2   &  2.7  & 73.3 & 50.9 & 33.3 & &  6.0 & 79.7 & 58.2 & 46.0 \\
NGC\,5728         & S2   & 20.2  & 79.0 & 63.1 & 56.9 & & 31.0 & 100  & 71.9 & 68.9 \\
ESO\,137-G034     & S2   & 15.2  & 72.8 & 53.8 & 59.6 & & 24.4 & 78.0 & 61.7 & 67.1 \\
NGC\,6552         & S2   & 45.6  & 91.2 & 77.2 & 73.8 & & 51.0 & 100  & 76.2 & 73.4 \\
IC\,5063          & S2   & 67.6  & 97.9 & 91.7 & 91.6 & & 76.4 & 100  & 93.0 & 91.3 \\
NGC\,7172         & S2   & 39.2  & 72.4 & 57.3 & 55.8 & & 45.7 & 76.6 & 63.4 & 62.7 \\
NGC\,7319         & S2   & 42.1  & 100  & 93.0 & 96.4 & & 57.4 & 100  & 91.3 & 95.0 \\
		\hline \hline
	\end{tabular}
\end{table}

For each model and spectrum, a set of variations of the baseline model is tested. 
\begin{itemize}
\item AGN dust model template: we use the AGN dust model with the only inclusion of foreground extinction by dust grains using the {\sc zdust} component \citep[][]{Pei92}, already included as a multiplicative component within the XSPEC software. {We use the extinction curve for the Small-Magellanic Cloud (SMC), although the extinction curves of the Galactic ISM or the Large-Magellanic Cloud produce similar best fits. The main difference using mid-infrared data is the resulting value for the color excess, $\rm{E(B-V)}$, which is significantly lower for the extinction curve of the SMC.} The free parameter is the color excess, $\rm{E(B-V)}$. The number of free parameters of the AGN dust models varies from four to seven parameters plus the normalization. The free parameters of each of the AGN dust models are described in \citet{Gonzalez-Martin23}. The main impact on the resulting spectrum of this extinction model is the attenuation of the near-infrared emission and the inclusion of silicate absorption features. 

\item AGN dust model plus circumnuclear contribution: we add the circumnuclear spectrum obtained for each MIRI/MRS observation as an additive model to account for any contribution of the host within the PSF spectrum. This could be particularly relevant for distant or intrinsically faint AGN, where the host galaxy might have a non-negligible contribution within the PSF extraction. All the circumnuclear spectra of objects in our AGN collection were converted into a model using the same methodology applied to the AGN dust models. The free parameter of the circumnuclear template is its normalization. 

\item AGN dust model plus absorption features: the water ice at around $\rm{\sim 5.8-6.2\mu m}$ and aliphatic hydrocarbon ices at $\rm{\sim 6.85}$ and $\rm{\sim 7.27\mu m}$ have been detected in extragalactic sources, mainly in sources with signatures of buried nuclei \citep[][Ramos Almeida et al. submitted]{Spoon01,Spoon22,Garcia-Bernete24A,Alonso-Herrero24}. We also included a test where five absorption features are included systematically in the analysis as Gaussian components with negative amplitudes. Three Gaussian components (at 5.83, 6.0, and 6.2$\rm{\mu m}$) mimic the water ice feature, and the other two Gaussian components are included to reproduce the aliphatic hydrocarbon ice features (at 6.85 and 7.27$\rm{\mu m}$). Centroids are fixed to these wavelengths, and widths can vary in the range 0.1-0.25$\rm{\mu m}$.

\item AGN dust model plus absorption features and circumnuclear contribution: This is the most complex template used, adding the absorption features as Gaussians and the host-galaxy template to each AGN dust model tested.  
\end{itemize}

%The AGN spectra in our sample show several emission lines, which cannot be included in the spectral fitting procedure. 

Since we are interested in fitting the AGN dust continuum, the emission lines (associated with winds or star-forming processes) are excluded by ignoring the specific wavelength range. Table\,\ref{tab:emissionlines} includes the list of emission lines excluded from the spectral fit. We used a width of 0.1, 0.16, and 0.24$\rm{\mu m}$ below 8, in the range of 8-12, and beyond 12$\rm{\mu m}$, respectively, to exclude these lines. PAH features are not excluded as they aid in finding the residual circumnuclear host-galaxy contribution within the PSF spectra. 

\begin{figure}
\includegraphics[width=0.49\columnwidth,trim={0.45cm 1cm 0.4cm 0cm},clip]{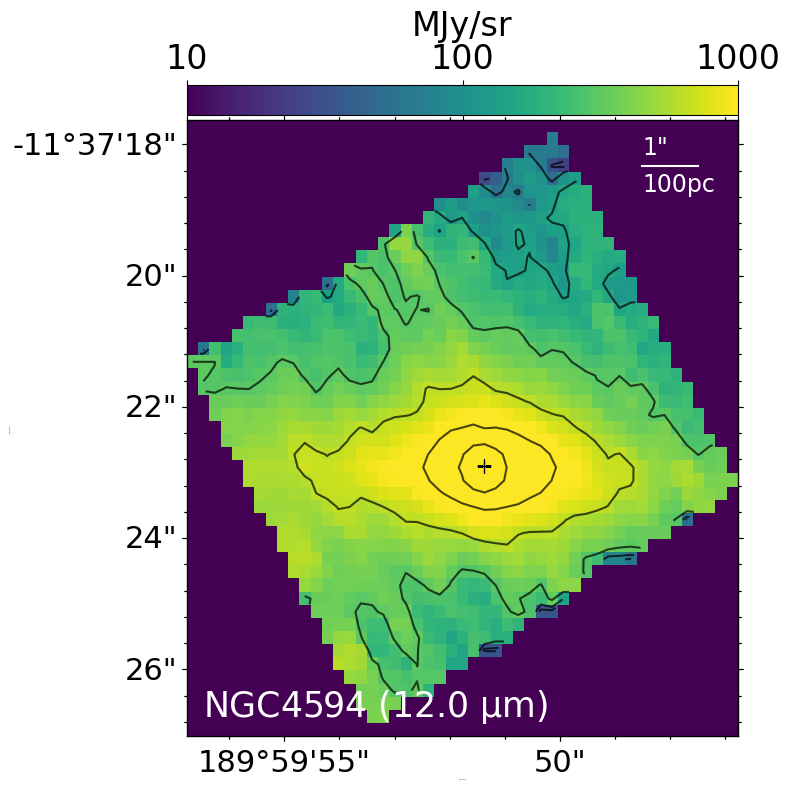}
\includegraphics[width=0.49\columnwidth,trim={0.45cm 1cm 0.cm 0cm},clip]{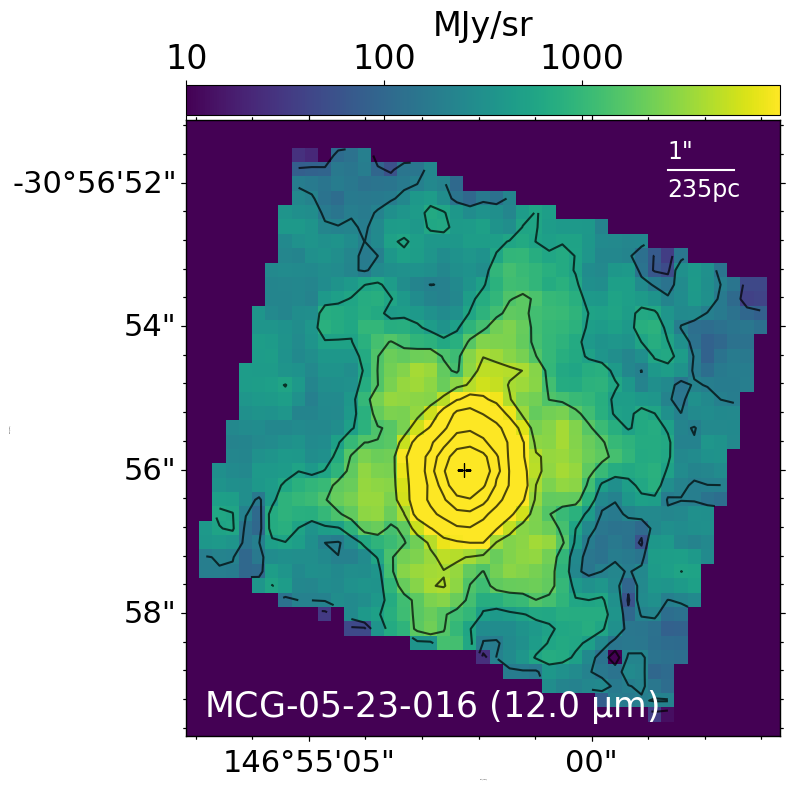} %{0cm 0cm 0cm 1.76cm}
\includegraphics[width=0.49\columnwidth,trim={0.45cm 1cm 0.4cm 2.95cm},clip]{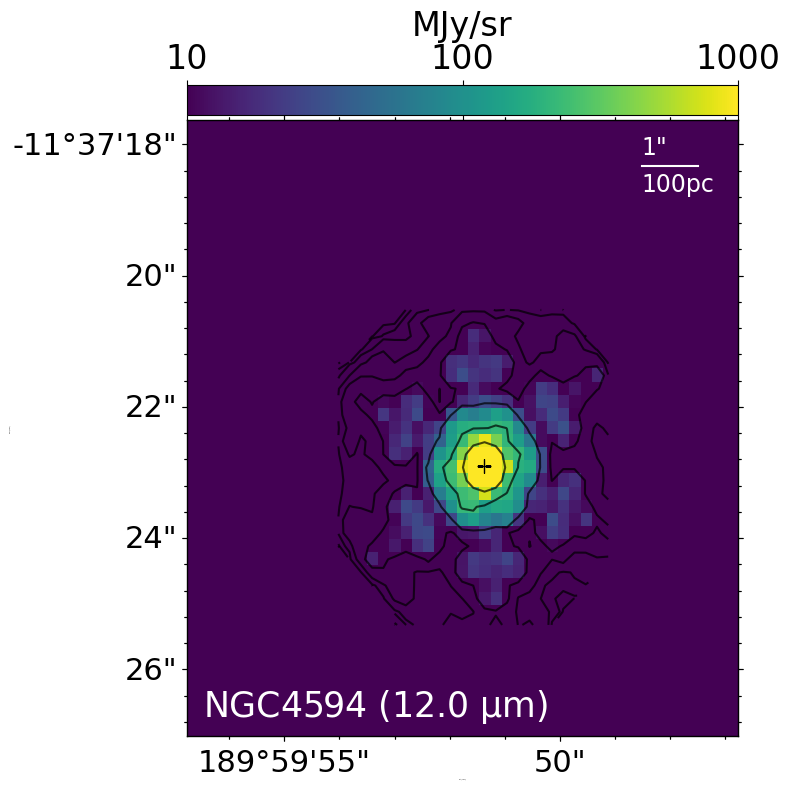}
\includegraphics[width=0.49\columnwidth,trim={0.45cm 1cm 0cm 2.95cm},clip]{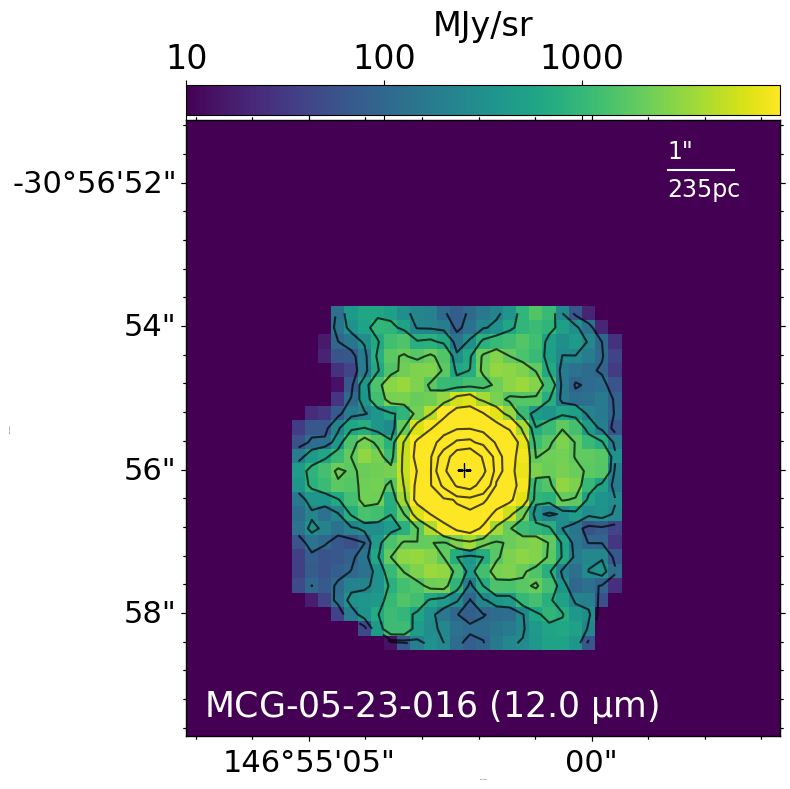}
\includegraphics[width=0.49\columnwidth,trim={0.45cm 0cm 0.4cm 2.95cm},clip]{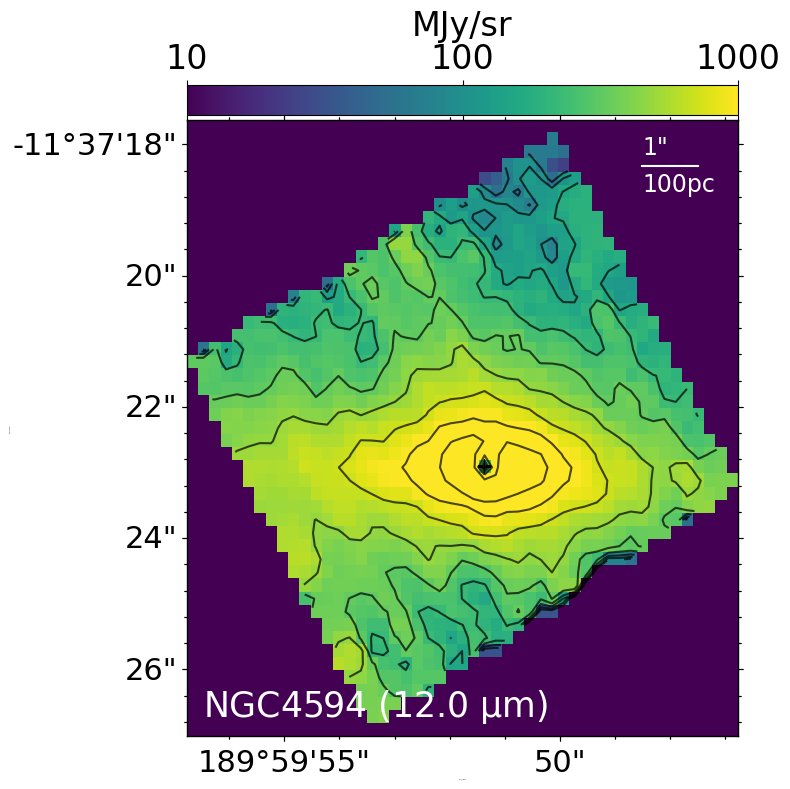}
\includegraphics[width=0.49\columnwidth,trim={0.45cm 0cm 0.cm 2.95cm},clip]{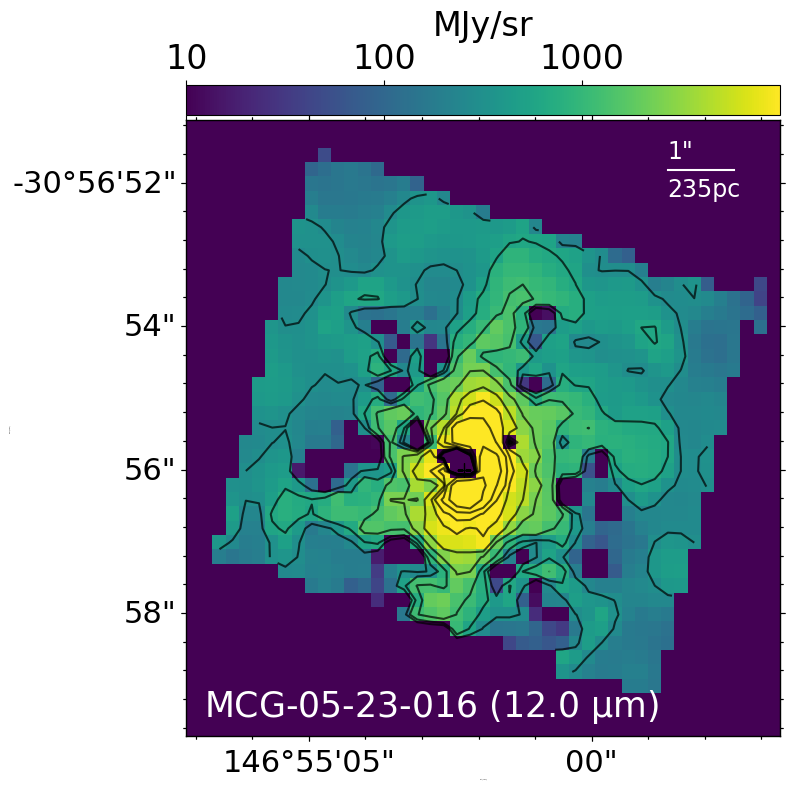}
\caption{MIRI/MRS images at 12\,$\rm{\mu m}$ for the low-luminosity AGN NGC\,4594 (left column) and the intermediate luminosity AGN MCG\,-05-23-016 (right column). From top to bottom: original image, best-fit PSF, and circumnuclear map at 12\,$\rm{\mu m}$. Black contours indicate emission above $\rm{3\sigma}$ level up to the maximum of the image (10 level). The color map is set to a logarithmic scale. \label{fig:NGC4594_MCG}}
\end{figure}

\begin{figure*}
\includegraphics[width=0.49\columnwidth,trim={0.2cm 0cm 0cm 0cm},clip]{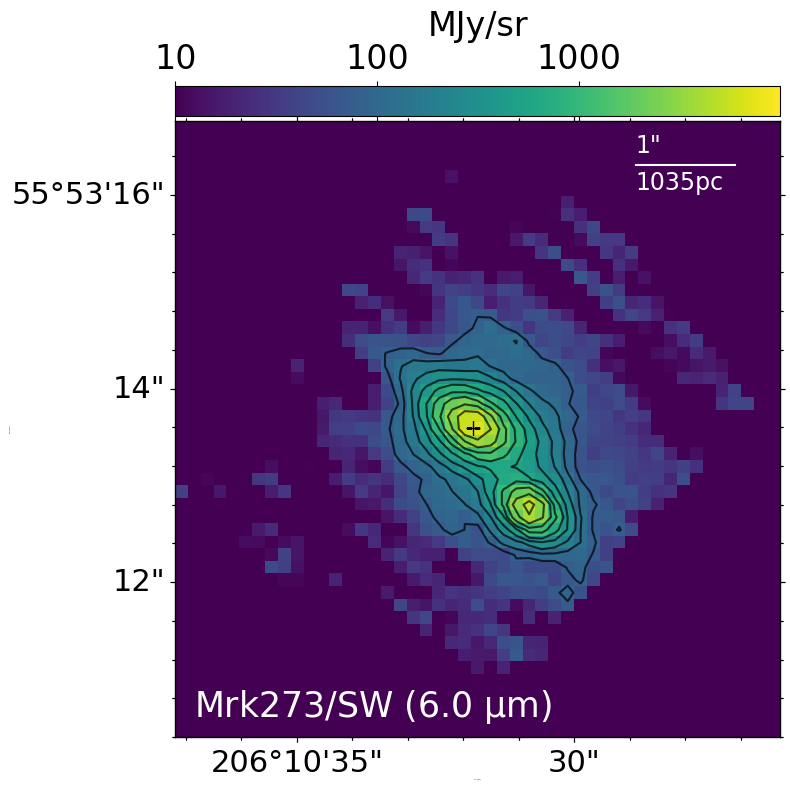}
\includegraphics[width=0.39\columnwidth,trim={4.15cm 0cm 0cm 0cm},clip]{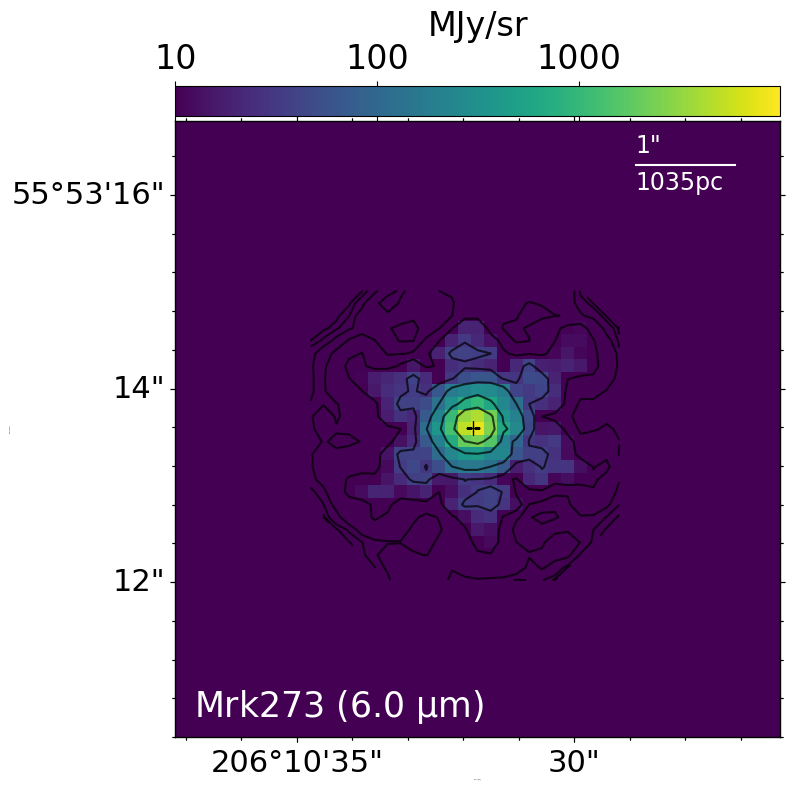}
\includegraphics[width=0.39\columnwidth,trim={4.15cm 0cm 0cm 0cm},clip]{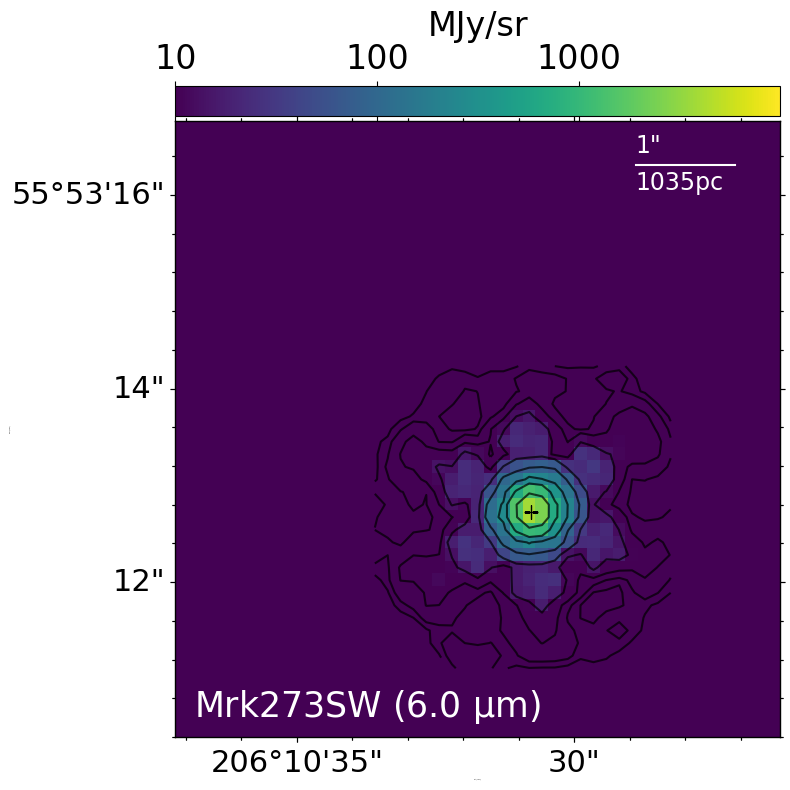}
\includegraphics[width=0.39\columnwidth,trim={4.15cm 0cm 0cm 0cm},clip]{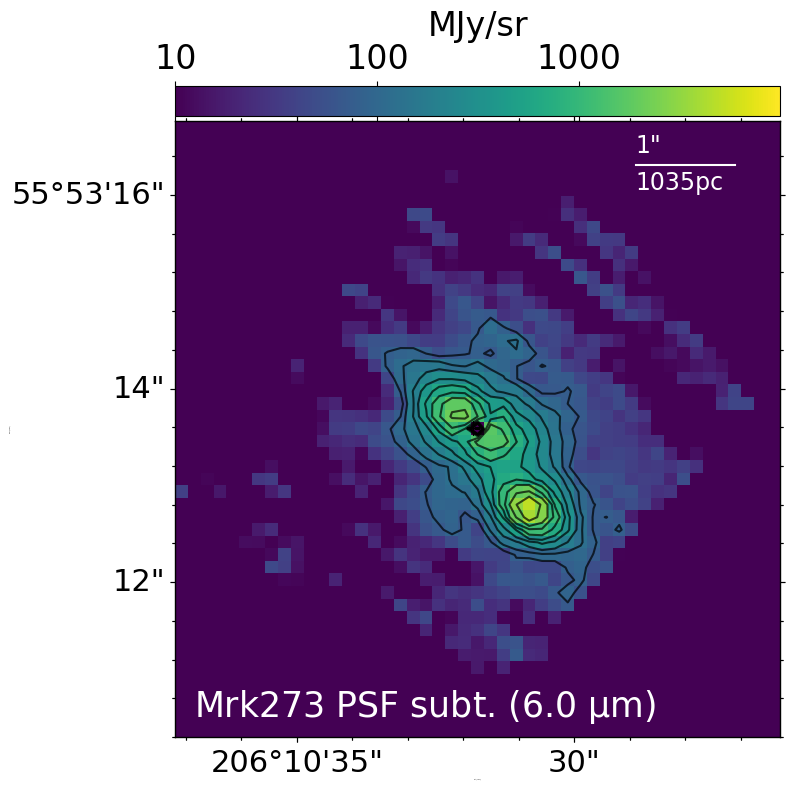}
\includegraphics[width=0.39\columnwidth,trim={4.15cm 0cm 0cm 0cm},clip]{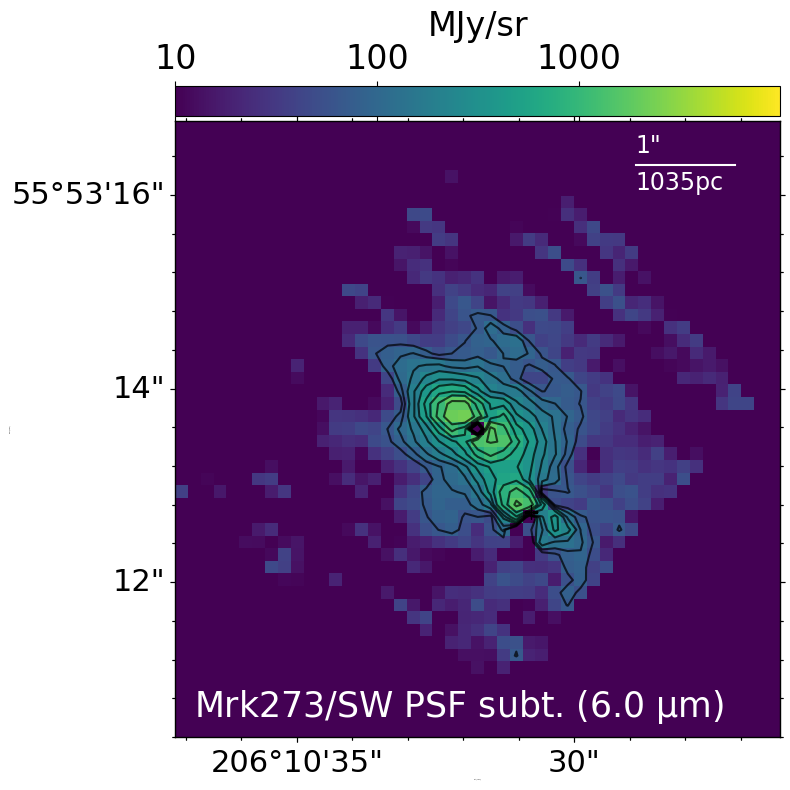}
\includegraphics[width=0.49\columnwidth,trim={0.2cm 0cm 0cm 2.cm},clip]{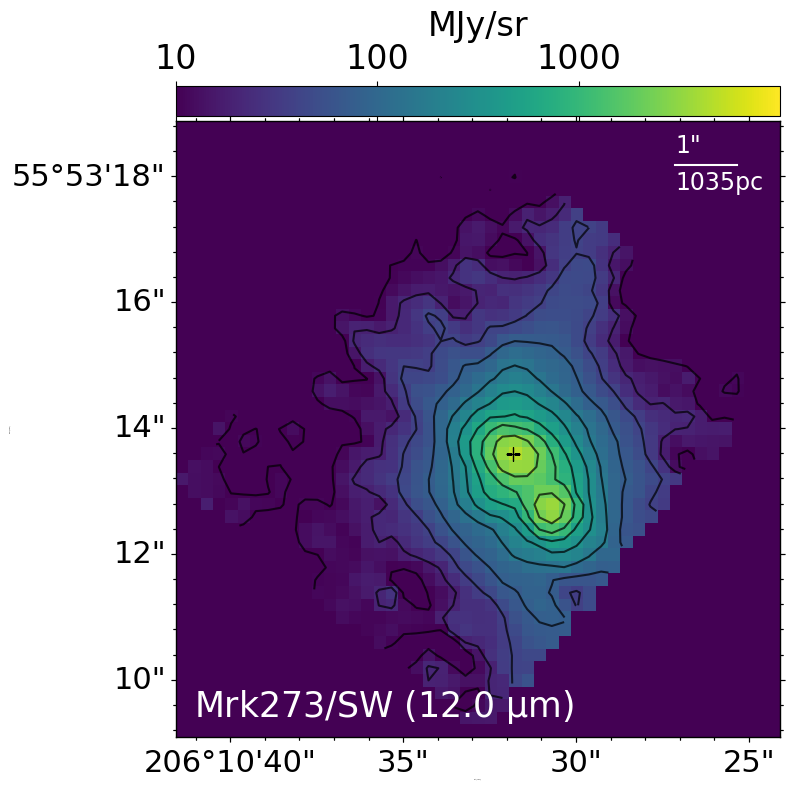}
\includegraphics[width=0.39\columnwidth,trim={4.15cm 0cm 0cm 2.cm},clip]{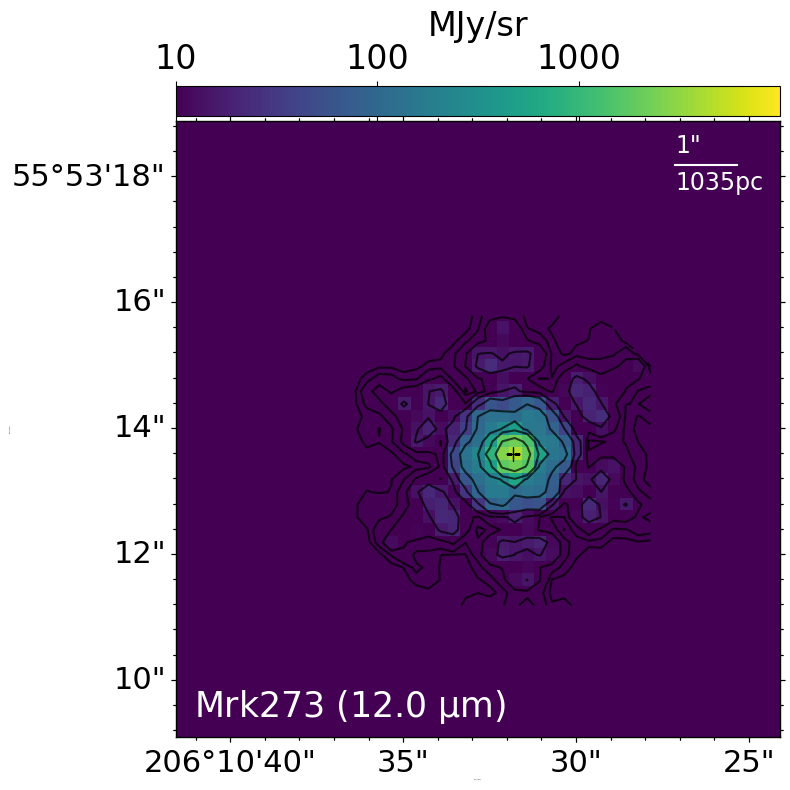}
\includegraphics[width=0.39\columnwidth,trim={4.15cm 0cm 0cm 2.cm},clip]{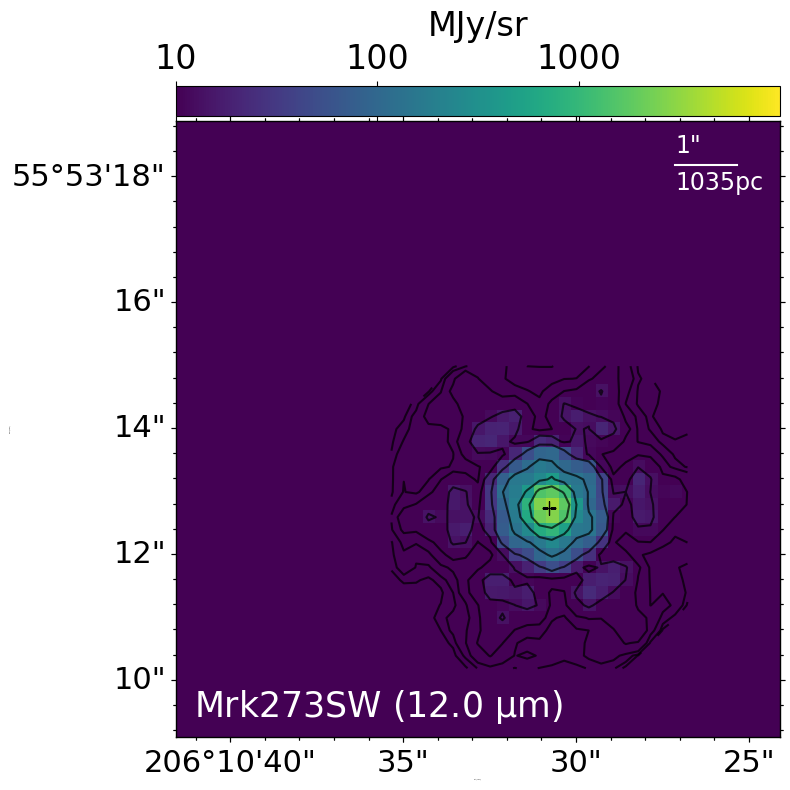}
\includegraphics[width=0.39\columnwidth,trim={4.15cm 0cm 0cm 2.cm},clip]{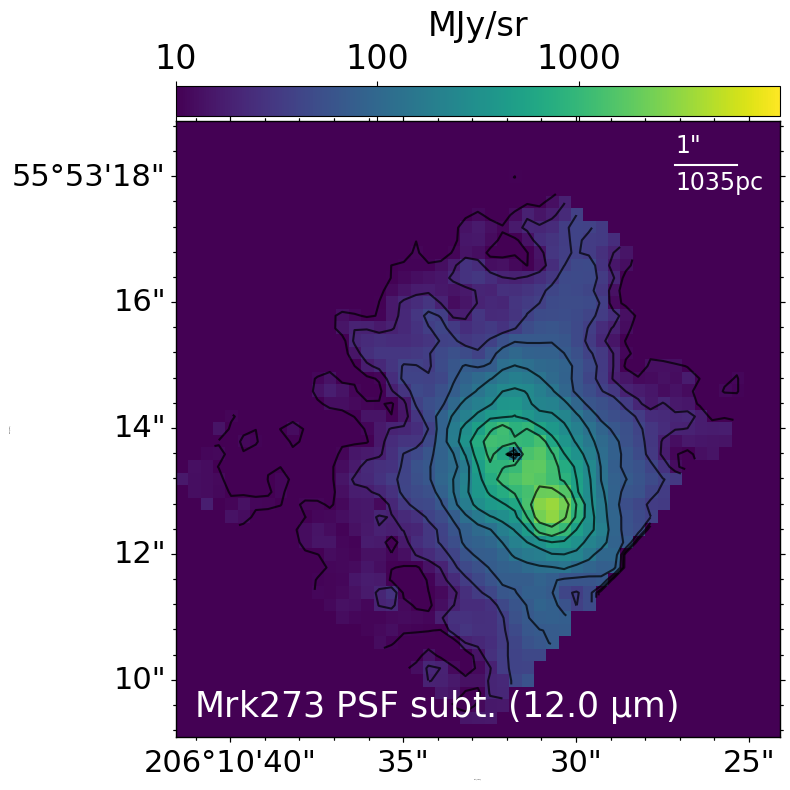}
\includegraphics[width=0.39\columnwidth,trim={4.15cm 0cm 0cm 2.cm},clip]{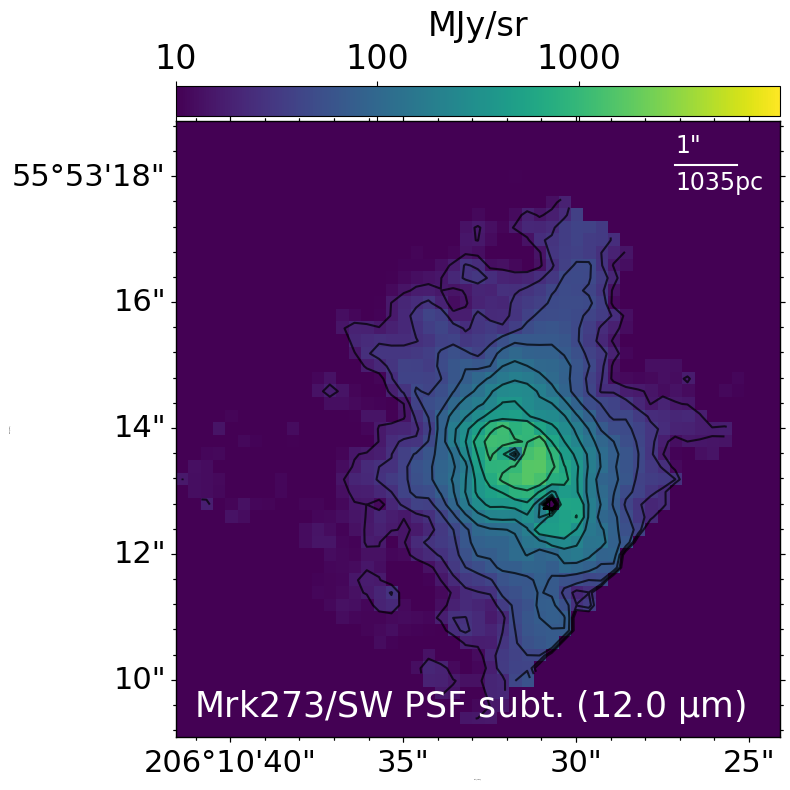}
\includegraphics[width=0.49\columnwidth,trim={0.2cm 0cm 0cm 2.cm},clip]{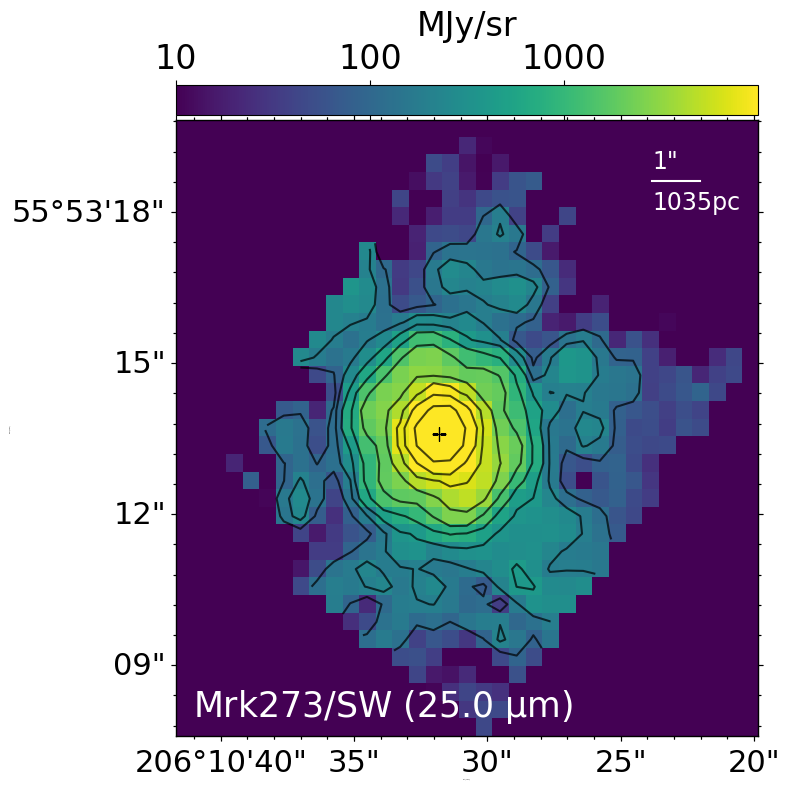}
\includegraphics[width=0.39\columnwidth,trim={4.15cm 0cm 0cm 2.cm},clip]{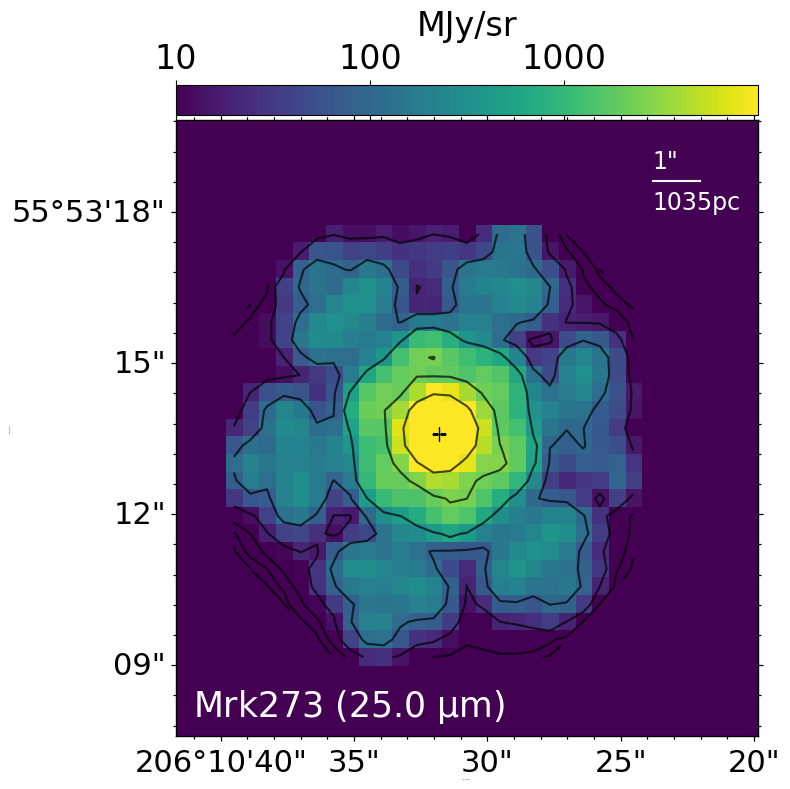}
\includegraphics[width=0.39\columnwidth,trim={4.15cm 0cm 0cm 2.cm},clip]{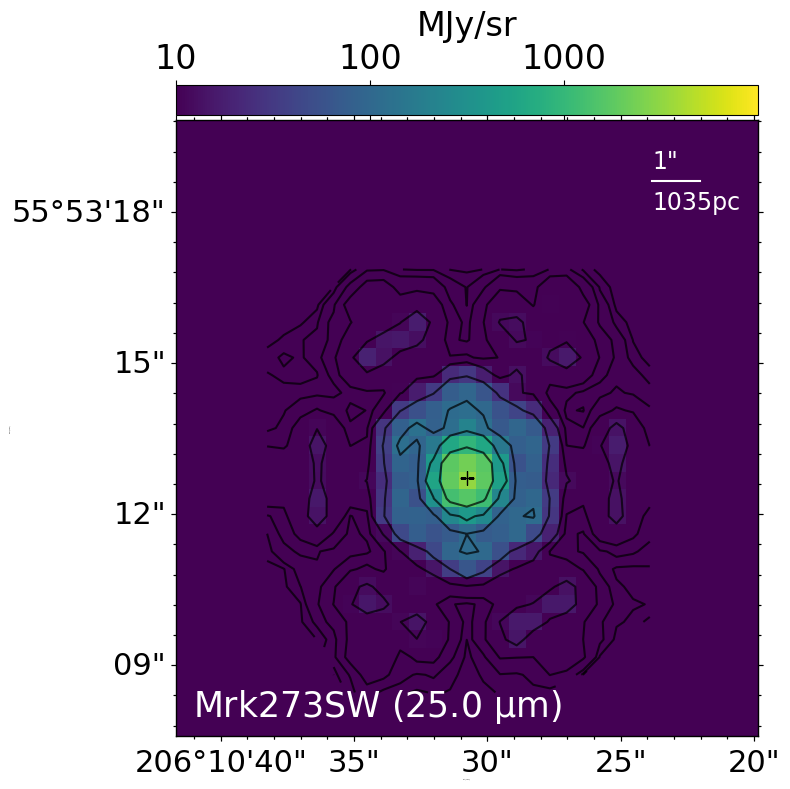}
\includegraphics[width=0.39\columnwidth,trim={4.15cm 0cm 0cm 2.cm},clip]{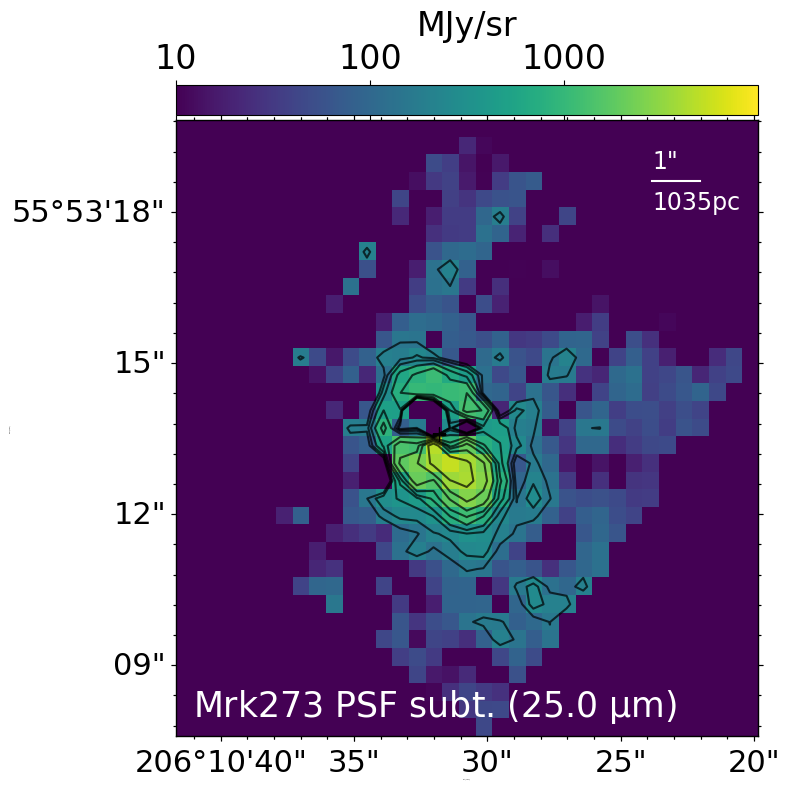}
\includegraphics[width=0.39\columnwidth,trim={4.15cm 0cm 0cm 2.cm},clip]{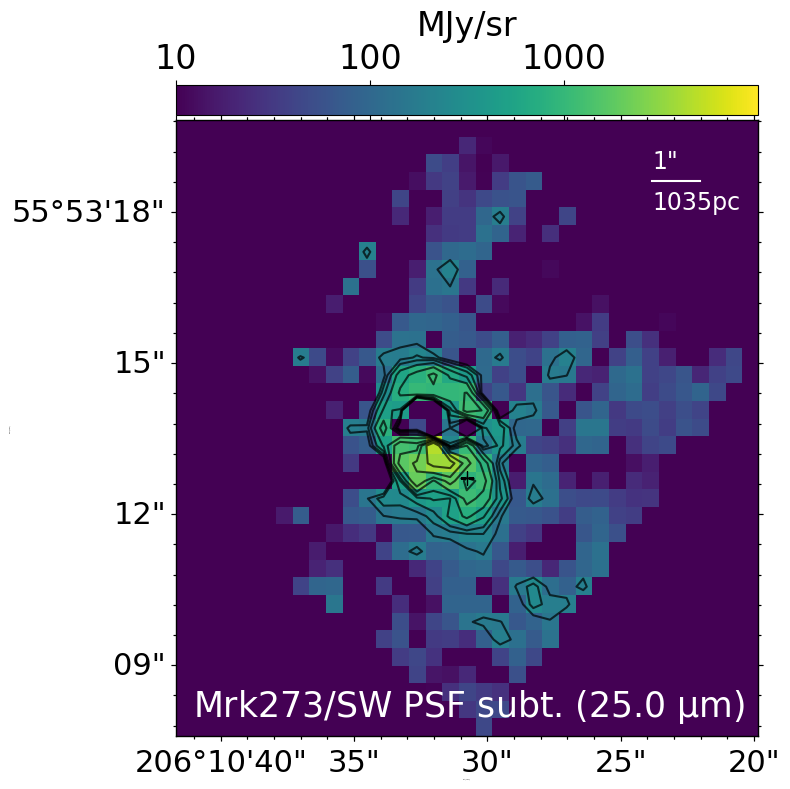}
\caption{MIRI/MRS images at 6, 12, and 25 $\rm{\mu m}$ for Mrk\,273/Mrk\,273SW (Col.\,1); best PSF for Mrk\,273 (Col.\,2) and Mrk\,273SW (Col.\,3); circumnuclear map after the subtraction of the nucleus of Mrk\,273 (Col.\,4); circumnuclear map after subtracting Mrk\,273 and Mrk\,273SW nuclei (Col.\,5). Black contours show any detection above the 3$\sigma$ level. Object name and wavelength are written in the bottom-left corner of each panel. The color map is set to a logarithmic scale.\label{fig:mrk273}}
\end{figure*}

\section{Results}\label{sec:results}

\begin{figure*}
\includegraphics[width=1.0\columnwidth,trim={1.3cm 0.cm 3.cm 1.75cm},clip]{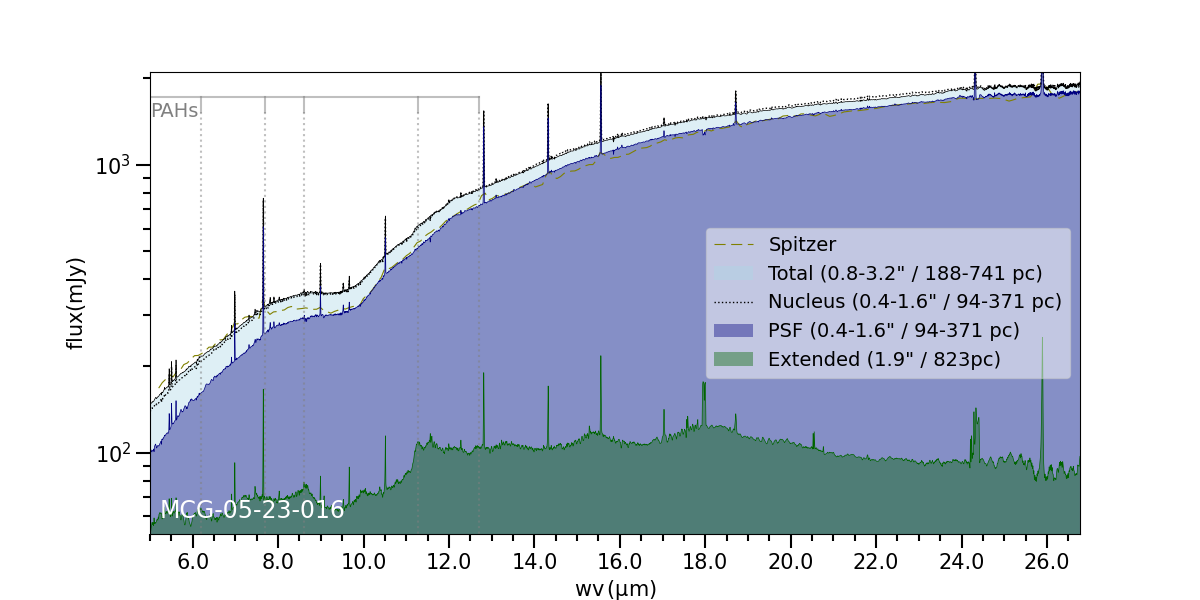}
\includegraphics[width=1.0\columnwidth,trim={1.3cm 0.cm 3.cm 1.75cm},clip]{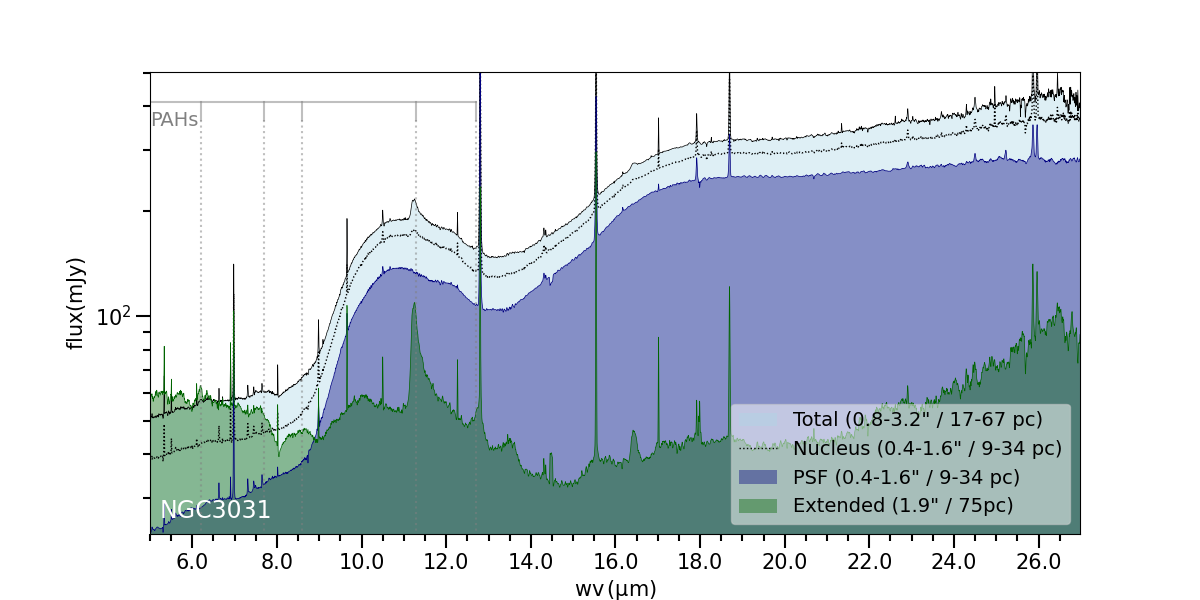} \\
\includegraphics[width=1.0\columnwidth,trim={1.3cm 0.cm 3.cm 1.75cm},clip]{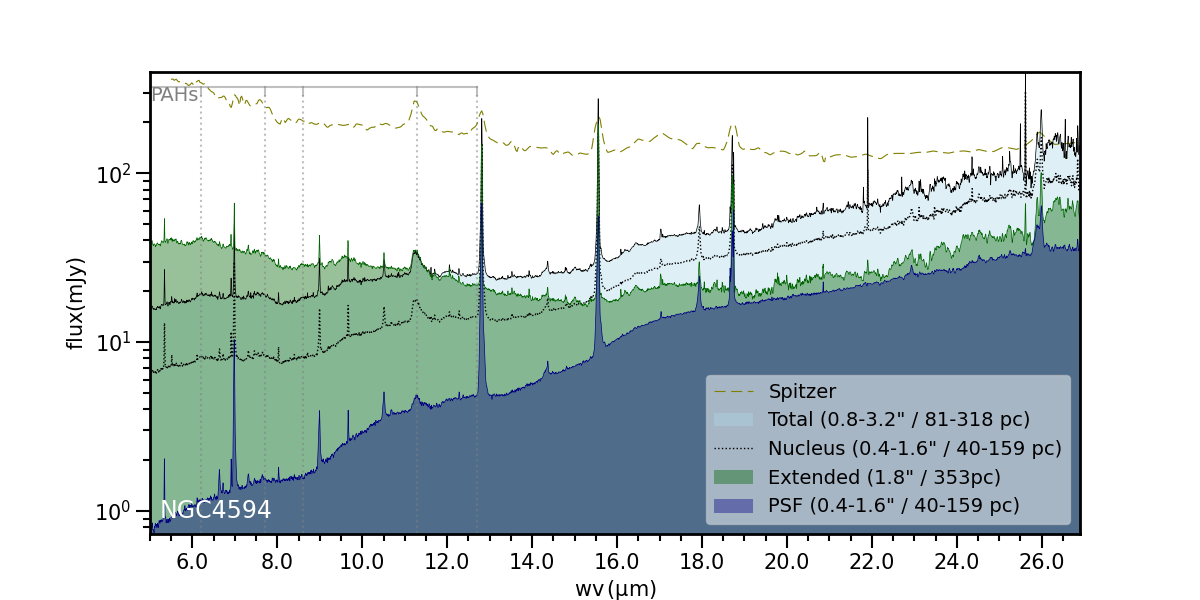}
\includegraphics[width=1.0\columnwidth,trim={1.3cm 0.cm 3.cm 1.75cm},clip]{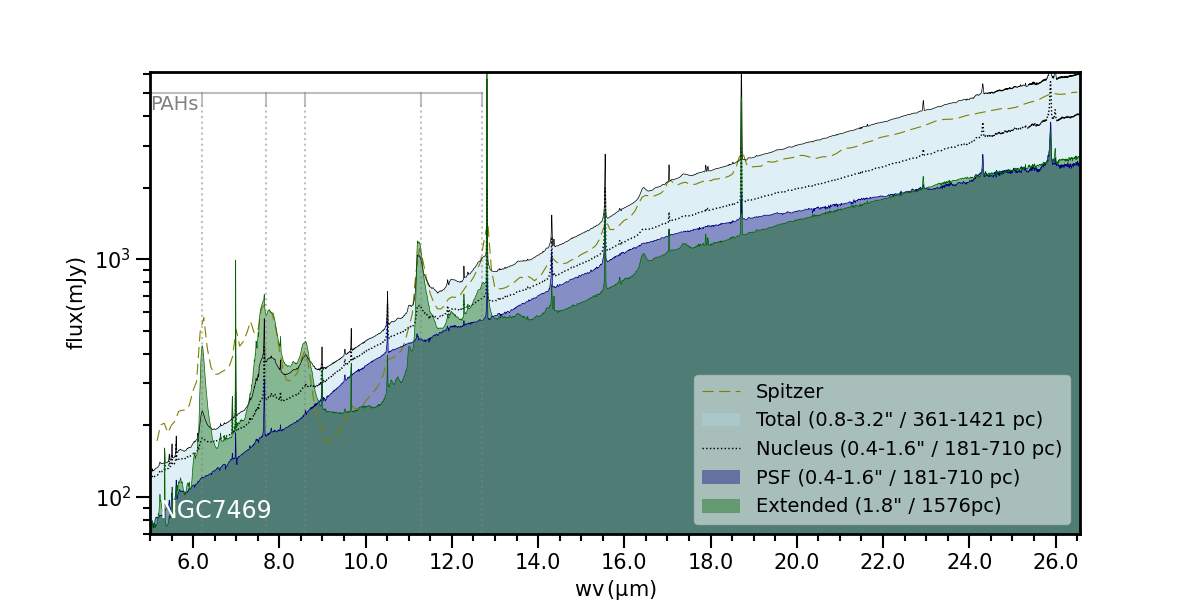} \\
\includegraphics[width=1.0\columnwidth,trim={1.3cm 0.cm 3.cm 1.75cm},clip]{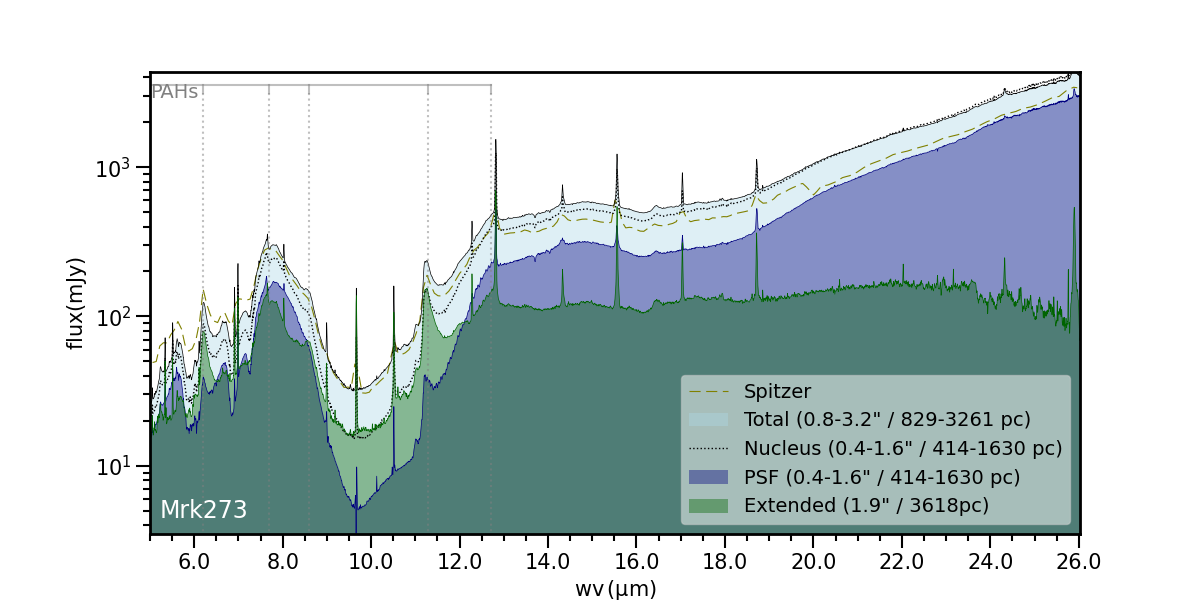} 
\includegraphics[width=1.0\columnwidth,trim={1.3cm 0.cm 3.cm 1.75cm},clip]{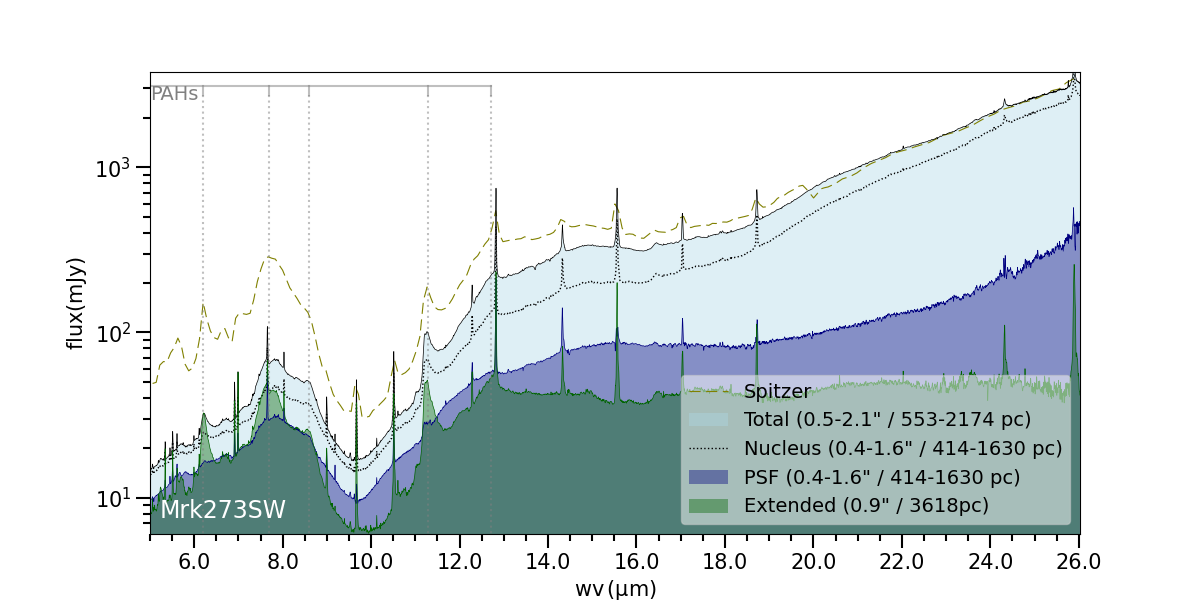}
\caption{Decomposed spectra of MCG-05-23-016 (top-left), NGC\,3031 (top-right), NGC\,4594 (middle-left), NGC\,7469 (middle-right), Mrk\,273 (bottom-left), and Mrk\,273SW (bottom-right). Each panel shows the total (solid-black line filled in light blue), nuclear (dotted-black line), PSF (solid-blue line filled in blue), and circumnuclear (solid-green line filled in green) spectra. When available, \emph{Spitzer}/IRS spectrum is shown with a long-dashed green line. Legends show the spatial scales for each spectral extraction radius. Appendix\,\ref{app:decomposedspectra} shows the full AGN collection figures.  \label{fig:spectra}}
\end{figure*}

\subsection{Accuracy of the decomposition using MRSPSFisol}

Table\,\ref{tab:percentage} shows the percentages (range, average, and at 12\,$\rm{\mu m}$) of PSF contribution to the total and nuclear spectra, respectively. Fig.\,\ref{fig:NGC4594_MCG} shows two examples of the decomposition made at 12$\rm{\mu m}$ for NGC\,4594 (left panels) and MCG-05-23-016 (right panels). In both targets, an oversubtraction is visible at the central pixel (bottom panels). This is because the simulated PSF shape is not yet a perfect match with the observations, being more sharp-pointed than the actual \emph{JWST} PSF. NGC\,4594 is one of the objects with the lowest percentage of nuclear flux within the MRS FOV in our AGN collection (with percentages that can be as low as 5\% and 11\% of the total and nuclear spectra, respectively, see Table\,\ref{tab:percentage}). The extended emission in this object matches the bulge of this galaxy, also known as the Sombrero galaxy \citep[][]{Goold24}. A slight over-subtraction of the AGN PSF might result in the circumnuclear map (bottom-left panel). This might explain the residual signatures of PAHs in the PSF spectrum. This method allows for isolating the nuclear flux in a spectrum entirely dominated by the circumnuclear emission otherwise (see below). The nuclear PSF of MCG\,05-23-016 dominates at $\rm{12 \mu m}$ (86\%, see Table\,\ref{tab:percentage}). It contributes above 70\% to the nuclear spectrum for the entire mid-infrared wavelength range \citep[see also][]{Esparza-Arredondo24}. 
The original image reveals the PSF spikes (top-right panel). After removing the PSF from this source, extended residual emission that could not be inferred from the original image is detected, almost elongated in a North-South direction. The level of accuracy of the decomposition also reflects on the accurate subtraction of the PSF wings, which no longer appear in the circumnuclear map (bottom-right panel). 

In our sample, the merging system Mrk\,273/Mrk\,273SW deserves particular attention due to its complexity in decomposing the data cubes. The data cubes for the Mrk\,273/Mrk\,273SW system show a double AGN within the same FOV, which is already reported at X-rays \citep[][]{Liu19}. This makes the decomposition complex, with code adjustments needed to perform the decomposition. First, the nuclear position is set manually based on the position reported by NED. The determination of the center of brightness finds a position in a region between the two nuclei and, therefore, is not accurate for either of the two nuclei. MRSPSFisol works well for Mrk\,273  (the brightest of the two sources). However, although using a fixed position, the decomposition of Mrk\,273SW is not possible in the original data cube because, at long wavelengths, this source is embedded in the PSF emission of Mrk\,273. To sort out this issue, once the decomposition is applied for Mrk\,273, the output of the circumnuclear map is used to decompose the image now centered at the position of Mrk\,273SW. This results in a final circumnuclear map decontaminated from the two nuclei. Although the process could benefit from more complex modeling (e.g., with two PSFs modeled simultaneously), this simple approach gives satisfactory results, as shown in Fig.\,\ref{fig:mrk273}. This figure shows the decomposition at three wavelengths (6, 12, and 25\,$\rm{\mu m}$, from top to bottom rows). The two nuclei are visible at 6 and 12\,$\rm{\mu m}$ (left-top and left-middle panels), while at 25\,$\rm{\mu m}$ both nuclei are blended into a single source due to the lower resolution at this wavelength range. Columns 2 and 3 show the PSF map of Mrk\,273 and Mrk\,273SW at these wavelengths. The nuclear flux of Mrk\,273 dominates at 25$\rm{\mu m}$ compared with that obtained for Mrk\,273SW. This is not the case for shorter wavelengths with similar fluxes at both wavelengths. This is also visible from Table\,\ref{tab:percentage}, with a percentage of PSF contribution ranging 6-78 and 10-80\%  for Mrk\,273SW and Mrk\,273, respectively. Column 4 and 5 in Fig.\,\ref{fig:mrk273} shows the resulting circumnuclear map when Mrk\,273 and Mrk\,273$+$Mrk\,273SW are removed, respectively. The MRSPSFisol tool can remove any residual of the PSF wings at 25\,$\rm{\mu m}$, with a circumnuclear map showing emission from the host at the three wavelengths.   

Using the byproducts of MRSPSFisol we isolate the nuclear mid-infrared spectrum from that of the host. Fig.\,\ref{fig:spectra} shows the resulting spectra for MCG\,-05-23-016, NGC\,3031, NGC\,4594, NGC\,7469, and Mrk\,273/Mrk\,273SW. Although the spatial resolution achieved by the total spectrum considered here is similar to that of the \emph{Spitzer}/IRS spectra, we find flux calibration issues for some sources. For instance, the \emph{Spitzer}/IRS spectrum (long-dashed green line) is clearly below the total spectrum (solid black line filled in light blue) for NGC\,7469 for wavelengths above 8$\rm{\mu m}$. The discussion below shows that the \emph{JWST} fluxes agree with those obtained from ground-based observations. Thus, we conclude that this cross-calibration issue is unrelated to the \emph{JWST} calibration.

In all the cases, a non-negligible contribution of extended circumnuclear emission is found (green-filled spectra). However, the relative amount of circumnuclear emission compared to that of the PSF emission (dark-blue filled spectra) varies from object to object (see Table\,\ref{tab:percentage}). The nuclear PSF entirely dominates the total emission of MCG\,05-23-016 (average PSF contribution of 88\%). The opposite behavior is found for NGC\,4594, where most of the flux within the FOV at short wavelengths comes from circumnuclear contributors (average PSF contribution of 38\%). Interestingly, the seven type-1 AGN in our collection have more than 80\% PSF contributing within the nuclear extraction at 12$\rm{\mu m}$ (last column in Table\,\ref{tab:percentage}). Only three (MCG\,05-23-016, IC\,5063, and NGC\,7319) out of the 14 type-2 AGN show more than 80\% of the PSF contribution, reinforcing the need to isolate the nucleus for type-2 AGN. The shape of the PSF spectrum substantially changes from the best nuclear isolation (black dotted spectrum) obtained from the original data cube (before MRSPSFisol is applied), which shows strong signs of circumnuclear contribution for many objects. In particular, the strength of the silicate emission features of NGC\,4594 increases for the PSF spectrum compared to the nuclear spectrum. It is also clear that the long-wavelength slopes change (see, e.g., NGC\,3031 and NGC\,7469). These aspects are crucial for the spectral fitting technique, reinforcing the need for proper isolation before attempting to infer the AGN dust properties with this technique. 

PAH features are suppressed from the PSF spectra of all the sources in our AGN collection (see Appendix\,\ref{app:decomposedspectra}). However, it is entirely removed only in five cases (e.g., NGC\,3031, Fig.\,\ref{fig:spectra}), with a residual of PAH features in most targets (e.g., NGC\,4594, Fig.\,\ref{fig:spectra}). This reinforces the need to further isolate the AGN dust continuum from circumnuclear contributors spectroscopically. On the other hand, PAH features appear in the circumnuclear map when nearly detected in the total/nuclear spectra \citep[e.g., see the spectrum of MCG\,-05-23-016 in Fig.\,\ref{fig:spectra}, for more discussion in PAH features of MCG\,-05-23-016 see][]{Zhang24}. We also note that the spectral shapes of the host galaxies in our AGN collection vary. This is further discussed in Section\,\ref{sec:discussion}.

\begin{figure}
\includegraphics[width=1.\columnwidth,trim={0.2cm 5.2cm 0cm 0.cm},clip]{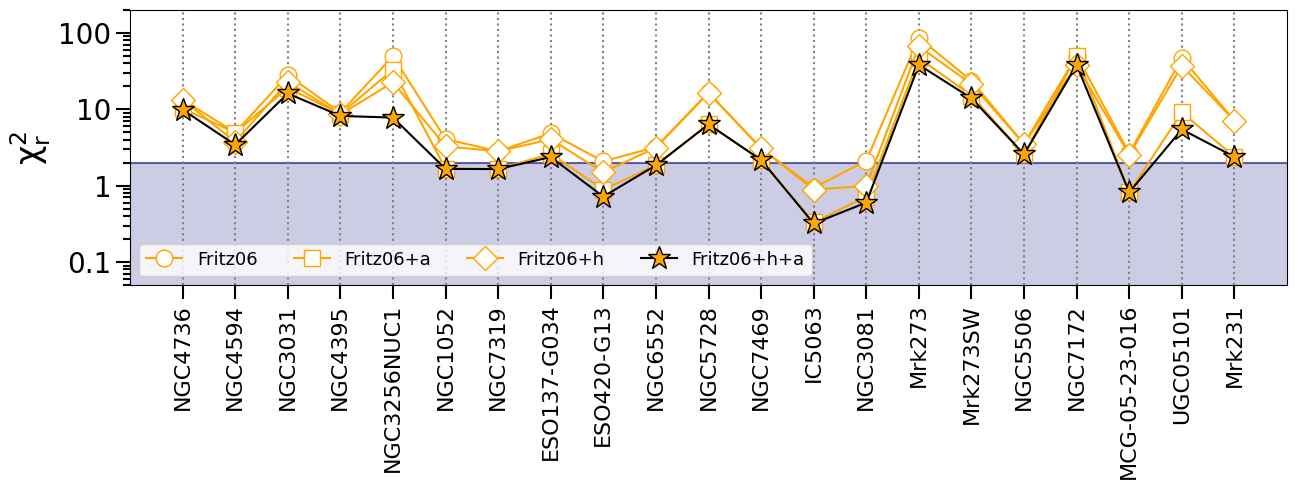}
\includegraphics[width=1.\columnwidth,trim={0.2cm 5.2cm 0cm 0.cm},clip]{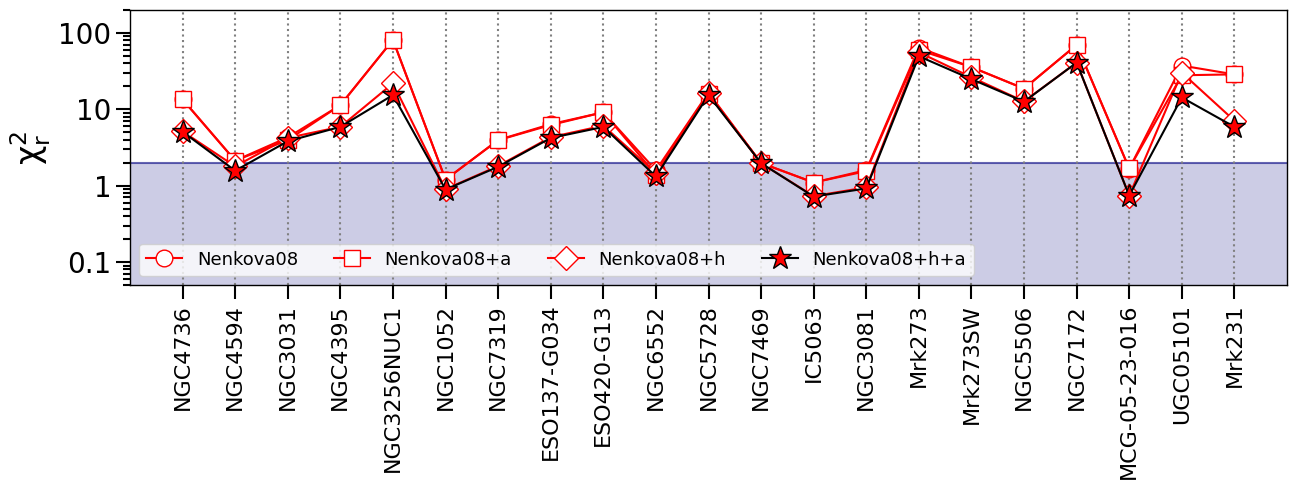}
\includegraphics[width=1.\columnwidth,trim={0.2cm 5.2cm 0cm 0.cm},clip]{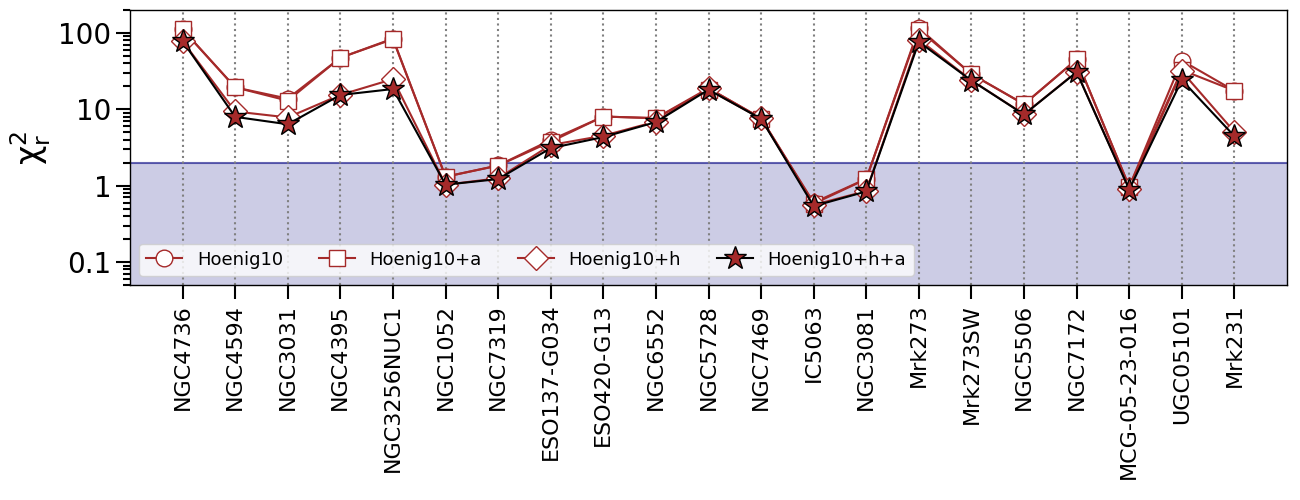}
\includegraphics[width=1.\columnwidth,trim={0.2cm 5.2cm 0cm 0.cm},clip]{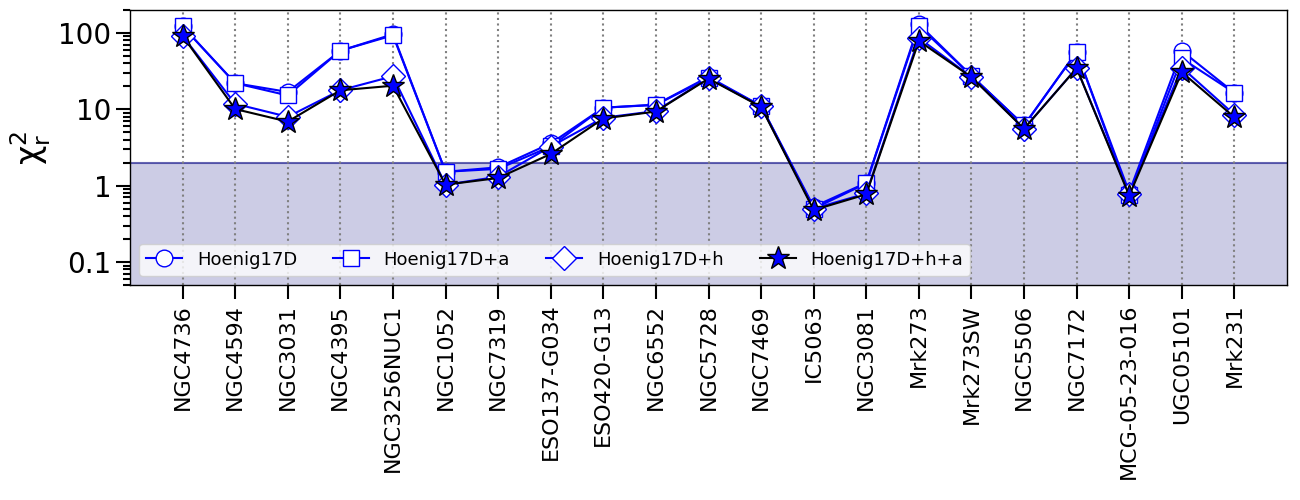}
\includegraphics[width=1.\columnwidth,trim={0.2cm 5.2cm 0cm 0.cm},clip]{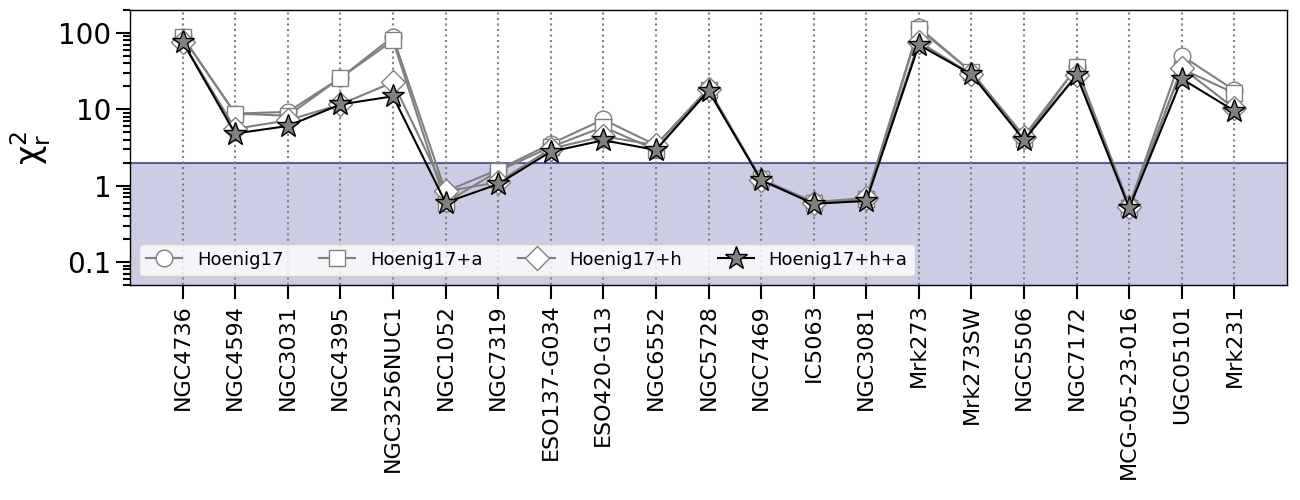}
\includegraphics[width=1.\columnwidth,trim={0.2cm 5.2cm 0cm 0.cm},clip]{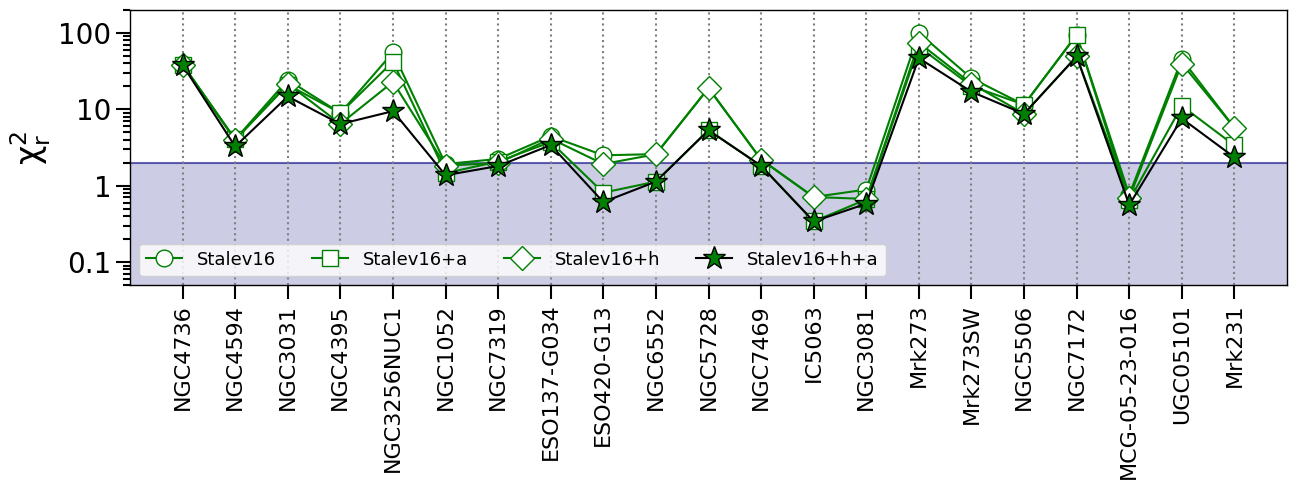}
\includegraphics[width=1.\columnwidth,trim={0.2cm 0.5cm 0cm 0.cm},clip]{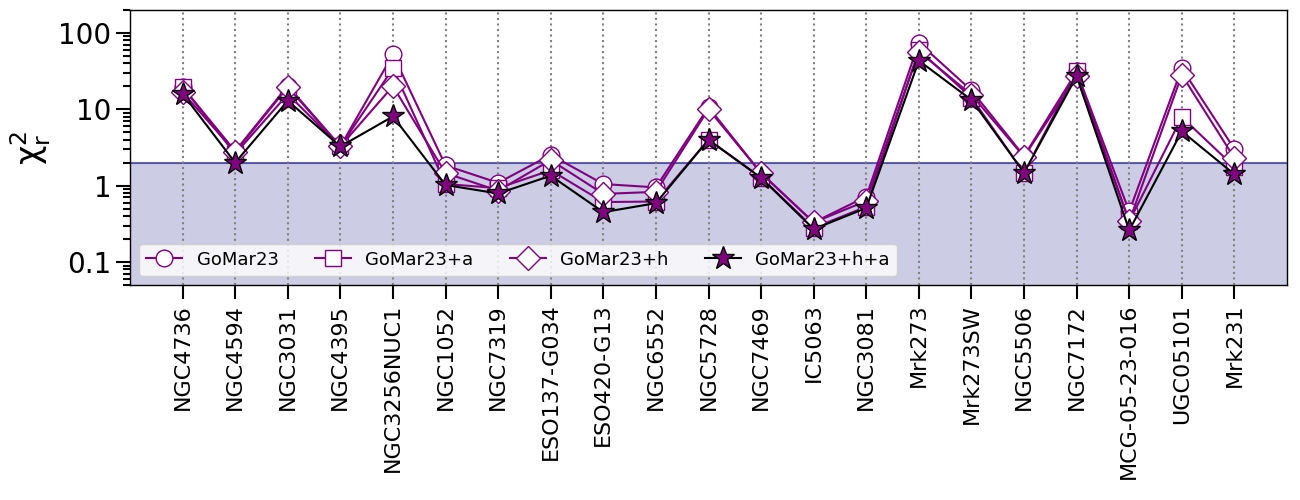}
\caption{Distribution of reduced $\rm{\chi^2}$ per object. Panels 1$-$7 show the results for each model and compare the $\rm{\chi^2_{r}}$ when adding absorption features (``+a", squares), host galaxy contribution (``+h", diamonds), and absorption features and host galaxy (``+a+h", stars) with the simplest fit to the AGN dust model (circles). Objects are sorted by X-ray luminosity (lowest values toward the left and highest values toward the right, X-ray luminosities in the x-axis in the top panel). \label{fig:ChiLum}}
\end{figure}

\begin{figure*}
\includegraphics[width=2.\columnwidth,trim={0.2cm 0.4cm 0cm 0.cm},clip]{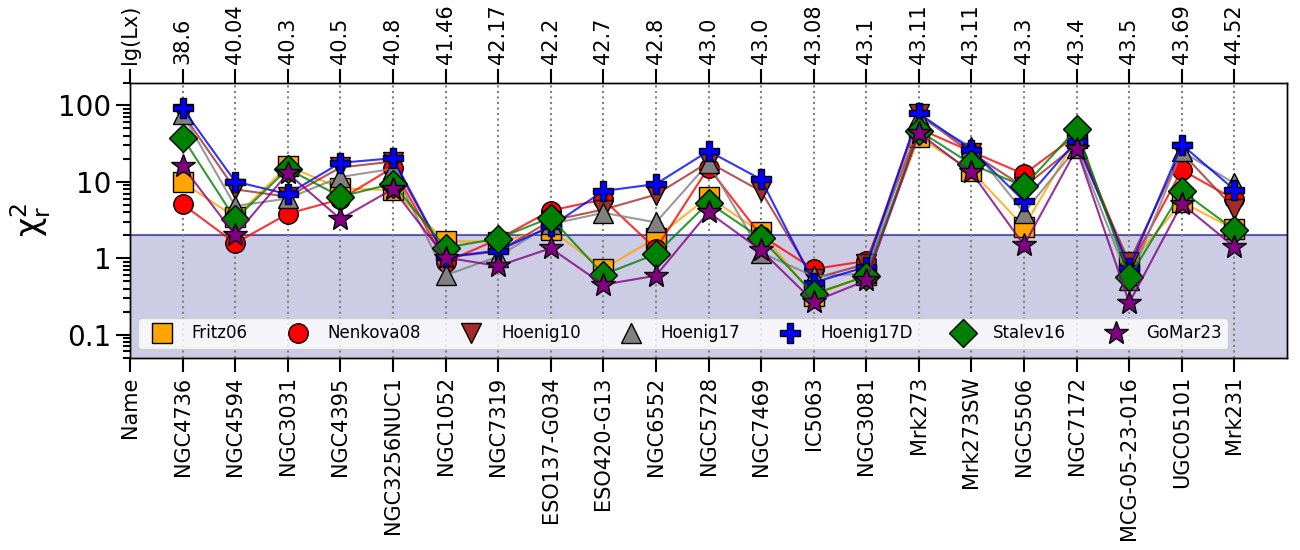}
\includegraphics[width=2.\columnwidth,trim={0.2cm 0.5cm 0cm 0.cm},clip]{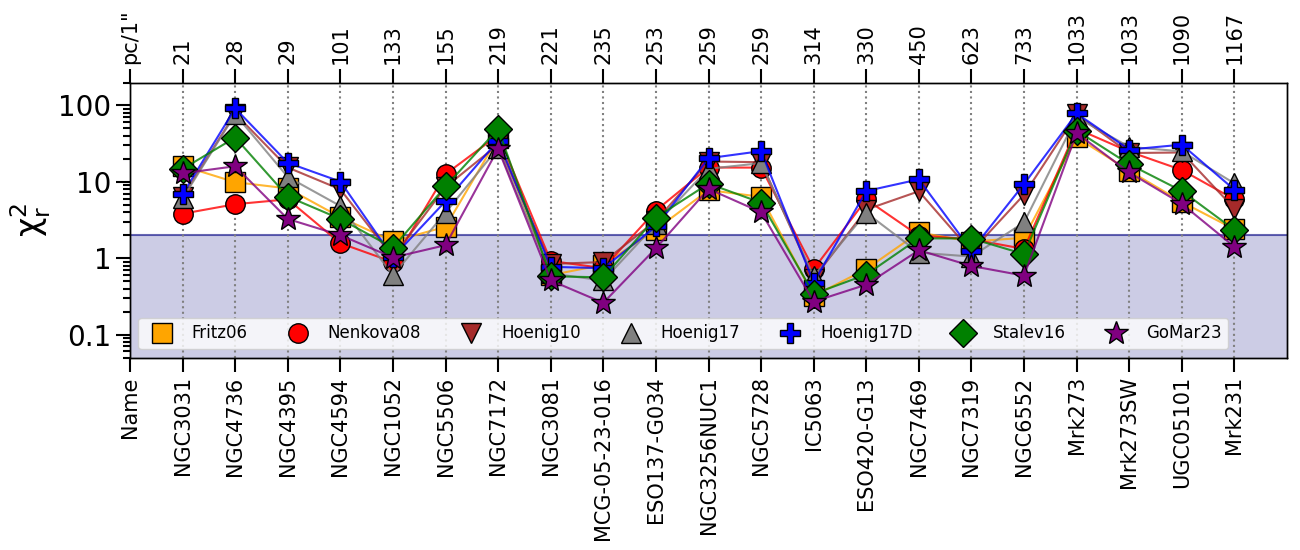}
\caption{Distribution of reduced $\rm{\chi^2}$ for each AGN dust model when adding host contributions and absorption features. {Top panel: objects sorted out by the luminosity. Bottom panel: objects sorted out by the spatial scale.} \label{fig:ChiDist}}
\end{figure*}

\begin{table*}
	\centering
    \footnotesize
    \renewcommand{\tabcolsep}{0.07cm}
\caption{Columns: 1. Object name, 2. best-fit model, 3. the need for host galaxy contribution marked with ticks,  4. the need for absorption features is marked with ticks, 5. $\rm{E(B-V)}$ for the foreground extinction obtained with the best fit,  {6. fraction of flux from host galaxy at 12$\rm{\mu m}$ needed for the best fit ($\rm{f_{host}}$)}, and 7. reduced $\rm{\chi^2}$ for the best, 8. good fits are marked with ticks, 9. good fits with $\rm{\chi^2}/dof < 2$, 10. maximum gran size obtained for GoMar23 when this model produces good fits. Models in Col.8 are named as follows: F06 for Fritz06; N08 for Nenkova08; H10 for Hoenig10; H17 for Hoenig17; H17D for Hoenig17D; S16 for Stalev16; and GM23 for GoMar23 (see Table\,\ref{tab:models} for more details in these models).
}

\label{tab:bestfitmodel}
\begin{tabular}{lcccccccll}
\hline \hline
{Name} & {Best Model} & {Host} & {Absorpt.} & {E(B-V)} & $\rm{f_{Host}}$ &  {$\rm{\chi^2/dof}$} &  {Goodness} & {Other models with $\rm{\chi^2/dof <2}$} & {$\rm{P_{max} (\mu m)}$} \\
(1) & (2) & (3) & (4) & (5) & (6) & (7) & (8) & (9) & (10) \\
% \colnumbers 
\hline
NGC1052        &   Hoenig17  &                &                 & 0.45$\rm{\pm}$0.01    &      & 0.84   &   $\tikzcmark$ & F06, N08, H10, H17D, S16, GM23 & $\rm{<0.01}$\\
ESO420-G13     &   GoMar23   &  $\tikzcmark$  &  $\tikzcmark$   & $\rm{<0.01}$          & 0.53 & 0.45   &   $\tikzcmark$ & F06, S16 & $\rm{0.549\pm0.001}$ \\
UGC05101       &   GoMar23   &  $\tikzcmark$  &  $\tikzcmark$   & 1.81$\rm{\pm}$0.01    & 0.60 & 5.15   &                &  &  \\ % $\rm{0.023\pm 0.001\mu m}$\\
MCG-05-23-016  &   GoMar23   &  $\tikzcmark$  &  $\tikzcmark$   & 0.09$\rm{\pm}$0.01    & 0.53 & 0.26   &   $\tikzcmark$ & F06, N08, H10, H17, H17D, S16 & $\rm{8.7\pm 0.1}$\\
NGC3031        &   Nenkova08 &  $\tikzcmark$  &  $\tikzcmark$   & 0.847$\rm{\pm}$0.002  & 0.51 & 3.82   &                &  &\\
NGC3081        &   GoMar23   &  $\tikzcmark$  &  $\tikzcmark$   & 0.094$\rm{\pm}$0.003  & 0.57 & 0.51   &   $\tikzcmark$ & F06, N08, H10, H17, H17D, S16 & $\rm{0.10\pm0.01}$\\
NGC3256NUC1    &   Fritz06   &  $\tikzcmark$  &  $\tikzcmark$   & 4.28$\rm{\pm}$0.01    & 0.74 & 7.79   &                &  & \\
% NGC3256NUC2    &   GoMar23   &  $\tikzcmark$  &  $\tikzcmark$   &&   2.55   &              &   & \\
NGC4395        &   GoMar23   &                &                 & $\rm{<0.01}$          &      & 3.27    &                &  & \\ % $\rm{0.025\pm0.001\mu m}$\\
NGC4594        &   Nenkova08 &  $\tikzcmark$  &                 & 0.57$\rm{\pm}$0.01    & 0.53 & 1.79   &   $\tikzcmark$ & GM23 & $\rm{<0.01}$\\
NGC4736        &   Nenkova08 &  $\tikzcmark$  &  $\tikzcmark$   & $\rm{<0.01}$          & 0.62 & 5.09   &                & \\
Mrk231         &   GoMar23   &  $\tikzcmark$  &  $\tikzcmark$   & 0.428$\rm{\pm}$0.003  & 0.53 & 1.41   &   $\tikzcmark$ & & $\rm{0.50\pm0.01}$\\
Mrk273SW       &   GoMar23   &  $\tikzcmark$  &  $\tikzcmark$   & 1.677$\rm{\pm}$0.003  & 0.61 & 13.3   &                & &  \\ % $\rm{0.50\pm0.01\mu m}$\\
Mrk273         &   Fritz06   &  $\tikzcmark$  &  $\tikzcmark$   & $\rm{<0.01}$          & 0.77 & 38.2   &                & & \\
NGC5506        &   GoMar23   &                &  $\tikzcmark$   & 0.736$\rm{\pm}$0.003  &      & 1.47    &   $\tikzcmark$ & & $\rm{0.25\pm 0.01}$ \\
NGC5728        &   GoMar23   &  $\tikzcmark$  &  $\tikzcmark$   & 0.248$\rm{\pm}$0.003  & 0.57 & 4.00   &                & &  \\ % $\rm{0.46\pm 0.01\mu m}$ \\
ESO137-G034    &   GoMar23   &  $\tikzcmark$  &  $\tikzcmark$   & 0.12$\rm{\pm}$0.01    & 0.60 & 1.35   &   $\tikzcmark$ & & $\rm{0.706\pm0.004}$ \\
NGC6552        &   GoMar23  &  $\tikzcmark$  &  $\tikzcmark$   & 0.037$\rm{\pm}$0.003   & 0.53 & 0.59   &   $\tikzcmark$ & F06, N08 & $\rm{>9.98}$  \\
IC5063         &   GoMar23 &                &                 & 0.636$\rm{\pm}$0.003    &      & 0.33    &   $\tikzcmark$ & F06, N08, H10, H17, H17D, S16 & $\rm{1.000\pm0.003}$   \\
NGC7172        &   GoMar23 &  $\tikzcmark$  &  $\tikzcmark$   & 4.46$\rm{\pm}$0.01      & 0.64 & 27.1   &                & & \\ %$\rm{0.50\pm 0.01\mu m}$  \\
NGC7319        &   GoMar23 &  $\tikzcmark$  &  $\tikzcmark$   & 0.56$\rm{\pm}$0.01      & 0.54 & 0.79   &   $\tikzcmark$ & F06, N08, H10, H17, H17D, S16 & $\rm{2.16\pm 0.01 }$  \\
NGC7469        &   Hoenig17  &  $\tikzcmark$  &                 & 0.137$\rm{\pm}$0.002  & 0.51 & 1.18   &   $\tikzcmark$ & S16, GM23 & $\rm{0.050\pm 0.001}$ \\ \hline \hline
	\end{tabular}
\end{table*}

\begin{figure*}
\includegraphics[width=0.67\columnwidth,trim={0.2cm 0.1cm 0cm 0.65cm},clip]{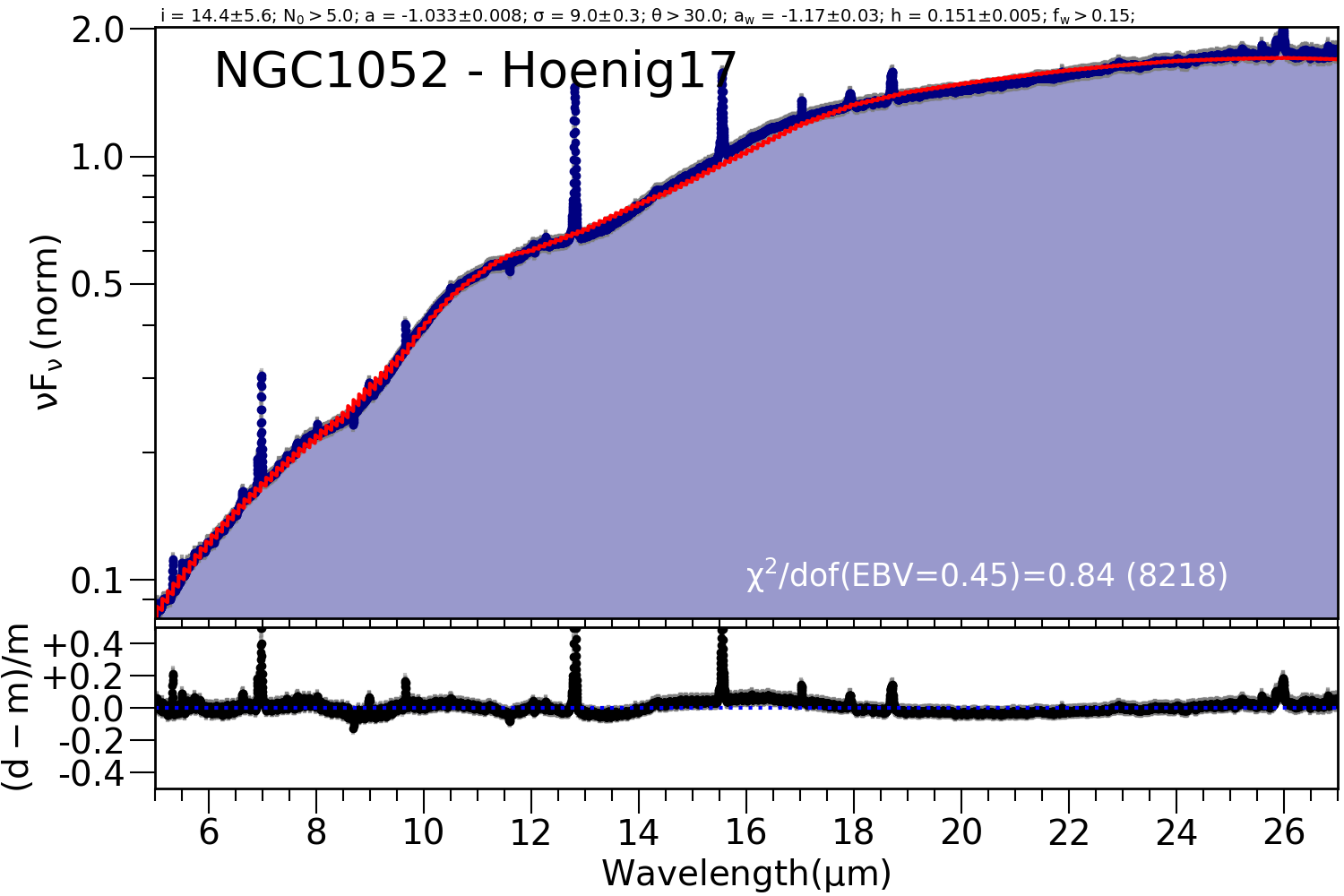}
\includegraphics[width=0.67\columnwidth,trim={0.2cm 0.1cm 0cm 0.65cm},clip]{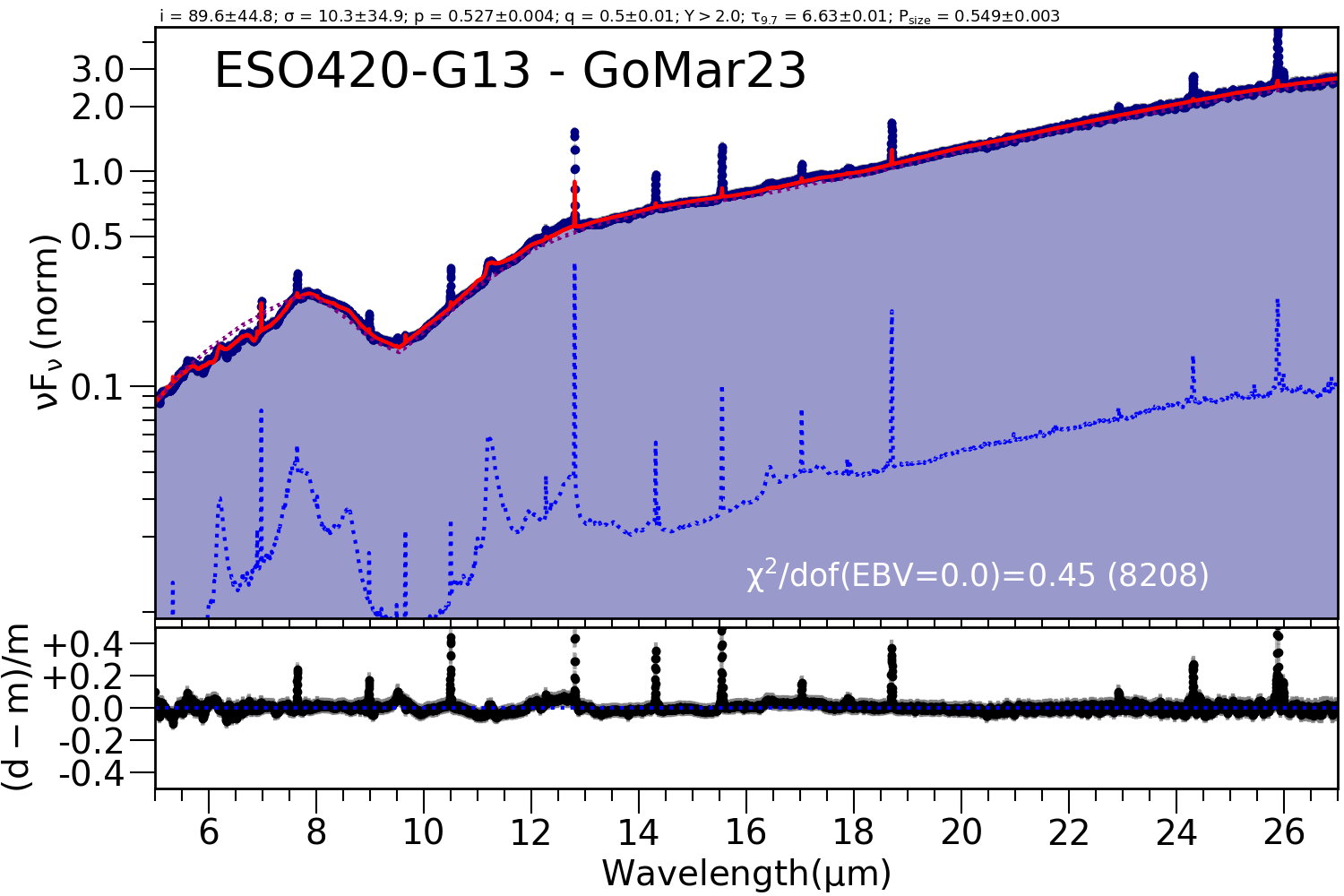}
\includegraphics[width=0.67\columnwidth,trim={0.2cm 0.1cm 0cm 0.65cm},clip]{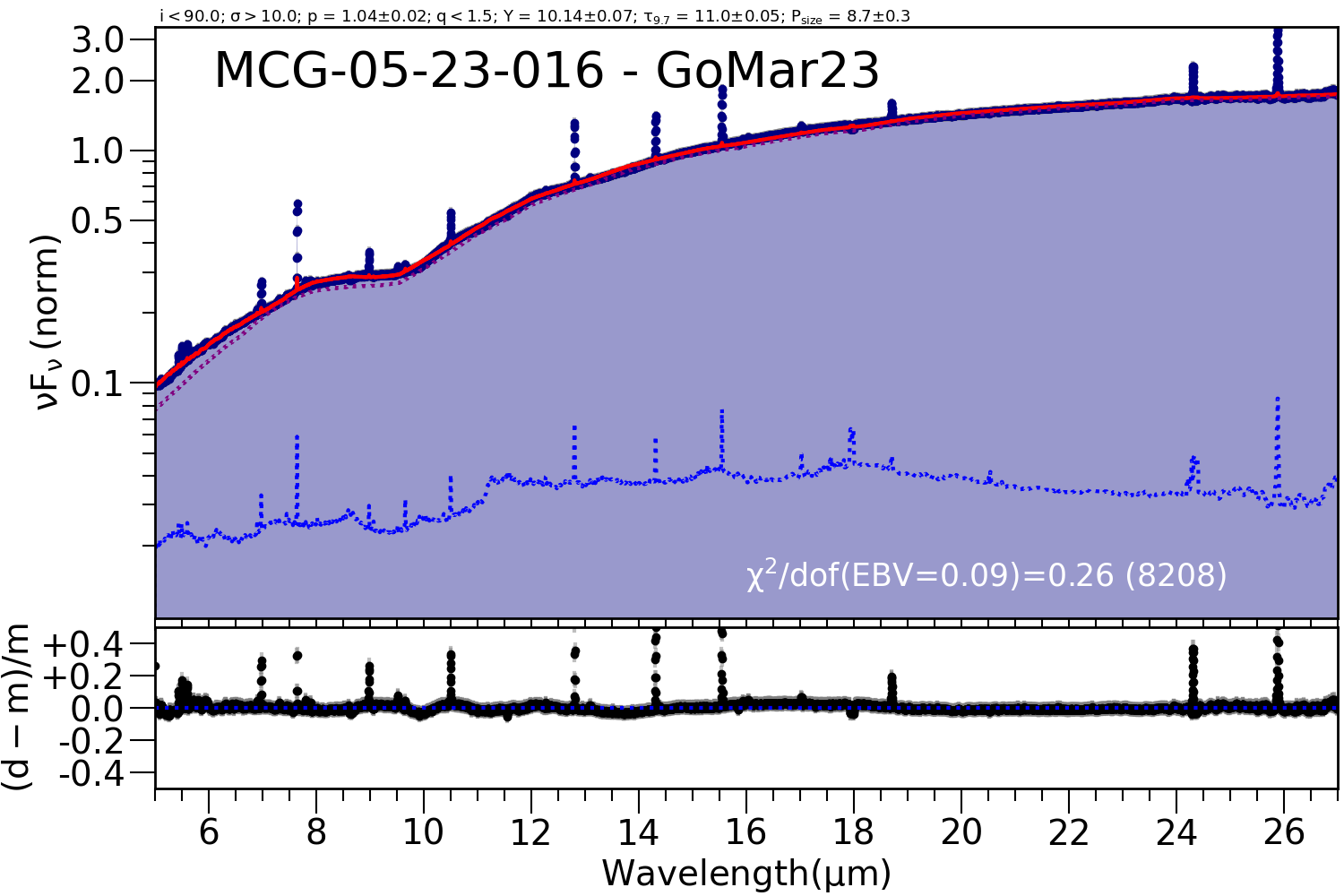} \\
\includegraphics[width=0.67\columnwidth,trim={0.2cm 0.1cm 0cm 0.65cm},clip]{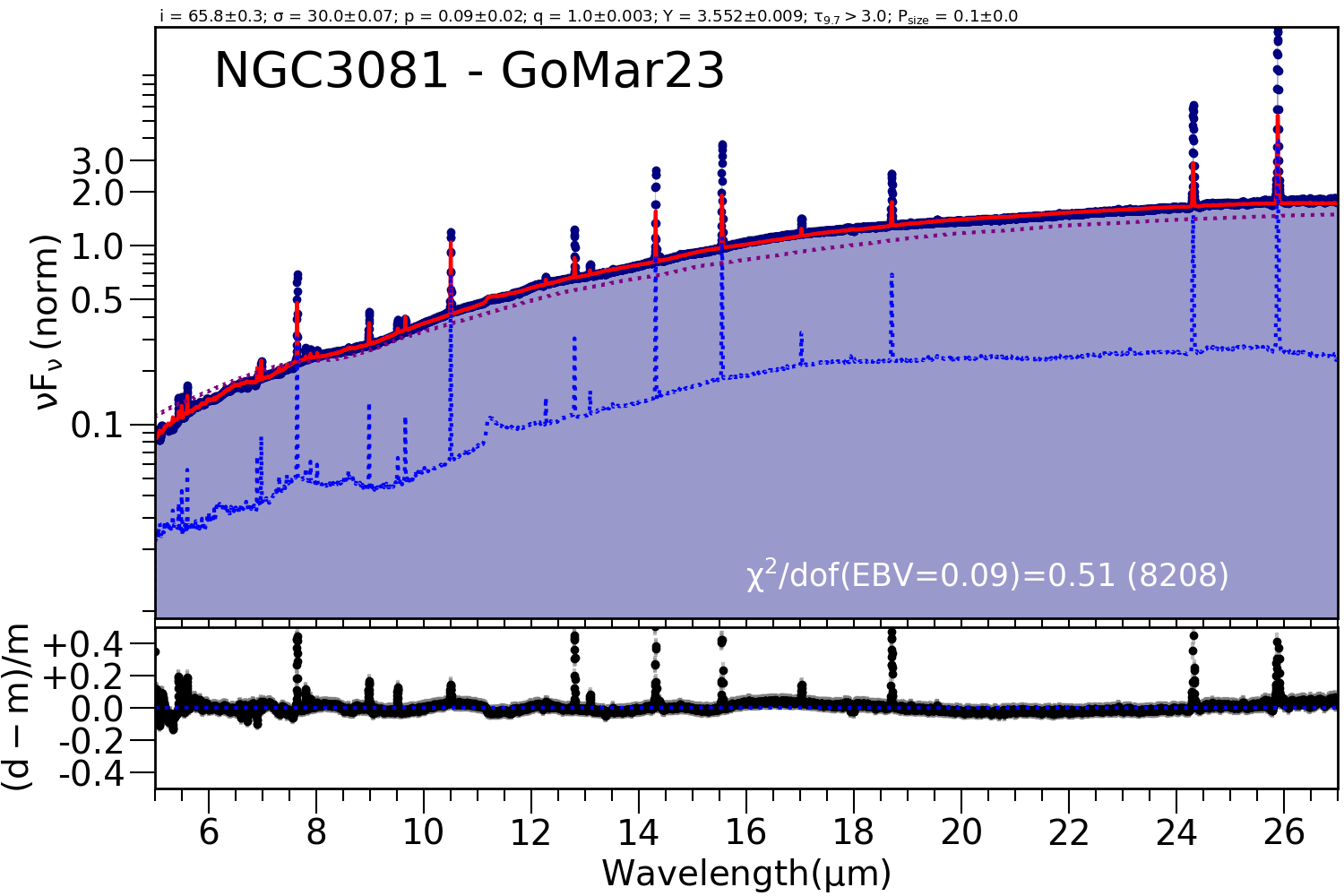}
\includegraphics[width=0.67\columnwidth,trim={0.2cm 0.1cm 0cm 0.65cm},clip]{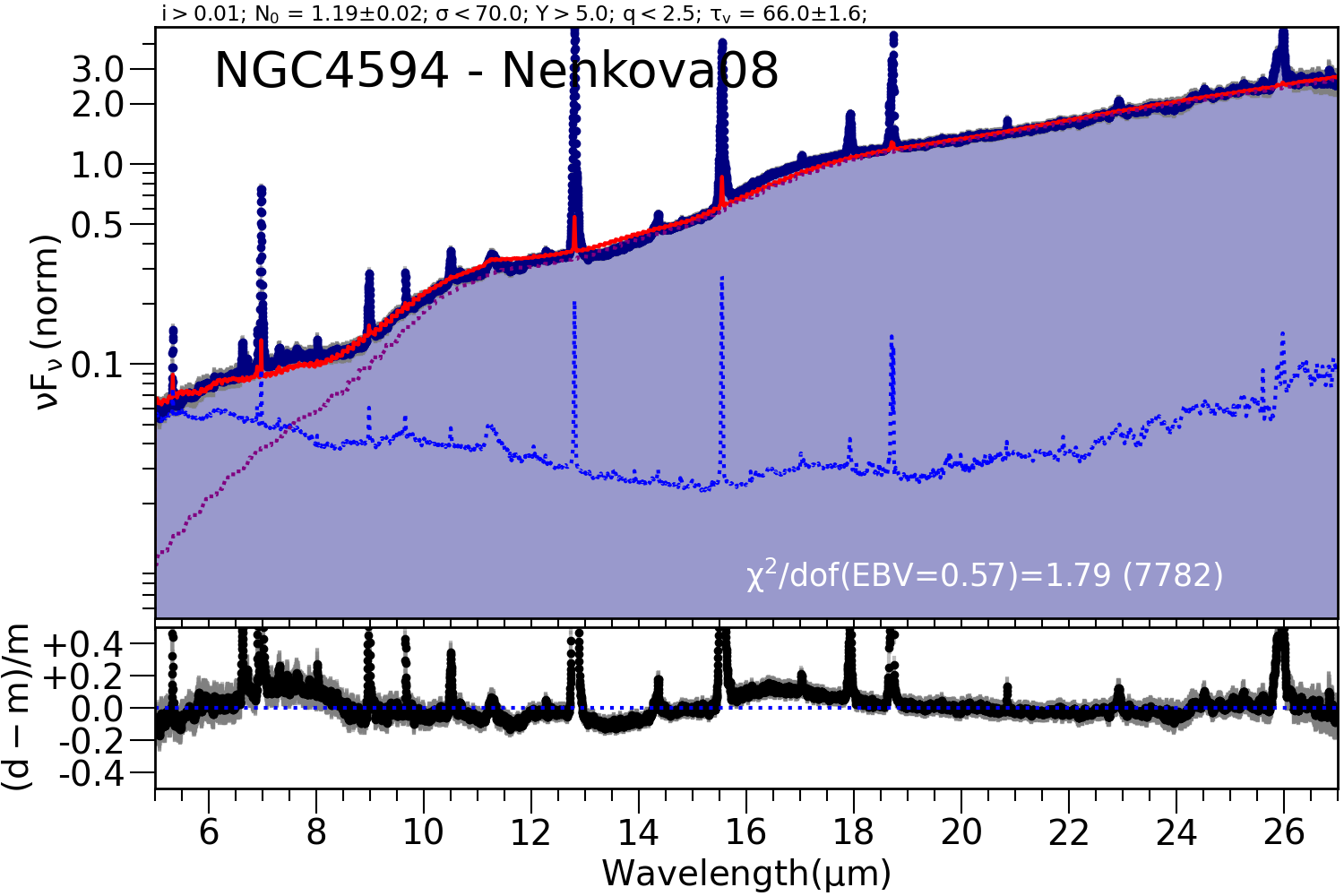}
\includegraphics[width=0.67\columnwidth,trim={0.2cm 0.1cm 0cm 0.65cm},clip]{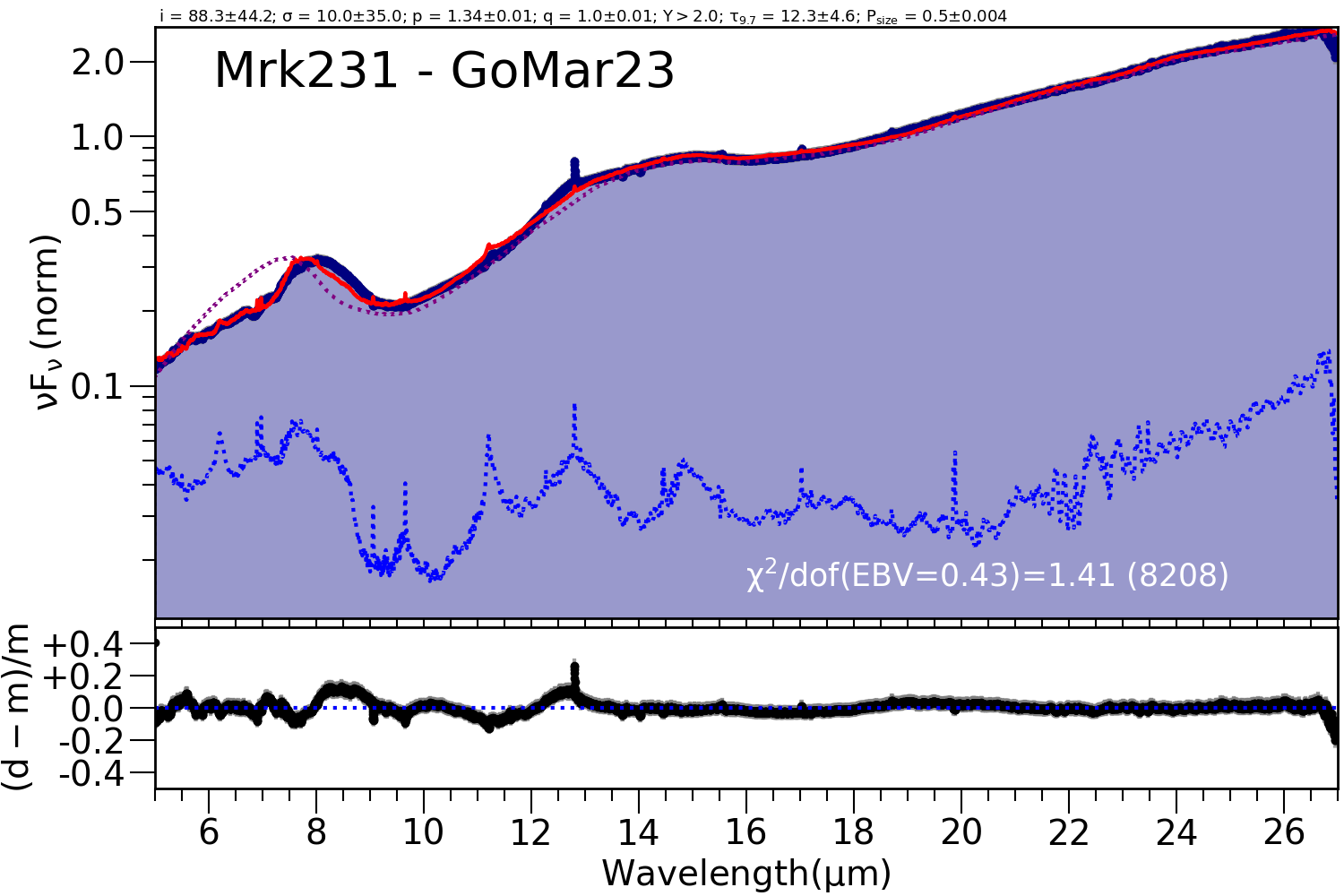} \\
\includegraphics[width=0.67\columnwidth,trim={0.2cm 0.1cm 0cm 0.65cm},clip]{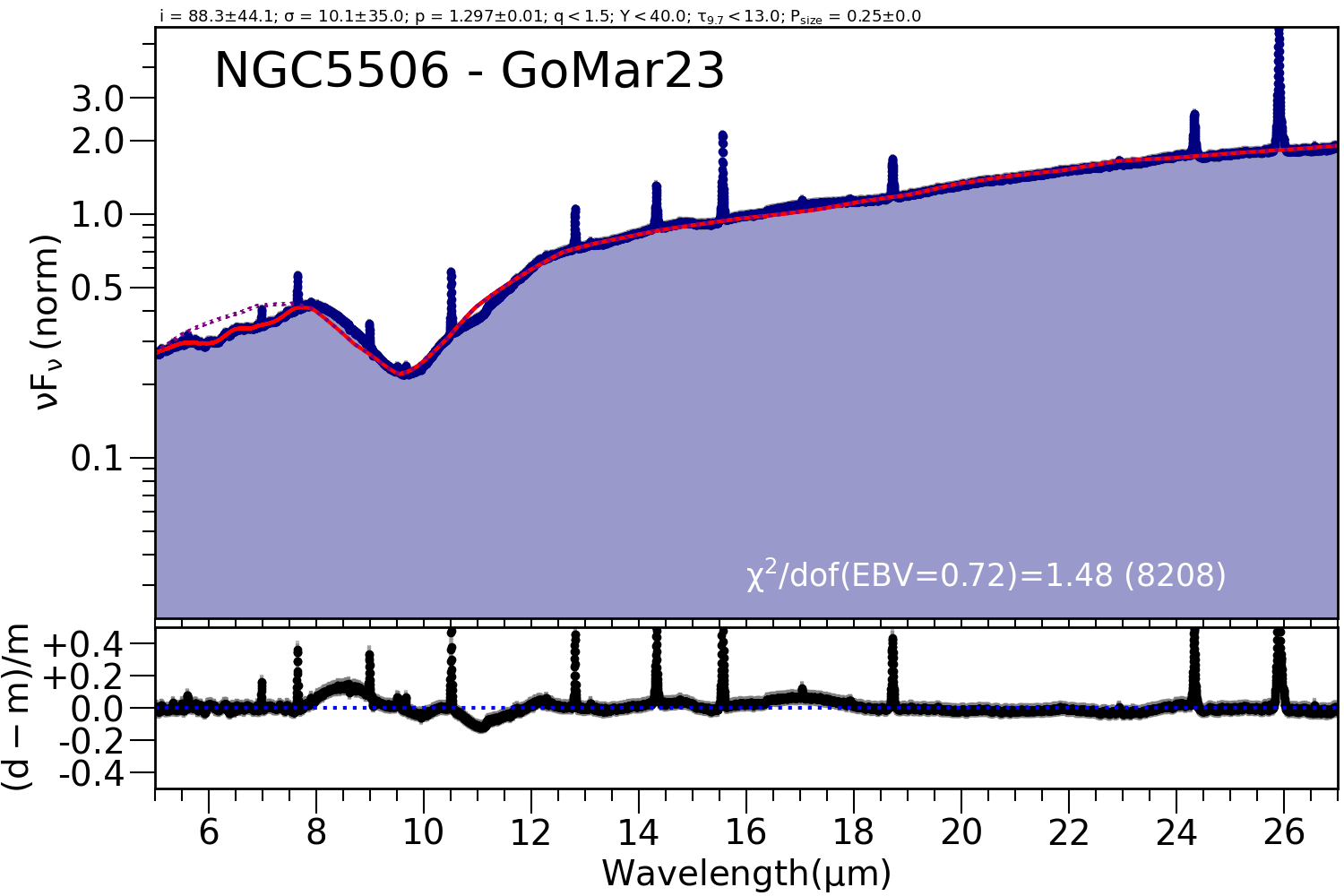}
\includegraphics[width=0.67\columnwidth,trim={0.2cm 0.1cm 0cm 0.65cm},clip]{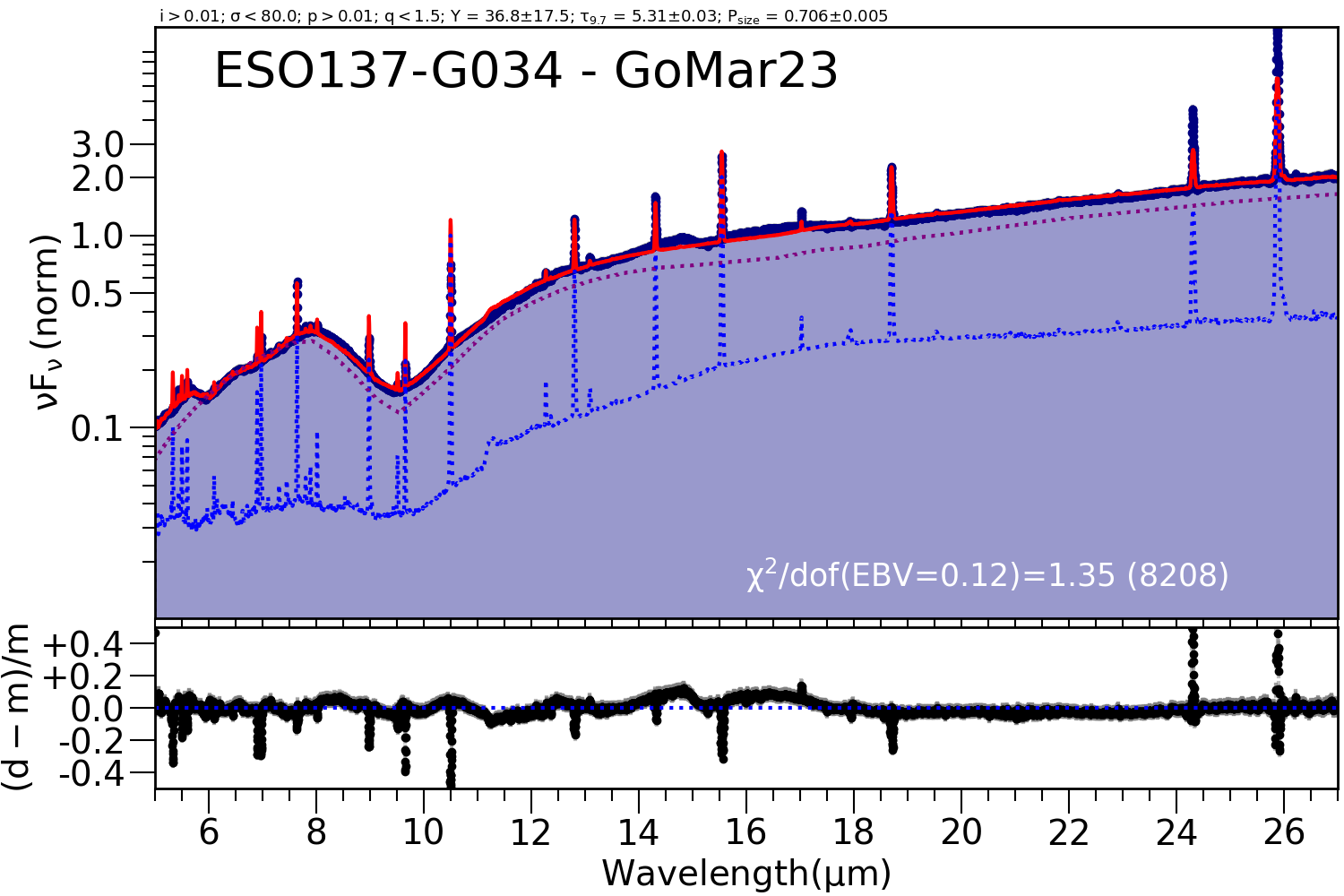}
\includegraphics[width=0.67\columnwidth,trim={0.2cm 0.1cm 0cm 0.65cm},clip]{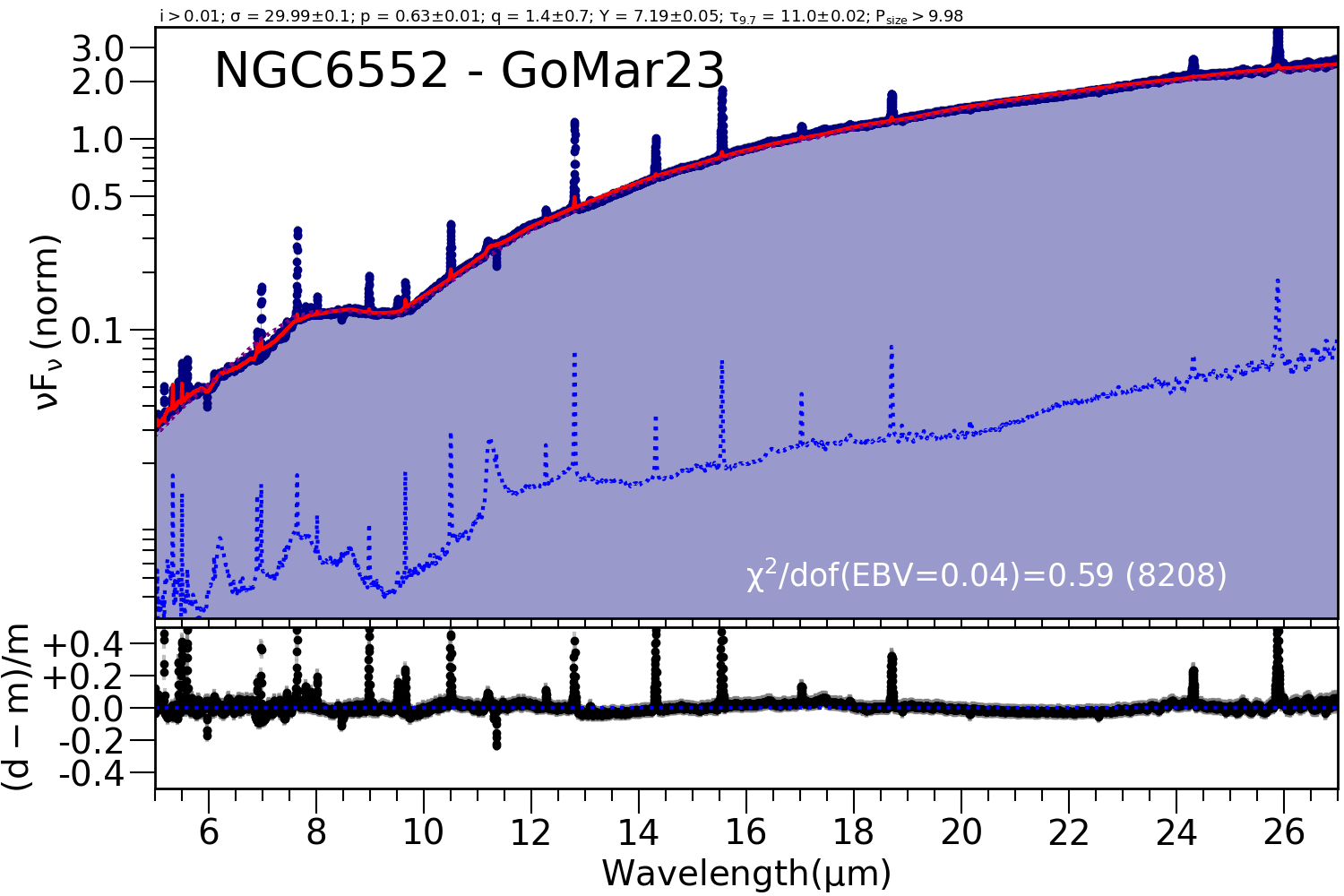} \\
\includegraphics[width=0.67\columnwidth,trim={0.2cm 0.1cm 0cm 0.65cm},clip]{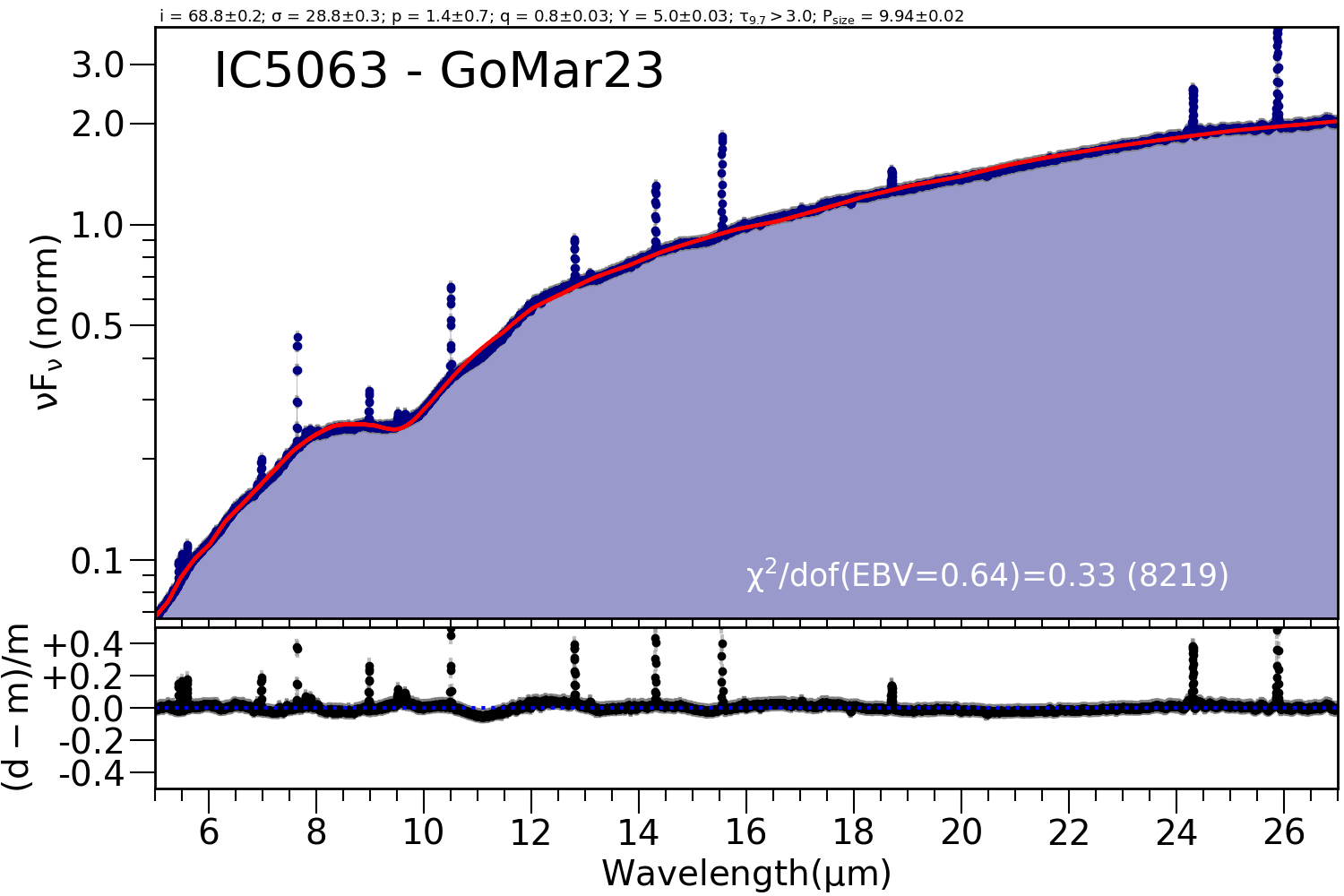}
\includegraphics[width=0.67\columnwidth,trim={0.2cm 0.1cm 0cm 0.65cm},clip]{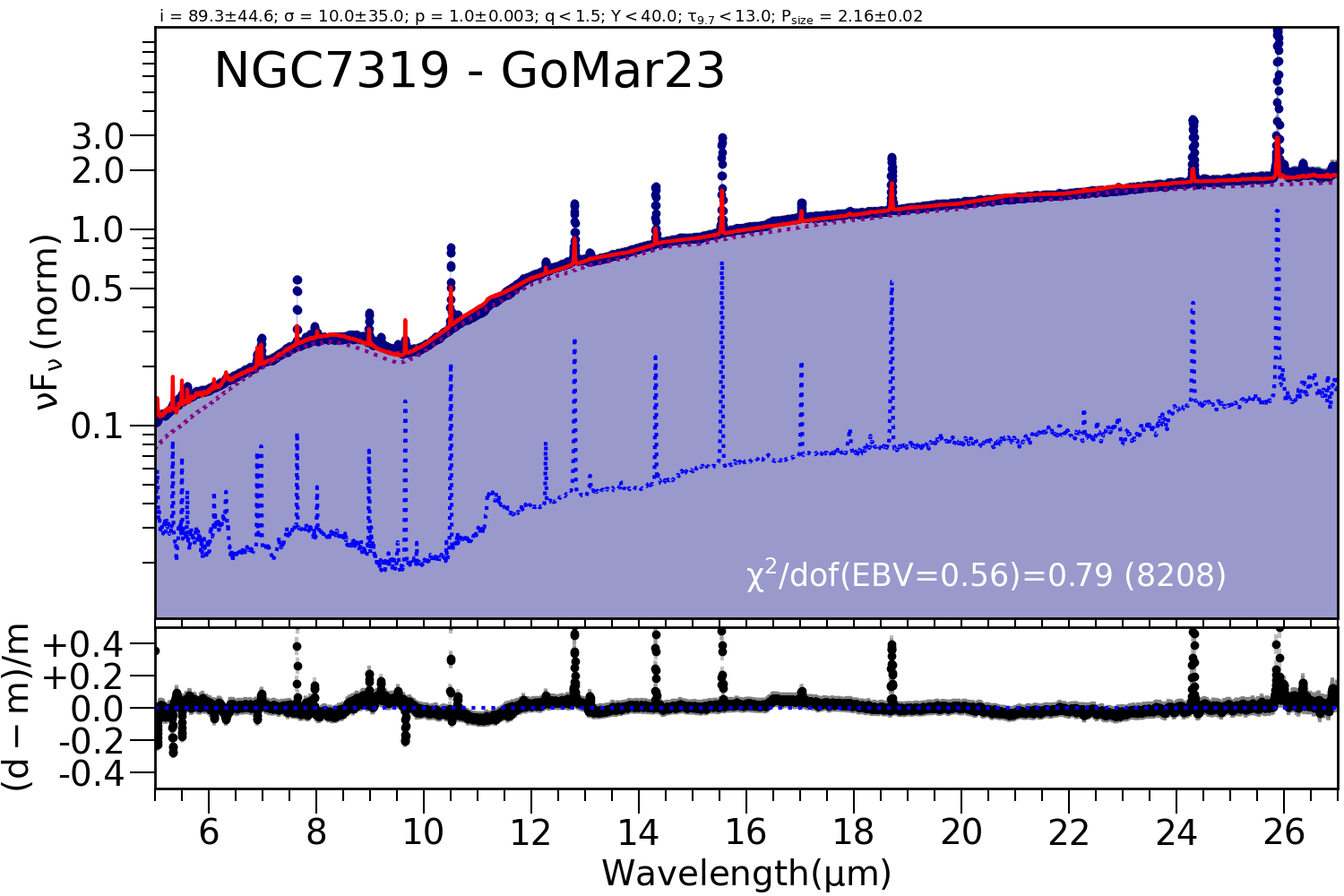}
\includegraphics[width=0.67\columnwidth,trim={0.2cm 0.1cm 0cm 0.65cm},clip]{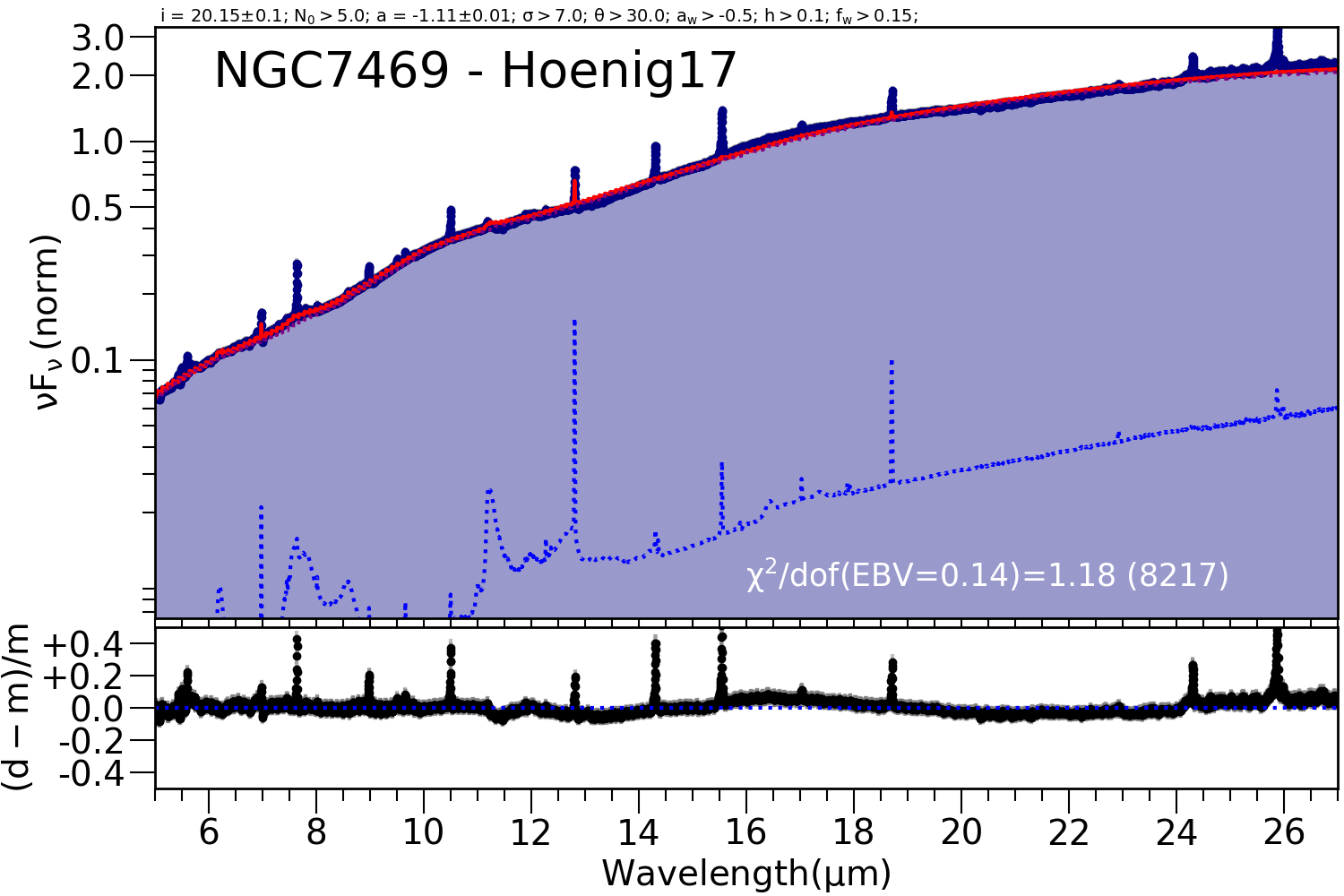}
\caption{Objects well fitted to our baseline models. The spectrum is shown in black and filled in with blue. The best fit is shown with a red continuum line, and the AGN dust model and host galaxy contribution are shown with purple and blue (dotted) lines, respectively. The bottom panels show the residuals. Note that emission lines are included here but excluded to perform the spectral fit. \label{fig:goodfits}}
\end{figure*}

\begin{figure*}
\includegraphics[width=0.67\columnwidth,trim={0.2cm 0.1cm 0cm 0.65cm},clip]{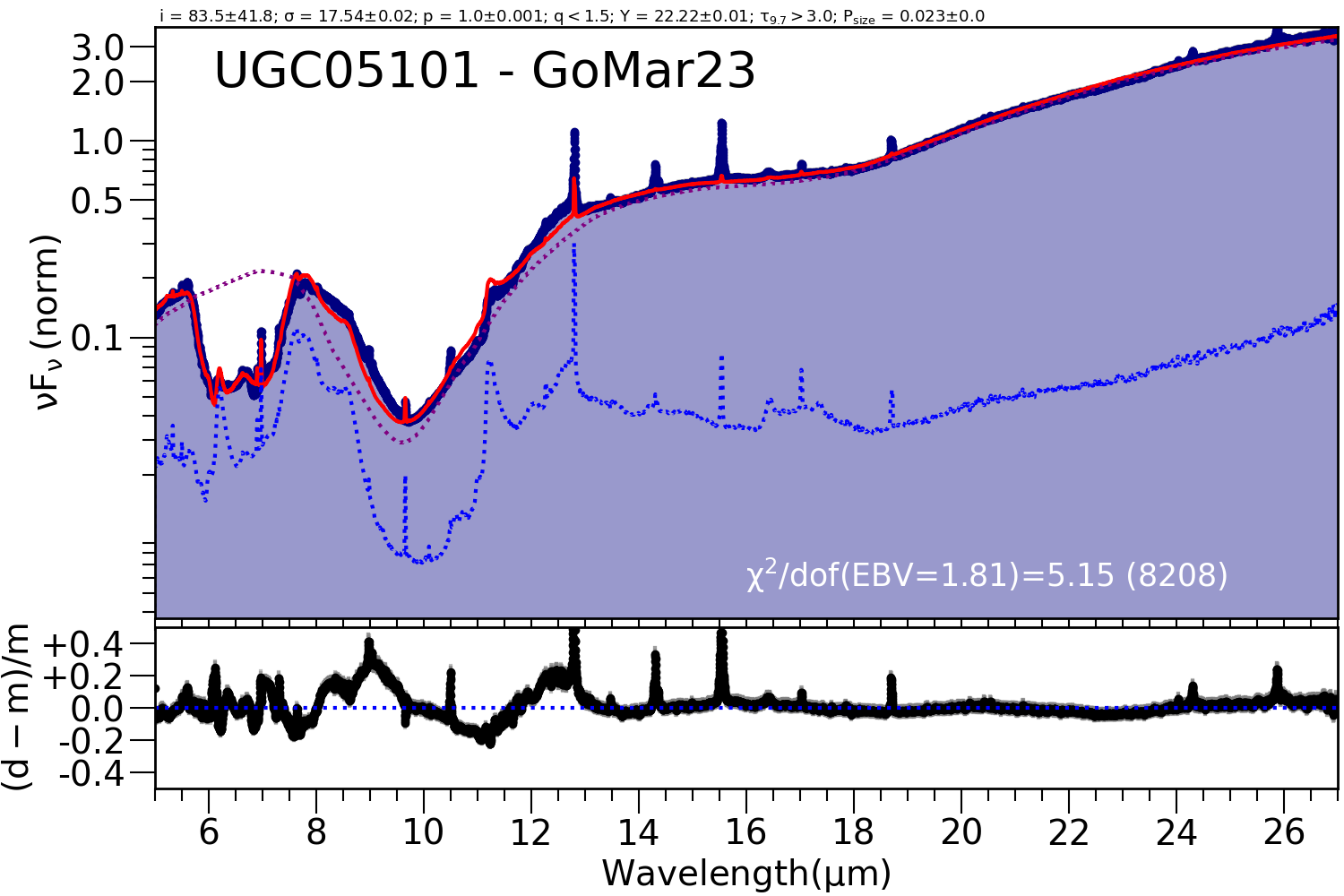}
\includegraphics[width=0.67\columnwidth,trim={0.2cm 0.1cm 0cm 0.65cm},clip]{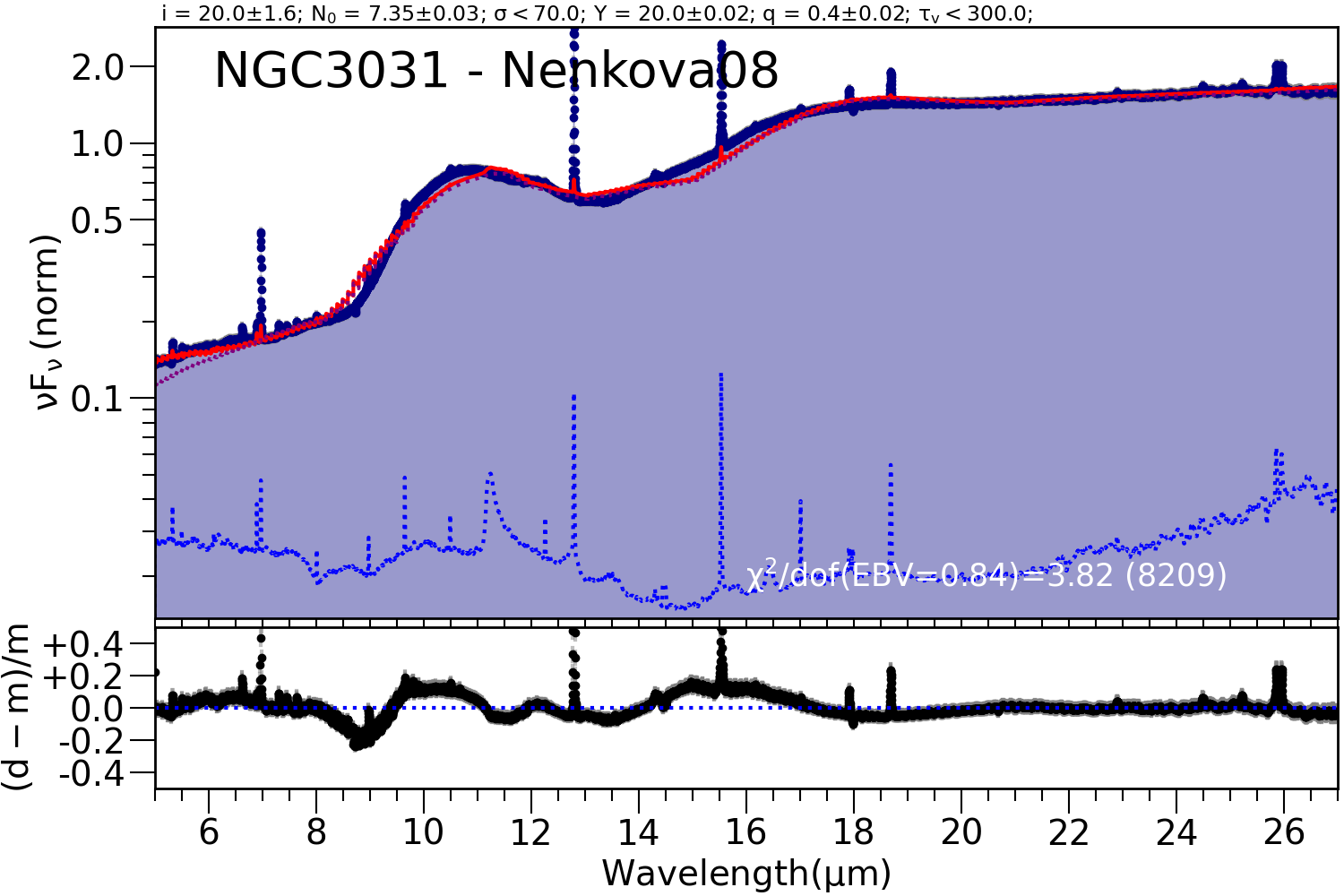}
\includegraphics[width=0.67\columnwidth,trim={0.2cm 0.1cm 0cm 0.65cm},clip]{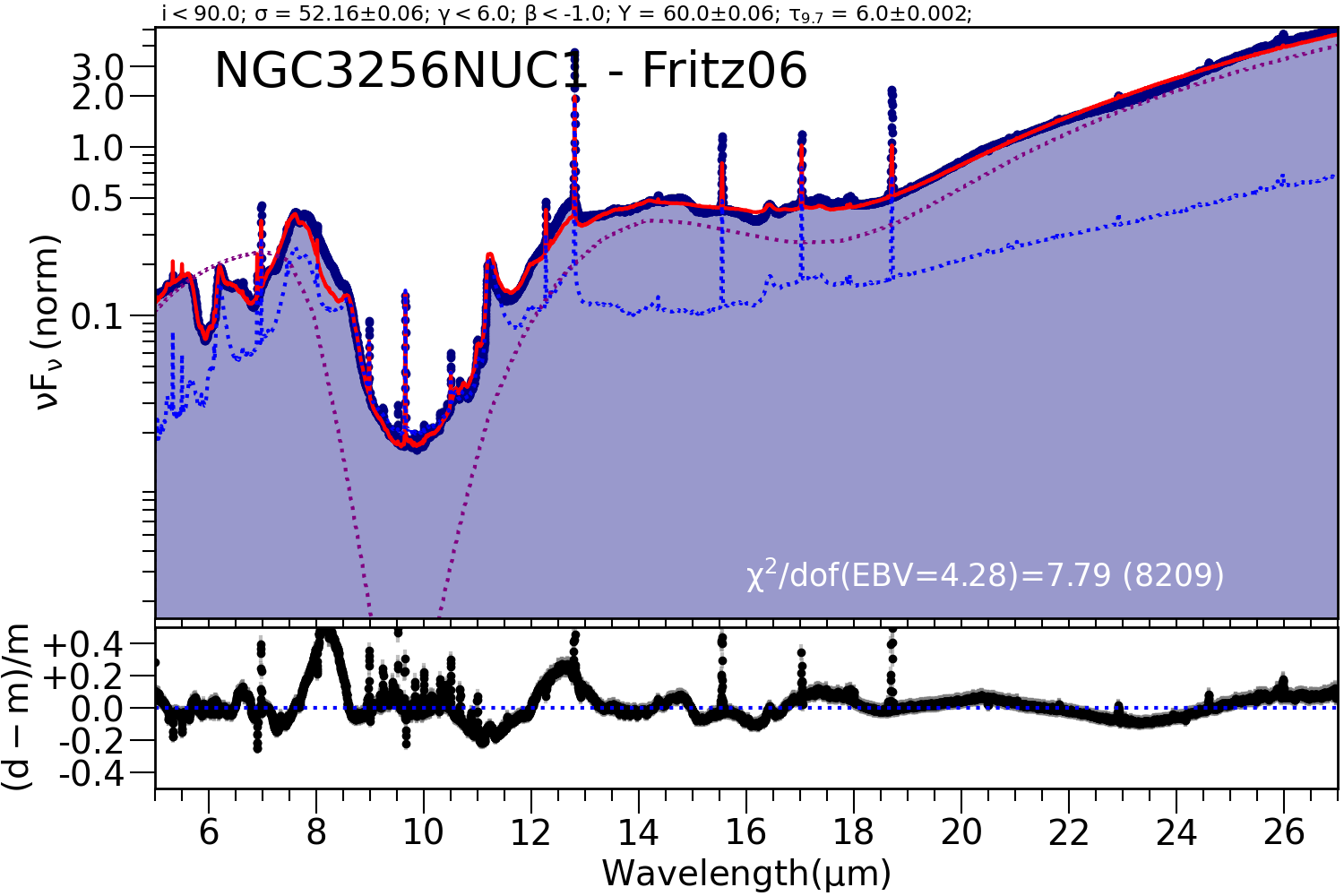} \\
\includegraphics[width=0.67\columnwidth,trim={0.2cm 0.1cm 0cm 0.65cm},clip]{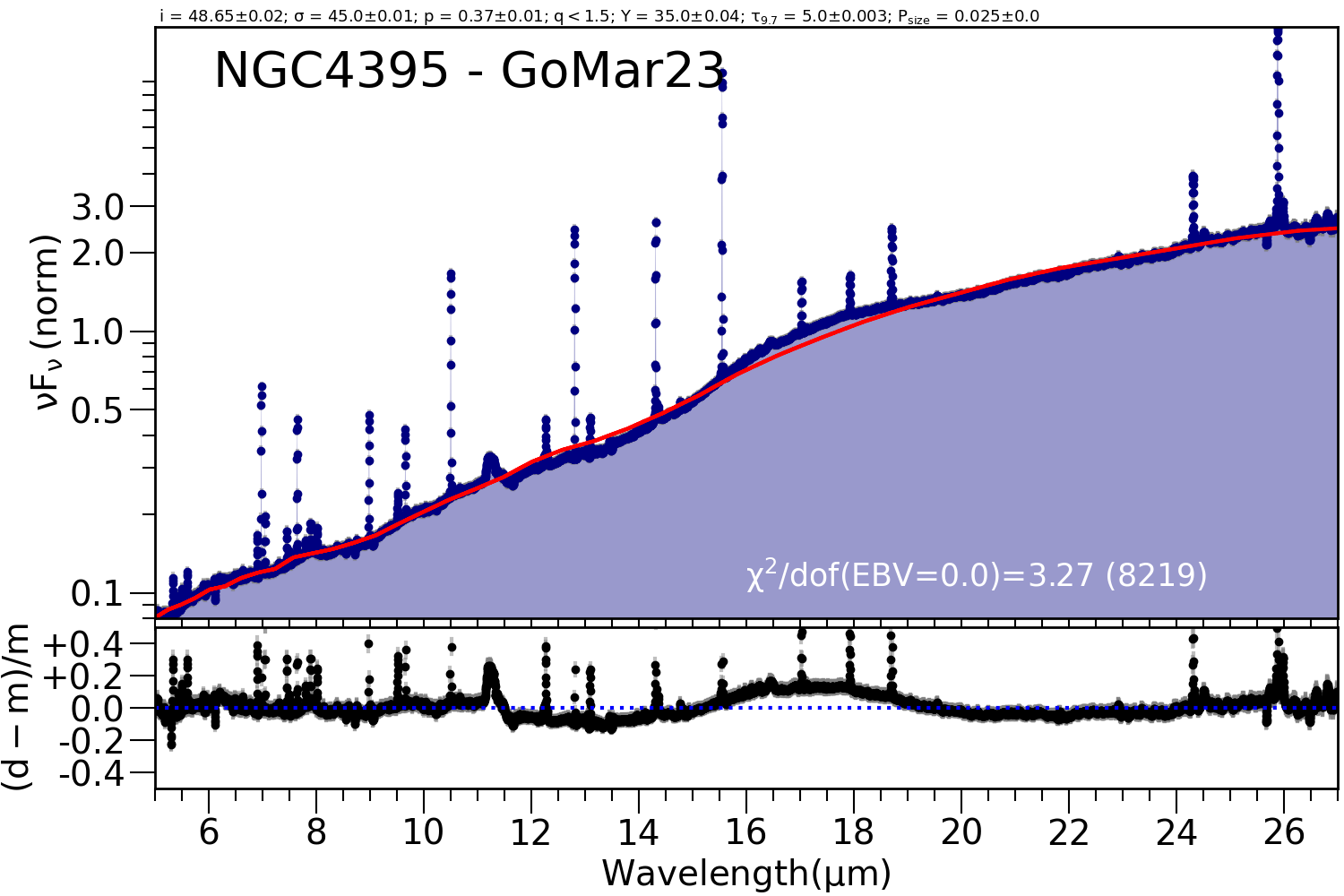}
\includegraphics[width=0.67\columnwidth,trim={0.2cm 0.1cm 0cm 0.65cm},clip]{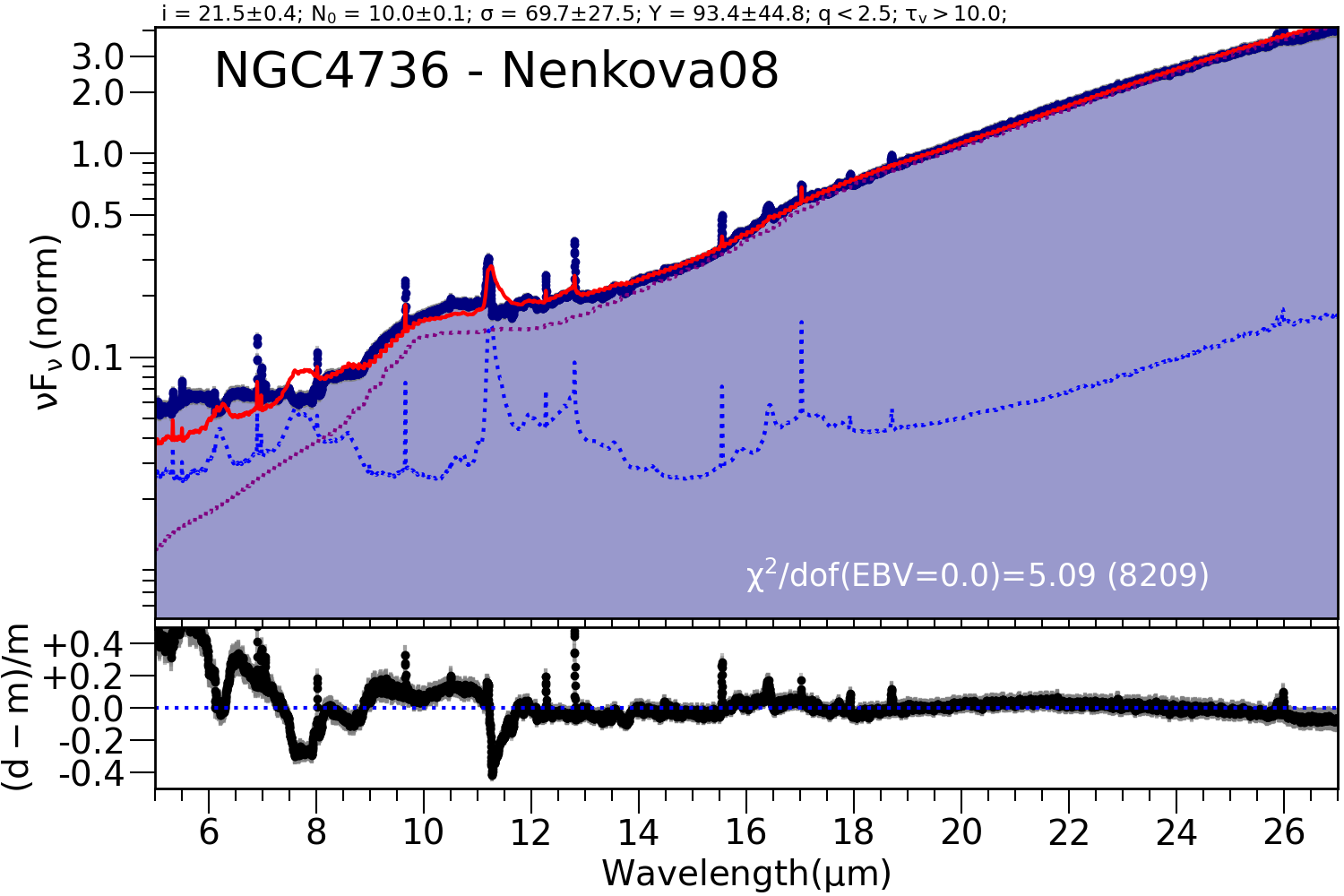} 
\includegraphics[width=0.67\columnwidth,trim={0.2cm 0.1cm 0cm 0.65cm},clip]{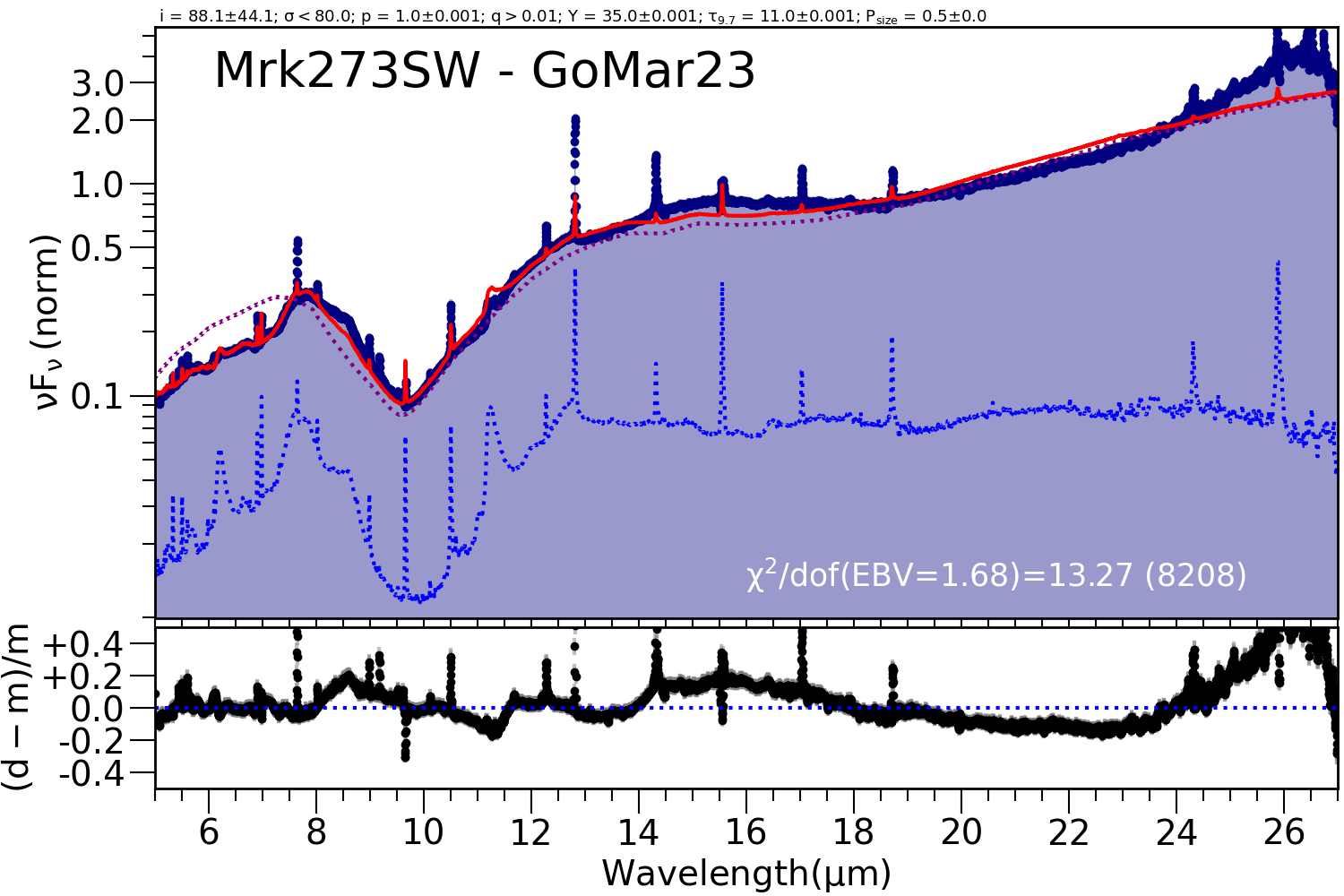} \\
\includegraphics[width=0.67\columnwidth,trim={0.2cm 0.1cm 0cm 0.65cm},clip]{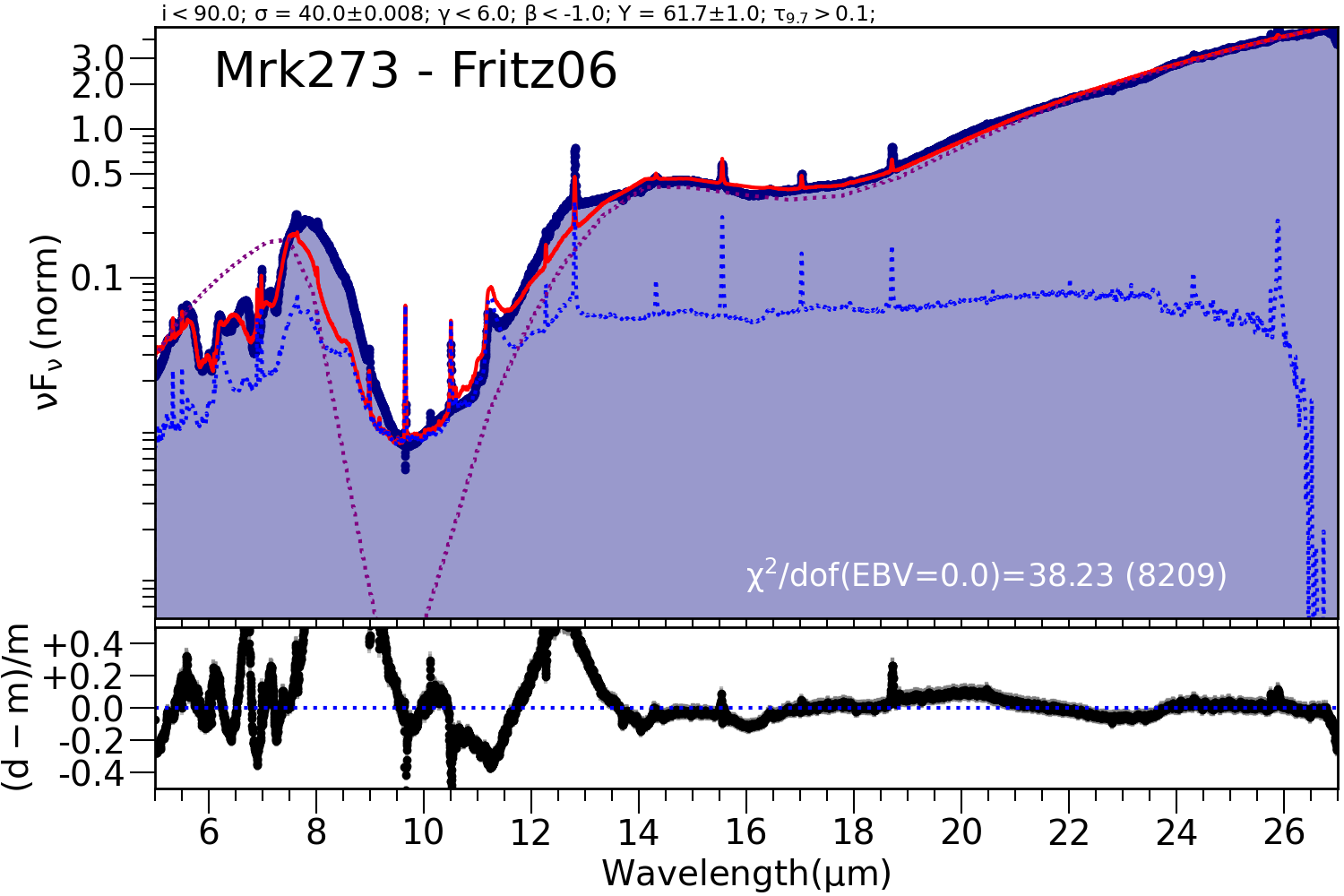}
\includegraphics[width=0.67\columnwidth,trim={0.2cm 0.1cm 0cm 0.65cm},clip]{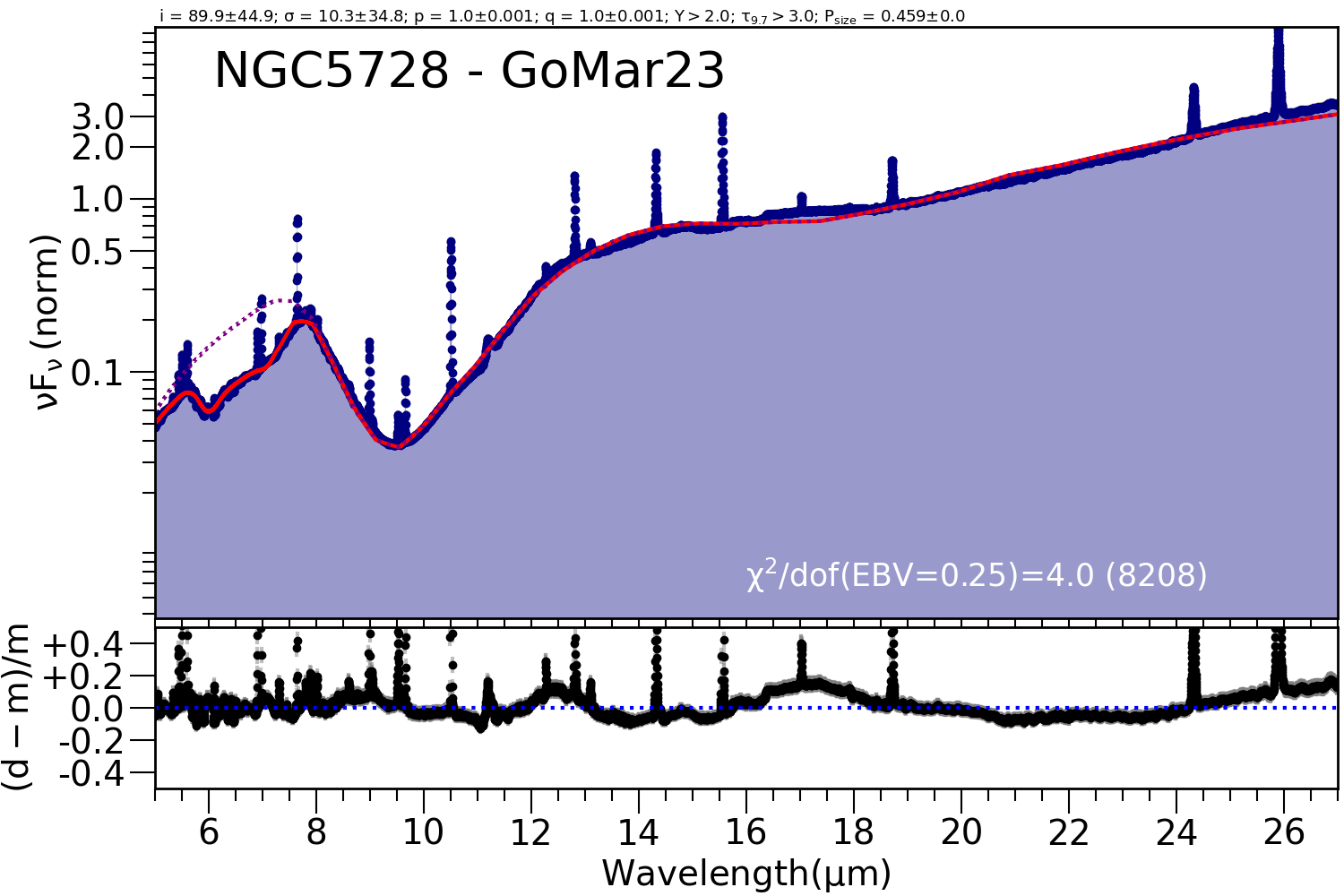} 
\includegraphics[width=0.67\columnwidth,trim={0.2cm 0.1cm 0cm 0.65cm},clip]{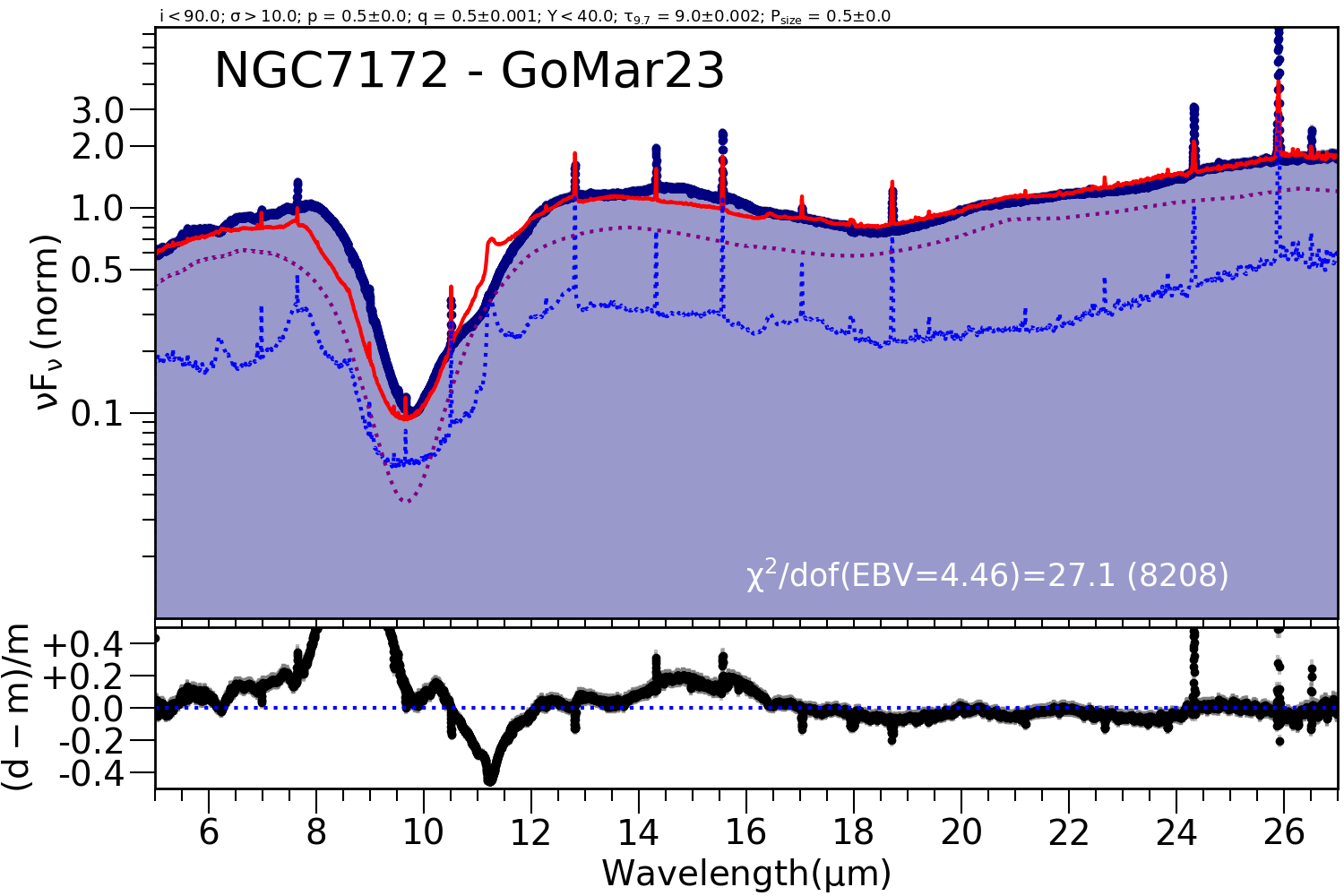}
\caption{Objects where a good fit was not found with any tested baseline models. The spectrum is shown in black and filled in with blue. The best fit is shown with a red continuum line, and the AGN dust model and host galaxy contribution are shown with purple and blue (dotted) lines, respectively. The bottom panels show the residuals. Note that emission lines are included here, but they are excluded to perform the spectral fit. \label{fig:poorfits}}
\end{figure*}

\subsection{Spectral fitting to AGN dust models}

Fig.\,\ref{fig:ChiLum} shows the resulting reduced $\rm{\chi^2}$ statistics for each of the models (from top to bottom), comparing the results for the various baseline models tested. Circles, squares, diamonds, and stars show the results for the AGN dust models alone, AGN dust model plus absorption features, AGN dust models plus host galaxy, and AGN dust models plus absorption features and host galaxy, respectively. {The top panel of Fig.\,\ref{fig:ChiDist} compares the results among the different models using the template that includes AGN dust models plus absorption features and host galaxy contributions sorted out by luminosity}. Objects showing $\rm{\chi_{r}^{2}<2}$ (shaded area in Figs.\,\ref{fig:ChiLum} and \ref{fig:ChiDist}) are marked with a tick in Col.\,8 of Table\,\ref{tab:bestfitmodel}. Objects in this plot are sorted out by their X-ray luminosities, lowest toward the left and highest toward the right. We also explore the effect of different spatial resolutions achieved by the data in Fig.\,\ref{fig:ChiDist} (bottom panel), sorting out the objects by the resolution power. 

The best performance is obtained with the latest AGN dust model, the two-phase torus model by \citet{Gonzalez-Martin23}, which best reproduced 12 out of the 21 AGN. Interestingly, the nine objects that cannot be reproduced by this model cannot be reproduced well with any other model. Mrk\,231, NGC\,5506, and ESO\,137-G034 can only be well reproduced with this model. Only NGC\,5506 is best fitted with the canonical value of 0.25$\rm{\mu m}$ for the maximum grain size of the grain size distribution \citep[used e.g. in ][]{Stalevski16}. Four objects require a smaller grain size, and seven need larger grains (see Col.\,10 in Table\,\ref{tab:bestfitmodel}).

The clumpy torus by \citet{Hoenig10} and the clumpy disk by \citet{Hoenig17} are able to explain five objects (NGC\,1052, MCG\,-05-23-016, NGC\,3081, IC\,5063, and NGC\,7319). Interestingly, all of the other models show $\rm{\chi^2_{r}<2}$ for these five objects. Except for MCG\,-05-23-016, all are in the middle range of luminosities with $\rm{L_{X}\sim 10^{41-43}erg\, s^{-1}}$ (see Table\,\ref{tab:collection}).

The clumpy disk$+$wind model by \citet{Hoenig17} can explain six spectra. This includes those spectra of objects well reproduced by the clumpy torus and disk models above with the addition of NGC\,7469. Although this model provides the best fit for this object, the smooth torus model by \citet{Fritz06}, the clumpy torus model by \citet{Nenkova08}, the two-phase torus model by \citet{Stalevski16}, and its variant with dust grain size as free parameter by \citet{Gonzalez-Martin23} also provide $\rm{\chi^2_{r}<2}$. These four models are also able to explain NGC\,6552. The smooth torus model by \citet{Fritz06}, the clumpy torus model by \citet{Nenkova08}, and the two-phase torus model by \citet{Stalevski16} show a slightly better performance, being able to explain the mid-infrared continuum of eight objects. In addition to the previous seven objects (i.e., NGC\,1052, MCG\,-05-23-016, NGC\,3081, IC\,5063, NGC\,6552, NGC\,7319, and NGC\,7469), the smooth torus model by \citet{Fritz06} and the two-phase torus model by \citet{Stalevski16} are able to explain ESO\,420-G13 while the clumpy torus model by \citet{Nenkova08} is able to fit NGC\,4594. 

We chose as a best-fit model the one providing the minimum reduced $\rm{\chi^2}$ for each target. Table\,\ref{tab:bestfitmodel} shows the best-fitting model for each target in Col.\,2, the foreground extinction required for the best fit in Col.\,5, and the resulting reduced $\rm{\chi^2}$ statistic in {Col.\,7}. We use f-test statistics to quantify if a more complex model is needed to reproduce the data. Using this test, Table\,\ref{tab:bestfitmodel} reports the need for host galaxy contribution in Col.\,3, the absorption features in Col.\,4, {and the fractional contribution of the host galaxy template to the best fit in Col.\,6.}  

Among the 12 objects with good fits, NGC\,4594 prefers the clumpy torus model by \citet{Nenkova08}, NGC\,1052 and NGC\,7469 are best fitted to the disk$+$wind model by \citet{Hoenig17}, and the other nine objects prefer the two-phase torus model by \citet{Gonzalez-Martin23} where the dust grain size is a free parameter. The best fits for these 12 AGN are shown in Fig.\,\ref{fig:goodfits}. Nine need host galaxy contribution (except for NGC\,1052, IC\,5063, and NGC\,5506). The absorption features are present in 8 of the 12 objects (exceptions are NGC\,1052, NGC\,4594, IC\,5063, and NGC\,7469). 

Nine AGN ($\rm{\sim 40\%}$ of the targets) cannot be reproduced with any baseline models tested. Fig.\,\ref{fig:poorfits} shows the fit with the minimum $\rm{\chi^2_{r}}$ for each object. NGC\,3031, NGC\,4395, and NGC\,5728 are clearly better fitted than the others with $\rm{3<\chi^2_r<4}$. For these objects, the main issue is the shape of the silicate features at 18$\rm{\mu m}$ (perhaps also in the 9.7$\rm{\mu m}$ feature). NGC\,4395 shows PAH features, but the circumnuclear spectrum quality is insufficient to obtain a good template for the host continuum (see Fig.\,\ref{fig:decompspec}). The other six sources show $\rm{\chi^2_r>5}$. They are best fitted with a strong host galaxy contribution, and five show deep silicate absorption features (except for NGC\,4736). Furthermore, three of them (UGC\,05101, NGC\,3256NUC1, and Mrk\,273) show deep absorption features at 6-7$\rm{\mu m}$. Our best fit fails at reproducing the spectral shape at $\rm{\sim 8\mu m}$ and around the [NeII] emission line at 12.8$\rm{\mu m}$. Optical and mid-infrared diagnostic emission-line diagrams as those presented by \citet{Martinez-Paredes23} (e.g. [NeIII]/[NeII] versus  [OIV]/[NeIII]) could help to understand if excitation due to far-UV emission produced by post-AGB stars or the AGN could contribute to the emission. For this purpose, AGN-free circumnuclear maps such as those produced here might be useful. On the other hand, this lack of emission at $\rm{\sim}$8 and 13$\rm{\mu m}$ could be due to a different chemical composition because adding amorphous olivine \citep[see][]{Reyes-Amador24} or christalline olivine \citep[][]{Tsuchikawa22} has been demostrated to help to reproduce the silicates. Interestingly, \citet{Tsuchikawa22} found that the mineralogical composition of silicate dust in heavily obscured AGN is similar to the circumstellar silicate ejected from mass-loss stars in our Galaxy. This might suggest that silicate in heavily obscured AGN is newly formed dust due to the recent circumnuclear starburst activity in these mergers. The wrong spectral shape at $\rm{\sim 13\mu m}$ could also be related to the libration mode of the water ice at 12-15$\rm{\mu m}$ \citep[][]{Garcia-Bernete24A,Garcia-Bernete24B}. 

{Fig.\,\ref{fig:ChiDist}} shows that tested baseline models are better suited for intermediate luminosities (exception is NGC\,5728), failing to reproduce high-luminosity infrared galaxies (Mrk\,273/Mrk\,273SW and UGC\,05101), or highly embedded with prominent nuclear dust lanes (NGC\,7172). Low-luminosity systems like NGC\,4736 or NGC\,3031, although better reproduced with the clumpy torus model \citet{Nenkova08} (see Fig.\,\ref{fig:ChiDist}, top panel), are not well fitted with any of the explored baseline models. This does not seem to be related to the different spatial scales involved, as depicted in the bottom panel of Fig.\,\ref{fig:ChiDist}, where we sort out our results from the highest spatial resolution (left) to the lowest (right).

\section{Discussion} \label{sec:discussion}

Until the advent of extremely large single-aperture telescopes, the AGN dust continuum morphology (usually referred to as AGN dusty torus) will not be unambiguously spatially resolved \citep[][]{Nikutta21A, Nikutta21B}, except for the brightest AGN with interferometric observations with VLTI/MIDI \citep[][]{Hoenig12, Burtscher13, Tristram14, Lopez-Gonzaga16, Hoenig17, Leftley18} and MATISSE \citep[][]{Isbell22, Gamez-Rosas22}, although at near-infrared GRAVITY observations have also helped to understand the geometry and size of the inner and hottest dust \citep[][]{GRAVITY20,GRAVITY24}. So far, interferometric observations at mid-infrared have been possible only in five of the AGN included in this analysis \citep[NGC\,1052, MCG-05-23-16, Fairall\,9, NGC\,5506, and NGC\,7469][]{Leftly19}. Although extended emission is present in many objects, the torus emission is mostly unresolved in the mid-infrared confined to the central parsecs \citep[][and references therein]{Alonso-Herrero21} except for a few cases where the extended emission dominates \citep[see Fig.\,3 in][]{Asmus16}. At the same time, sub-mm wavelengths show larger disks (up to 50\,pc) associated with colder dust and molecular gas \citep{Garcia-Burillo21}. \emph{JWST} does not provide the spatial resolution to image the dust even for the nearest AGN, except for the aperture masking interferometry (AMI) mode reaching 1\,pc scale images for nearby AGN at near-infrared (López-Rodriguez et al. in prep.). Fortunately, the spectral fitting technique has been proven helpful in obtaining essential information on the AGN dust where other techniques are not plausible \citep{Ramos-Almeida09, Martinez-Paredes21,Garcia-Bernete22B, Gonzalez-Martin23}. 
  
In this work, we used a collection of mid-infrared AGN spectra observed with MIRI/MRS to analyze how the spectral fitting technique can be applied to this data set to infer the AGN dust properties. This section discusses two main questions: 1. Are MIRI/MRS data cubes suitable for this technique? (section.\,\ref{sec:discussion_data}); and 2. What is missing from models to explain \emph{JWST} mid-infrared observations? (section.\,\ref{sec:discussion_models}).

\begin{figure}
\includegraphics[width=1.\columnwidth,trim={0cm 0cm 0cm 0cm},clip]{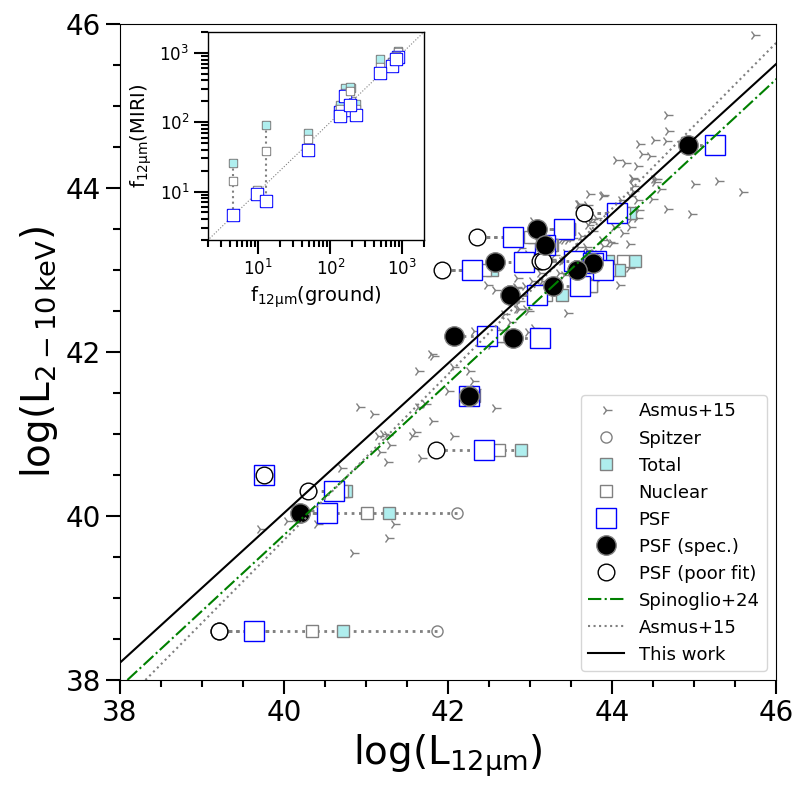}
\caption{X-ray 2-10 keV intrinsic luminosity versus the 12$\rm{\mu m}$ luminosity for the different apertures explored in this analysis: total (cyan squares), nuclear (small empty squares), and PSF (blue large squares). Large black circles show the final PSF flux obtained when the circumnuclear template decontaminates the PSF spectrum (filled and empty circles for good and poor spectral fits, respectively). Small and gray open circles show the luminosity obtained with \emph{Spitzer}/IRS spectra for comparison purposes. Green dot-dashed and gray diagonal dotted lines show the reported correlations by \citet{Spinoglio24} and \citet{Asmus15}, respectively. The correlation found in this work is shown with a continuum black line. Small gray blades show the AGN sample from \citet{Asmus15} using ground-based images. The top-left inset compares the 12$\rm{\mu m}$ fluxes (in mJy) obtained in this work with the ground-based available observations for our target collection. Several apertures of the same object are linked with horizontal dotted gray lines.\label{fig:LumLum}}
\end{figure}

\subsection{Suitability of MIRI/MRS for the SED fitting technique}\label{sec:discussion_data}

One of the main issues of using the SED fitting technique with \emph{JWST} mid-infrared observations is whether they can provide enough spatial resolution to isolate the bulk of the AGN-heated dust. As we show in Fig.\,\ref{fig:spectra} (see also Appendix\,\ref{app:decomposedspectra}), the nuclear spectra obtained with MIRI/MRS (black dotted line) show imprints of circumnuclear processes, such as PAH features or steep slopes associated to recent star-forming processes \citep{Gonzalez-Martin15}. Thanks to the IFU capabilities and superior sensitivity of MIRI/MRS, which allowed us to detect both nuclear and extended components at each slice of the data cubes, we developed the MRSPSFisol tool to produce isolated PSF and circumnuclear data cubes. To further investigate the isolation power of our method, we explore the well-known X-ray versus mid-infrared AGN correlation \citep{Lutz04,Ramos-Almeida07,Mason12,Gonzalez-Martin13,Asmus15,Spinoglio24}. Fig.\,\ref{fig:LumLum} shows this correlation for our AGN collection. The gain in tightness of the correlation when the nucleus is isolated is apparent, in particular for the low-luminosity AGN NGC\,4594 and NGC\,4736 (the two sources with the lowest X-ray luminosities in this plot), where the PSF extraction (large-empty squares with blue edges) shows an excellent agreement with the previously found correlations, compared with the total (small-blue squares) and the nuclear (small-empty squares). It is also interesting to note the agreement of PSF fluxes recovered with the decomposition technique and nuclear ground-based fluxes, when available (see inset panel in Fig.\,\ref{fig:LumLum}). This is particularly relevant for weak AGN where the nuclear extraction using the original MIRI/MRS data cubes (gray squares) is still contaminated by circumnuclear contributors. Note that the two targets with the lowest PSF contributions compared to the nuclear extraction, NGC\,4594 and NGC\,4736 (see Table\,\ref{tab:percentage}), were already reported as host-galaxy dominated using ground-based images by \citet{Mason12} with 20-30\% of PSF contribution, consistent with our results.  

We further isolated the AGN dust continuum through spectral fitting by including host-galaxy contributions to the model template. This host-galaxy contribution is non-negligible in 17 of the 21 AGN fitted. {When needed, the contribution of the host galaxy template at 12\,$\rm{\mu m}$ ranges from 50-74\% to that of the PSF spectrum (see Table\,\ref{tab:bestfitmodel}).} Fig.\,\ref{fig:LumLum} also illustrates it by showing the 12$\rm{\mu m}$ luminosity obtained from the best-fit AGN dust model for each target as large circles. For most objects, this is a minor correction, but for some of the objects for which we manage to find good fits (i.e., $\rm{\chi^2_{r}<2}$, black-filled circles), it further helps to find a good agreement with the expected relation. The Pearson value for the correlation found in this work (black line) using spectroscopic isolated values (black-filled circles) is $\rm{r=0.92}$ (p-value=$\rm{2\times 10^{-5}}$).

\subsection{Lessons learnt for new models}\label{sec:discussion_models}

The spectral fitting technique was initially applied using a single SED library of models, the so-called clumpy torus model developed by \citet{Nenkova08} \citep{Ramos-Almeida09, Ramos-Almeida11, Alonso-Herrero11, Audibert17, Garcia-Bernete19}. In recent years, several studies applied the same technique but compared different models, which permits us to test new geometries, dust size, and composition, including a \emph{disk+wind} geometry \citep[e.g.][]{Gonzalez-Martin19B, Martinez-Paredes21, Garcia-Bernete22B}. These works found that the \emph{disc+wind} model could better represent the geometry of the dust for high-luminosity AGN. Recently \cite{Gonzalez-Martin23} found that the \emph{two-phase} torus model (i.e., GoMar23) can reproduce $\rm{\sim85-90}$\% of a sample of nearby and luminous AGN observed with \emph{Spitzer} by allowing to choose the best grain size to fit the data (an aspect fixed in other SED libraries). 

So far, this technique has only been applied to single targets observed with \emph{JWST} \citep[][see also Section\,\ref{sec:intro}]{Garcia-Bernete24A,Garcia-Bernete24B}. This work is meant to start filling that gap. We find good fits for twelve AGN with \emph{JWST} (see Fig.\,\ref{fig:goodfits} and Table\,\ref{tab:bestfitmodel}). Among them, two (NGC\,1052 and NGC\,7469) prefer the \emph{disc+wind} model \citep{Hoenig17}. NGC\,1052 was already proposed as a polar-dust candidate using \emph{Spitzer} observations by \citet{Gonzalez-Martin19A} thanks to the relatively well-isolated spectrum with negligible host galaxy contributions. The PSF spectra of the two type-1 AGN NGC\,1052 and NGC\,7469 (dark-blue filled spectra in Fig.\,\ref{fig:goodfits}) look pretty similar and quite distinctive from the other sources: weak silicate emission features and a flattening of the spectrum beyond 20$\rm{\mu m}$. 

However, this polar dust emission cannot explain the mid-infrared continuum of all AGN. The clumpy torus model, widely used in pioneer works, is still the best at explaining the low-luminosity AGN NGC\,4594. Although not yielding an acceptable fit according to our criterium ($\rm{\chi^2_{r}<2}$), the other two AGN with the lowest X-ray luminosities (NGC\,3031 and NGC\,4736) are also better fit with the clumpy torus model when compared with the full set of libraries of models (see Table\,\ref{tab:bestfitmodel}). Furthermore, the model describing the larger number of good fits is the two-phase clumpy torus model presented by \citet{Gonzalez-Martin23}, which has the peculiarity that dust grain size distribution is allowed to vary beyond the canonical value of the ISM used in any other AGN dust model. As shown in Table\,\ref{tab:bestfitmodel} {(Col.\,10)}, only one object statistically prefers a grain size distribution with a maximum grain size consistent with the canonical value of 0.25$\rm{\mu m}$ used in most models. This confirms that, besides the geometry, other aspects of the dust need to be explored. This aligns with new results recently presented by \citet{Reyes-Amador24}, finding that the dust composition must also be carefully selected to better explain the silicate features.

\begin{figure}
\includegraphics[width=1.\columnwidth,trim={0.2cm 0.1cm 0cm 0.65cm},clip]{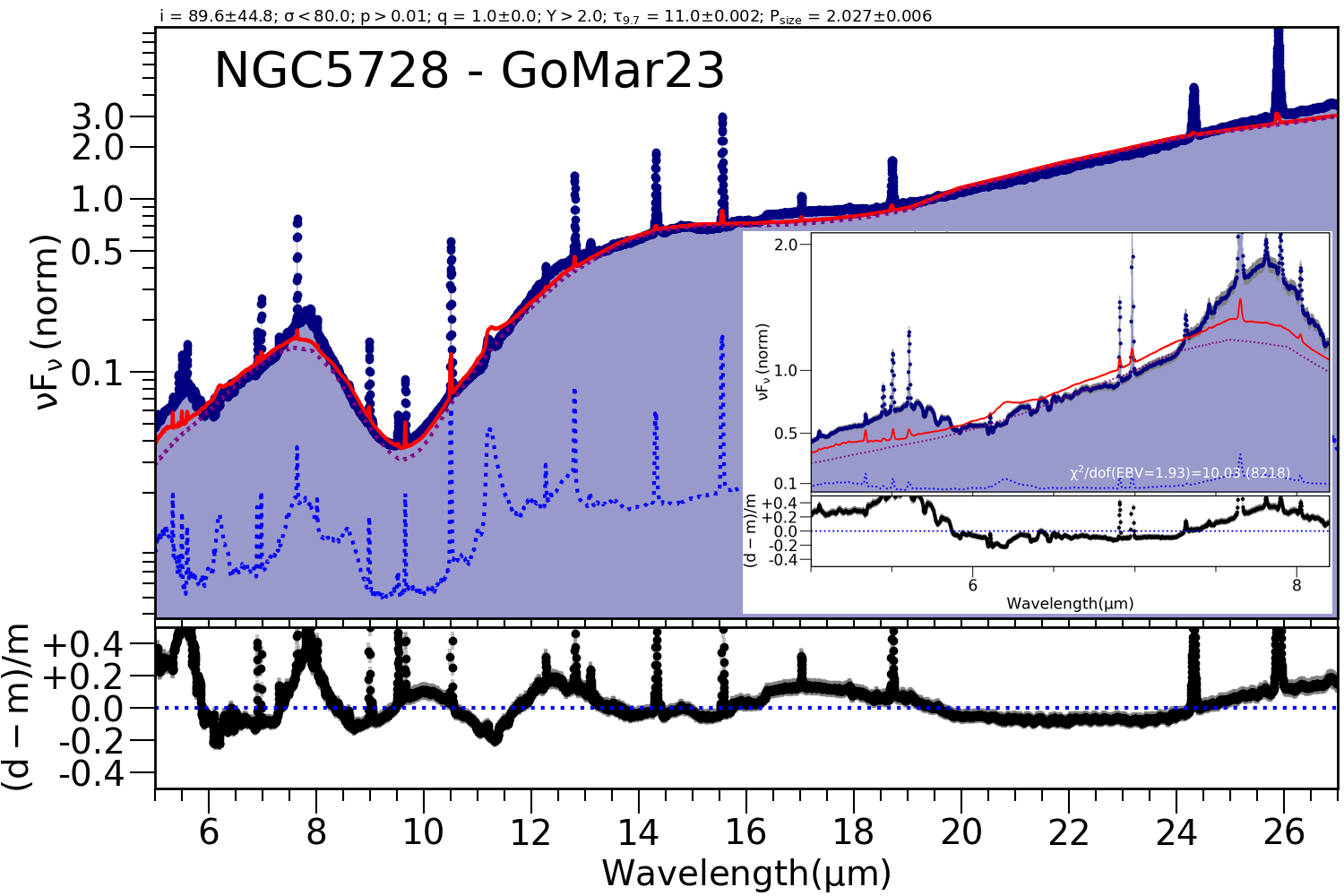} 
\includegraphics[width=1.\columnwidth,trim={0.2cm 0.1cm 0cm 0.65cm},clip]{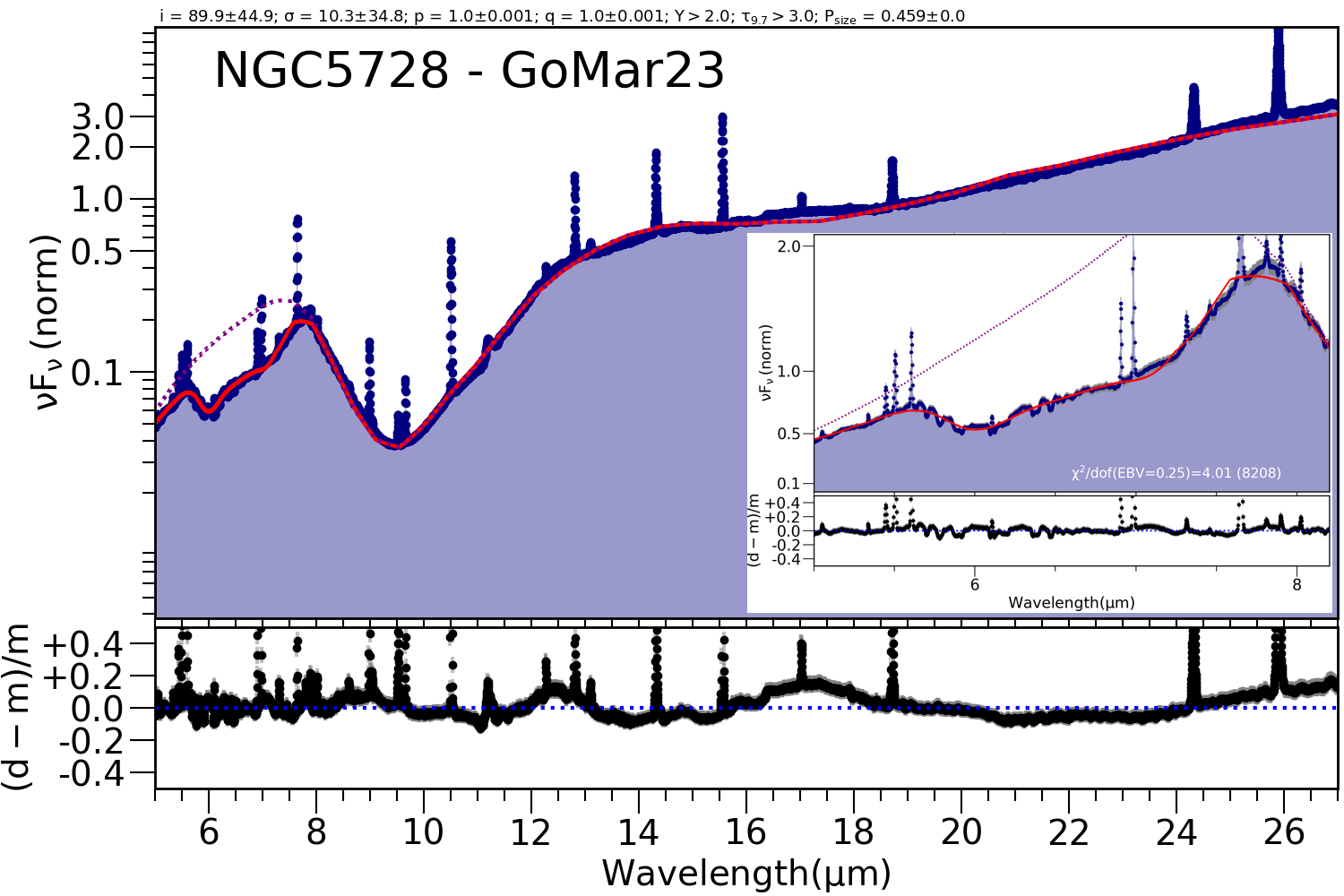} 
\caption{Spectral fit before (top) and after (bottom) the inclusion of water-ice and aliphatic hydrocarbon absorption features (located at 6-8\,$\rm{\mu m}$, see inset figure for a zoom in to this spectral range) to the final fit for NGC\,5728.  \label{fig:NGC5728water}}
\end{figure}

Poor spectral fits are still found for 9 of the 21 AGN collected. Among them, the five poorest spectral fits ($\rm{\chi^2_{r}>5}$) are found in mergers, prominent dust lanes, and/or deeply embedded systems (UGC\,05101, NGC\,3256NUC1, Mrk\,273, Mrk\,273SW, and NGC\,7172), although Mrk \,231 is also a deeply embedded system \citep{Alonso-Herrero24} which is well fitted to the two-phase flared-disk model by \citet{Gonzalez-Martin23}. These sources show strong silicate absorption features that models cannot fit. In particular, the absorption feature seems to be narrower and red-shifted than that in the model. This is clearly shown in the residuals of Mrk\,273, where two bumps are visible around the 9.7$\rm{\mu m}$ absorption feature at $\rm{\sim 8.5\mu m}$ and $\rm{\sim 12.5\mu m}$ (see the bottom-left panel in Fig.\,\ref{fig:poorfits}). The main suspects in adding deep silicate absorption features are the host galaxy star-forming contributions and foreground extinction. However, both are included in our fitting procedure. Although carefully selecting a host-galaxy template from the same observation, the properties of the circumnuclear environment could change in the close vicinity of the AGN. However, this issue should only play a minor role because the PSF luminosity for these objects agrees with X-ray luminosity (see Fig.\,\ref{fig:LumLum}), indicating that the same nuclear process dominates both. 
The derived $E(B-V)$ is high for four of these five objects ($E(B-V)>1$, shown in each panel of Fig.\,\ref{fig:poorfits}), but it does not reproduce the shape of the silicate feature. Foreground extinction by dust grains is included using the dust extinction law derived by \citet{Pei92} for the Milky Way and the Magellanic Clouds. 
New foreground extinction models might help to obtain better fits. However, it is worth noticing \citet{Garcia-Bernete22B} discarded the strong foreground extinction scenario as the primary source of the deep silicate absorption based on the PAH strength for NGC\,6552, NGC\,7319 and NGC\,7469. Therefore, new AGN dust models are needed to explain these deep silicate absorption features. One possibility to be explored is changing the chemical composition of silicate grains. Indeed, \citet{Gamez-Rosas22}, using MATISSE data of NGC\,1068 found that the silicate feature is better reproduced with olivine grains rather than the standard ISM dust mixture and \citet{Reyes-Amador24} recently found that a mix of periclase, porous alumina, and olivine grains of various sizes is a better representation of type-1 AGN. 
Furthermore, new models should include the chemistry to reproduce the water-ice and aliphatic hydrocarbon absorption features found in 16 of the 21 AGN \emph{JWST} spectra presented in this work, as suggested by \citet{Garcia-Bernete24A}. {As an example of that, Fig.\,\ref{fig:NGC5728water} shows the role of the inclusion of these absorption features to the final fit for NGC\,5728 with a zoom in to the 5-8$\rm{\mu m}$ region. It is clear how the inclusion of these absorption features not only improve the final fit (from $\rm{\chi^2_{r}=10}$ to $\rm{\chi^2_{r}=4}$) but also changes the fraction of the stellar component needed into the final fit (dotted blue line).} 

We need to be aware that we are currently fitting the spectra of the AGN for which the models were intended. According to \citet{Efstathiou95}, a successful AGN torus model must predict (moderate) absorption features at 10$\rm{\mu m}$ for edge-on views and very weak (or featureless) spectra for face-on views \citep[see also][]{Hao05}. This is based on early observations that showed that silicate emission features were weak in AGN \citep[][]{Roche91,Laor93}. Modelers have followed these indications from observers; all the objects showing a good fit in Fig.\,\ref{fig:goodfits} show either weak emission features or moderate absorption features. For instance, the clumpy torus models published by \citet{Nenkova08} were intended to reproduce spectra with these dust characteristics. The model created by \citet{Gonzalez-Martin23} is probably the first attempt to produce a model based on current observations. However, the sample used to test the AGN dust model excludes deeply embedded sources or sources where the star-forming process dominates the continuum. The lack of spatial resolution forces this pre-selection due to the relatively low spatial resolution of the \emph{Spitzer} observations used by \citet{Gonzalez-Martin23}.  
The way this could affect the results is unknown. We have developed models for what we thought was the AGN dust continuum. This seems to work well for intermediate luminosity AGN. However, it fails to reproduce low-luminosity and highly-embedded luminous AGN, as depicted in {Fig.\,\ref{fig:ChiDist}}. {Indeed, wind models should work for intermediate-to-high luminosity AGN and are not intended to reproduce low luminosities \citep{Elitzur16} or very high-luminosity (super-Eddington) systems \citep{Drewes25}.} Accurate templates for host galaxies as those produced here and new torus models better describing nearby AGN will be helpful for high-z studies where the same spatial decomposition will not be possible.

\section{Summary} \label{sec:summary}

This work is the first attempt to use MIRI/MRS observations to isolate the nuclear AGN dust continuum from the extended emission and to confront it with all available AGN dust libraries of models (i.e., torus models). We compiled MIRI/MRS observations of 21 nearby AGN observed with \emph{JWST} for that purpose. They are publicly available in the archive or obtained from the GATOS collaboration. 

We developed a tool, MRSPSFisol, to decompose MIRI/MRS data cubes into PSF and circumnuclear (extended) data cubes. Once the PSF continuum is morphologically isolated, we use the PSF spectrum to perform the spectral fitting to the AGN dust continuum. Four flavors of baseline models are tested: AGN dust model, AGN dust model with water-ice and aliphatic absorption features, AGN dust model and host galaxy using the circumnuclear spectrum as a template, and AGN dust models together with absorption features and host galaxy contribution. Altogether, we test 28 baseline models for each target (7 models and 4 baselines per model). Our main conclusions are:

%%% to the summary

\begin{itemize}

\item The AGN dust models can successfully reproduce 12 of the 21 AGN. The two-phase flared-disk model, which allows the dust grain size to vary, can reproduce all of them \citep{Gonzalez-Martin23}. %The smooth torus by \citet{Fritz06}, the clumpy torus model by \citet{Nenkova08}, and the two-phase flared-disk model by \citet{Stalevski16} fit eight objects. The clumpy disk$+$wind model by \citet{Hoenig17} fits six objects. The clumpy torus model \citet{Hoenig10} and the disk model by \citet{Hoenig17} fit five objects. 

\item Among the 12 objects with good fits, the LLAGN NGC\,4594 prefers the clumpy torus model \citep{Nenkova08}, we find indications of polar dust in NGC\,1052 and NGC\,7469 \citep[best fitted to the disk$+$wind model,][]{Hoenig17}, and the other nine objects prefer the two-phase torus model where the dust grain size is a free parameter \citep{Gonzalez-Martin23}. 

\item Among the 12 objects with good fits, a residual host galaxy contribution is needed to reproduce nine of the 12 objects, and dusty water-ice and aliphatic hydrocarbon absorption features are detected in eight objects. 

\item Nine AGN ($\rm{\sim 40\%}$ of the targets) cannot be accurately fitted to any of the models tested. They fail at reproducing deeply buried luminous and several low-luminosity AGN where these water-ice and aliphatic hydrocarbon absorption features are systematically detected.

\item MRSPSFisol works well for various morphologies, including targets with strong PSF contribution with negligible extended emission and targets with faint point-like emission embedded in a strong circumnuclear emission. 

\item The shape of isolated PSF spectra obtained with MRSPSFisol changes compared with the original nuclear spectra. They show shallower silicate absorption features and stronger silicate emission features, the PAH features are suppressed, and the overall slope of the spectra changes. 

\end{itemize}

New torus models based on the superior quality of \emph{JWST} observations, exploring, for instance, new dust chemistry, are needed to make further progress in the field.

\section*{Acknowledgements}

{The authors thank J. Leftley for his useful comments and suggestions as a reviewer of this manuscript.} OG-M thanks Luis Colina for his comments and suggestions that helped to improve the final manuscript. This work is based on the observations of NASA/ESA/CSA James Webb Space Telescope. The data were obtained from the Mikulski Archive for Space Telescopes at the Space Telescope Science Institute, which is operated by the Association of Universities for Research in Astronomy, Inc., under NASA contract NAS 5-03127 for \emph{JWST}. These observations are associated with program \#01670. Support for program \#01670 was provided by NASA through a grant from the Space Telescope Science Institute, which is operated by the Association of Universities for Research in Astronomy, Inc., under NASA contract NAS 5-03127. This research has used dedicated servers named Ocelote, Guajolote, and Ajolote, funded by the above projects and maintained at IRyA by Daniel Díaz-González, Miguel Espejel, Alfonso Ginori González, and Gilberto Zavala at IRyA-UNAM. All of them are gratefully acknowledged. 

This research is mainly funded by the UNAM PAPIIT project IN109123 and CONAHCyT Ciencia de Frontera project CF-2023-G-100 (PI OG-M). OG-M also acknowledges the Sabbatical program DGAPA/PASPA at UNAM. CRA, DE-A and AA acknowledge financial support from MICINN through the Juan de la Cierva program. DE-A and AA also acknowledge support from the Agencia Estatal de Investigaci\'on of the Ministerio de Ciencia, Innovaci\'on y Universidades (MCIU/AEI) under the grant ``Tracking active galactic nuclei feedback from parsec to kiloparsec scales'', with reference PID2022$-$141105NB$-$I00 and the European Regional Development Fund (ERDF). DE-A also acknowledges support from the Spanish Ministry of Science and Innovation through the Spanish State Research Agency (AEI-MCINN/10.13039/501100011033) through grants ``Participation of the Instituto de Astrof\'isica de Canarias in the development of HARMONI: D1 and Delta-D1 phases with references PID2019$-$107010GB100 and PID2022-140483NB-C21 and the Severo Ochoa Program 2020$-$2023 (CEX2019$-$000920$-$S)''. 
AA also acknowledges financial support from the European Union grant WIDERA ExGal-Twin, GA 101158446. IG-B is supported by the Programa Atracci\'on de Talento Investigador ``C\'esar Nombela'' via grant 2023-T1/TEC-29030 funded by the Community of Madrid. 
AA-H and LH-M acknowledge financial support from PID2021-124665NB-I00 funded by MCIN/AEI/10.13039/501100011033 and ERDF A Way of Making Europe. SG-B acknowledges support from the Spanish grant PID2022-138560NB-I00, funded by MCIN/AEI/10.13039/501100011033/FEDER, EU. 
EL-R thanks to the support of the NASA Astrophysics Decadal Survey Precursor Science (ADSPS) Program (NNH22ZDA001N-ADSPS) with ID 22-ADSPS22-0009 and agreement number 80NSSC23K1585. MP-S acknowledges support under grants RYC2021-033094-I and CNS2023-145506 funded by MCIN/AEI/10.13039/501100011033 and the European Union NextGenerationEU/PRTR. 
MP-S and EGA acknowledge funding support under grant PID2023-146667NB-I00 funded by the Spanish MCIN/AEI/10.13039/501100011033. EKSH, CP, and LZ acknowledge grant support from the Space Telescope Science Institute (ID: JWST-GO- 01670). EB acknowledges support from the Spanish grants PID2022-138621NB-I00 and PID2021-123417OB-I00, funded
by MCIN/AEI/10.13039/501100011033/FEDER, EU. MS acknowledges support from the Ministry of Science, Technological Development and Innovation of the Republic of Serbia (MSTDIRS) through contract no. 451-03-66/2024-03/200002 with the Astronomical Observatory (Belgrade). CR acknowledges support from Fondecyt Regular grant 1230345, ANID BASAL project FB210003 and the China-Chile joint research fund.

%%%%%%%%%%%%%%%%%%%%%%%%%%%%%%%%%%%%%%%%%%%%%%%%%%
\section*{Data Availability}

The data underlying this article are available in the MAST archive at https://dx.doi.org/10.17909/rm3a-jz95.



\bibliographystyle{mnras}
\bibliography{sample631}
% \bibliography{example} % if your bibtex file is called example.bib

%\begin{thebibliography}{99}
%\end{thebibliography}

%%%%%%%%%%%%%%%%%%%%%%%%%%%%%%%%%%%%%%%%%%%%%%%%%%

%%%%%%%%%%%%%%%%% APPENDICES %%%%%%%%%%%%%%%%%%%%%

\appendix

\section{Running the tool}\label{app:running}

The MRSPSFisol tool is accessible to any researcher interested upon request to o.gonzalez@irya.unam.mx. The MRSPSFisol tool runs through a list of targets in an input file the user should provide in the command line (see below). This input file should contain a line per target, including the following information separated by spaces: target name, proposal ID (PROPID), observations ID for the target (TARGETOBSID), observation ID for the background (BACKOBSID), RA, Dec, and redshift of the target. The MRSPSFisol tool expects the uncalibrated files to be located under the target name and the proposal ID (e.g., all uncalibrated files for NGC1052 are located in NGC1052/02016/) where the MRSPSFisol tool is running. Data should be downloaded from the STScI database before running the MRSPSFisol tool. Several inputs are also expected in the command line:

\begin{enumerate}
\item filename: file name where the list of targets is included. 
\item object number: row number of the target to be run within the input filename. If the number is negative, the MRSPSFisol tool will run the decomposition object-by-object basis in the same order provided through the input file.  
\item channel: channel to be run. The options are 1, 2, 3, 4, 1234, 234, 12, 34. Option 1234 should be written to run all the channels. 
\item band: grid to be chosen within the channels selected. Options are s, m, and l (for short, medium, and long, respectively)and sm, ml, and sml for several bands simultaneously. 
\item PSF: option to select between simulated (psf0) or empirical PSF (empiric). We encourage the users to use the simulated PSF option (see Section\,\ref{sec:decomposition}).  
\item initialization parameter: options are y and n. If set to ``y", it will start the analysis by cleaning any previous data cube generated. If this value is set to `n', it will use the previously found values. This option is mainly for test purposes, and it is expected to be combined with a wavelength range where the user might want to rerun the MRSPSFisol tool (following two options).
\item minimum wavelength: minimum wavelength range where the MRSPSFisol tool should be run to compute the decomposition. If this option is set to a negative number, the MRSPSFisol tool will start at the first slice of the data cube. 
\item maximum wavelength: maximum wavelength range where the MRSPSFisol tool should be run to compute the decomposition. If this option is set to a negative number, the MRSPSFisol tool will start at the first slice of the data cube. 
\item crowding: options are y and n. If set to ``y" it will assume that several sources with similar brightness are within the FOV. This will result in a location of the center of mass computed in a small area of the FOV centered at the position of the target given in the input file. It will also run the automatic extraction in half of the FOV used for general purposes. Please select n unless crowding is an issue with your target. 
\item recentering: option to select whether the MRSPSFisol tool computes the center of mass (y) or relies on the coordinates given by the user in the input file. We strongly suggest using the center of mass unless apparent difficulties in the morphology of the target prevent an accurate determination of the center of mass. 
\end{enumerate}

Note that the MRSPSFisol tool assumes there are 12 data cubes (one per channel and band) with a file name: `jw'+PROPID (e.g., 02016)+`0'+TARGETOBSID (e.g., 10) + `001\_miri\_ch'+CHANNEL (1-4)+`-'+BAND (short/medium/long) + `\_s3d\_sub.fits'. It also assumes the data cubes are background subtracted already. This is the general terminology of the outputs of \emph{JWST} MRSPSFisol tool with the addition of `\_sub' that we added when the background is subtracted from the target data cubes. The MRSPSFisol tool always goes from shorter to longer wavelength data cubes. Moreover, it does not work in parallel for several channels or bands because it is optimized to process several slices simultaneously.

\section{Decomposed spectra}\label{app:decomposedspectra}

This appendix includes all the decomposed spectra for our AGN collection. 

\begin{figure*}
\includegraphics[width=1.\columnwidth,trim={1.3cm 0.cm 3.cm 1.75cm},clip]{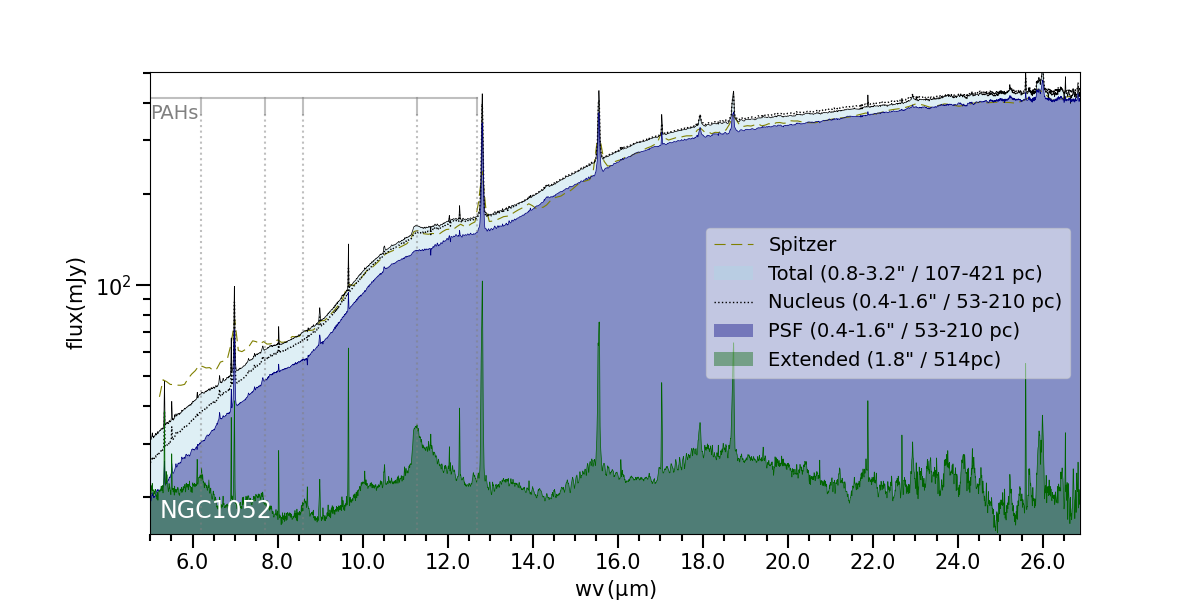}
\includegraphics[width=1.\columnwidth,trim={1.3cm 0.cm 3.cm 1.75cm},clip]{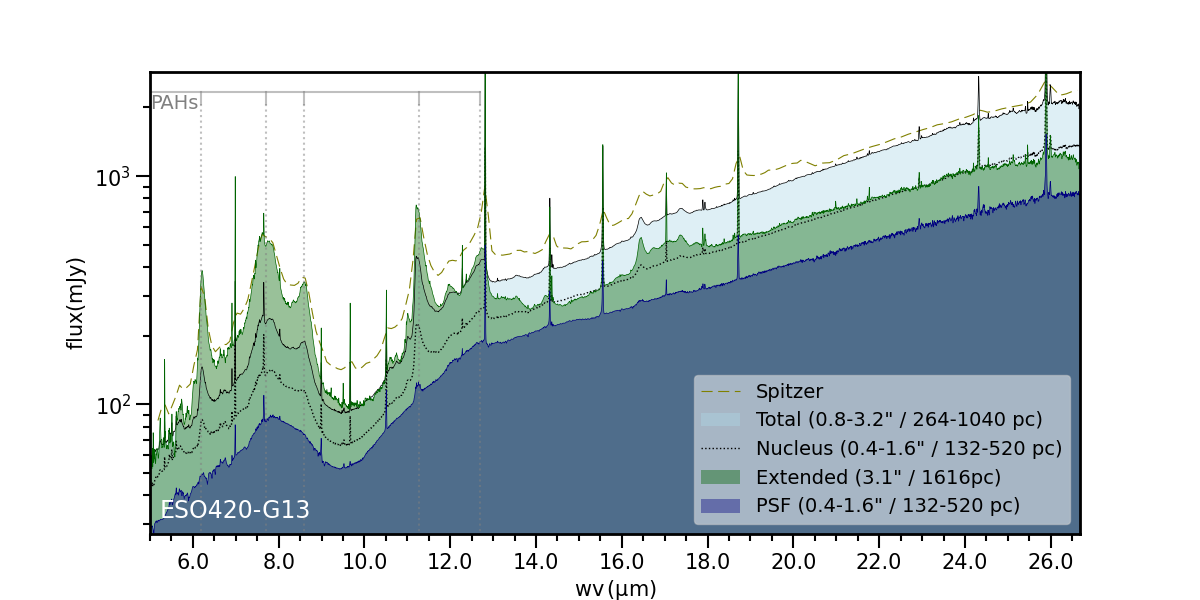} \\
\includegraphics[width=1.\columnwidth,trim={1.3cm 0.cm 3.cm 1.75cm},clip]{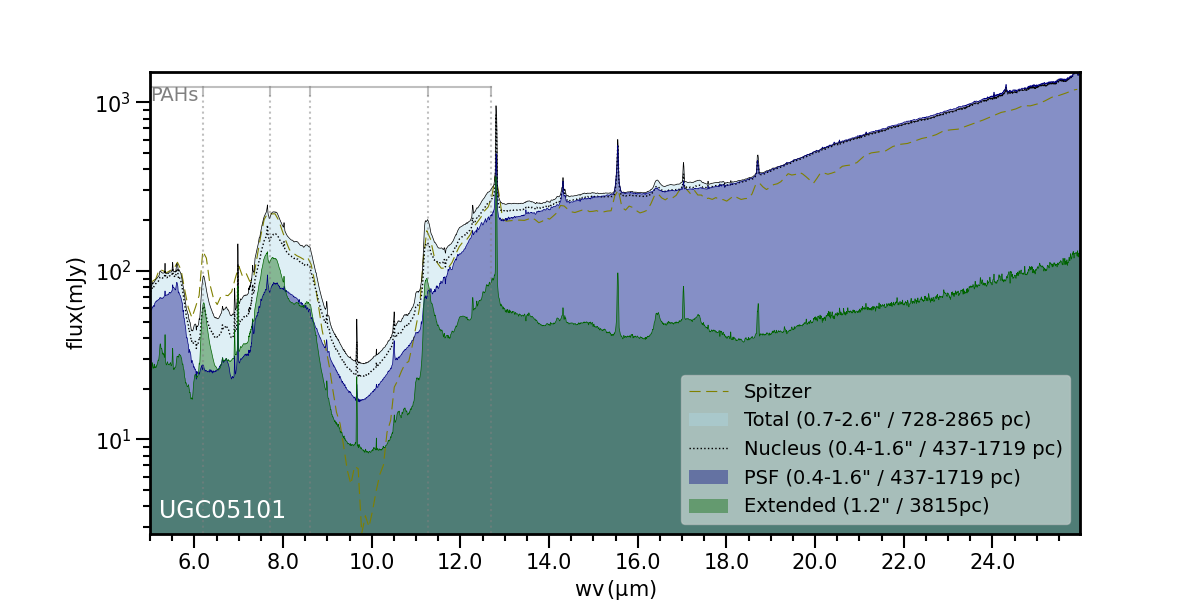}
\includegraphics[width=1.\columnwidth,trim={1.3cm 0.cm 3.cm 1.75cm},clip]{figures/MCG-05-23-016_SpecWV_P0.png} \\
\includegraphics[width=1.\columnwidth,trim={1.3cm 0.cm 3.cm 1.75cm},clip]{figures/NGC3031_SpecWV_P0.png} 
\includegraphics[width=1.\columnwidth,trim={1.3cm 0.cm 3.cm 1.75cm},clip]{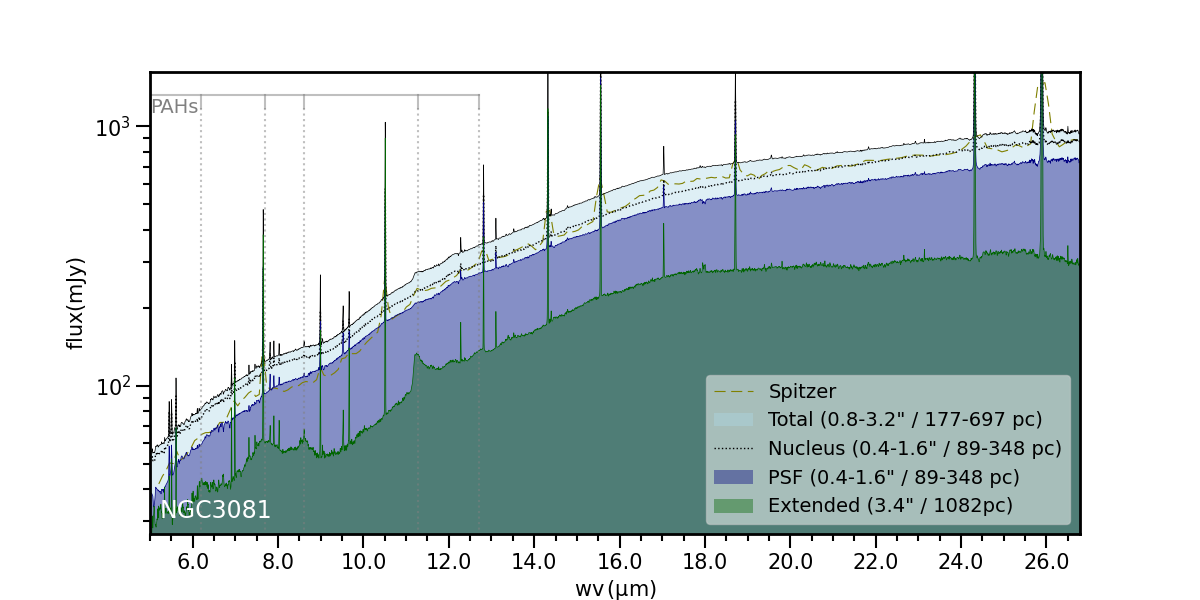} \\
\includegraphics[width=1.\columnwidth,trim={1.3cm 0.cm 3.cm 1.75cm},clip]{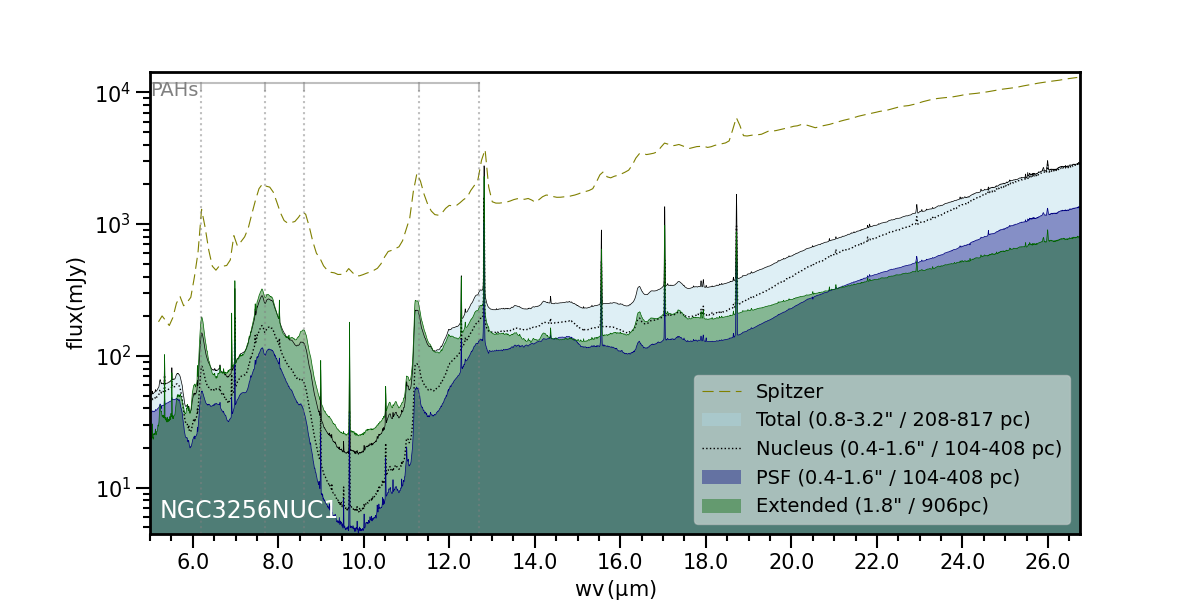} 
\includegraphics[width=1.\columnwidth,trim={1.3cm 0.cm 3.cm 1.75cm},clip]{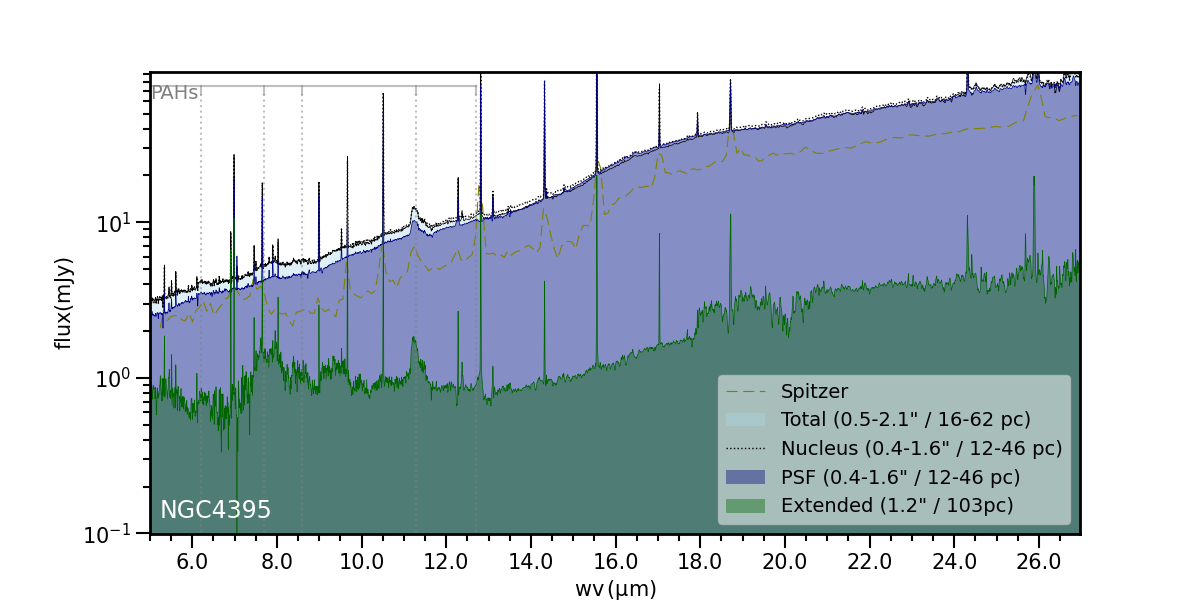}
\label{fig:decomposedspectra1}
\caption{Each panel shows the total (solid-black line filled in light blue), nuclear (dotted-black line), PSF (solid-blue line filled in blue), and circumnuclear (solid-green line filled in green) spectra. When available, \emph{Spitzer}/IRS spectrum is shown with the long-dashed green line. Legends show the spatial scales for each spectral extraction radius.}
\end{figure*}

\begin{figure*}
\includegraphics[width=1.\columnwidth,trim={1.3cm 0.cm 3.cm 1.75cm},clip]{figures/NGC4594_SpecWV_P0.png} 
\includegraphics[width=1.\columnwidth,trim={1.3cm 0.cm 3.cm 1.75cm},clip]{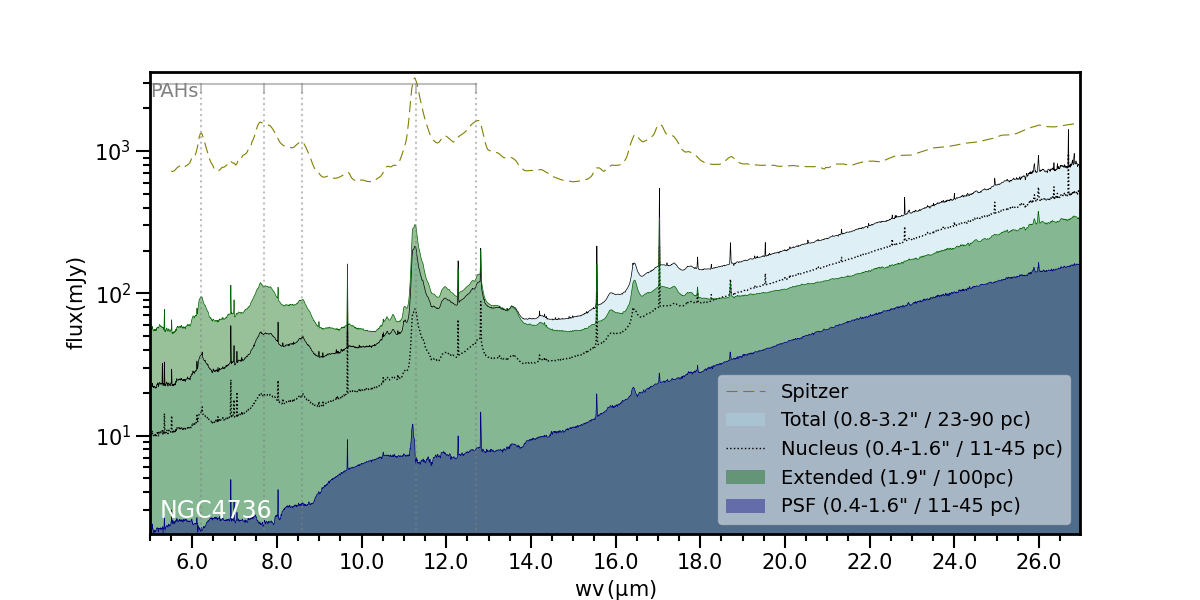} \\
\includegraphics[width=1.\columnwidth,trim={1.3cm 0.cm 3.cm 1.75cm},clip]{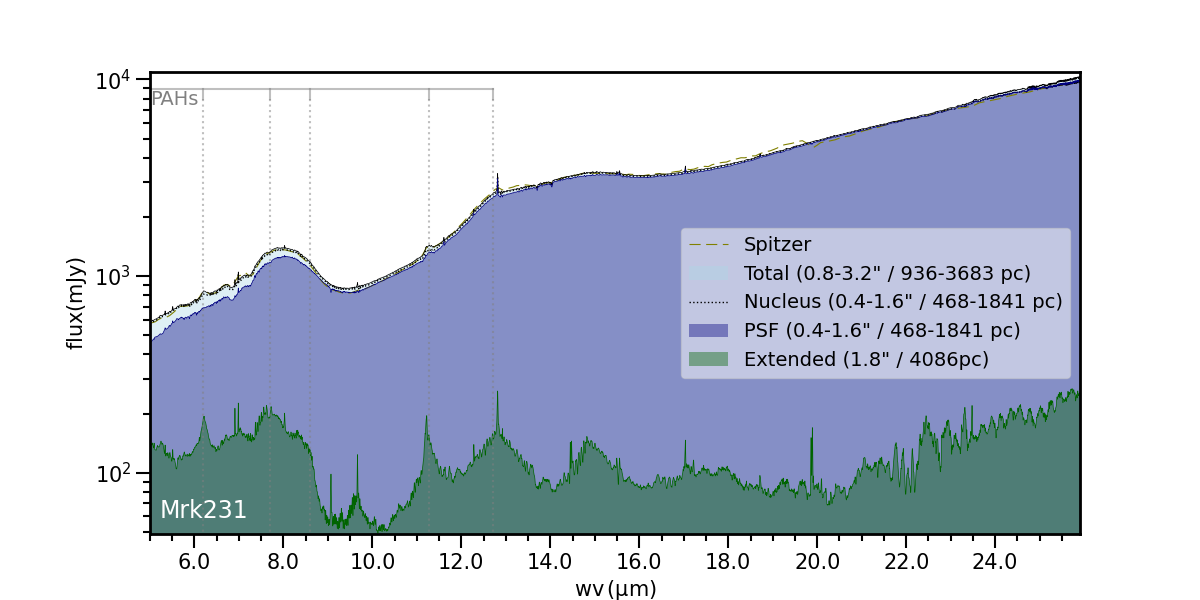} 
\includegraphics[width=1.\columnwidth,trim={1.3cm 0.cm 3.cm 1.75cm},clip]{figures/Mrk273SW_SpecWV_P0.png} \\ 
\includegraphics[width=1.\columnwidth,trim={1.3cm 0.cm 3.cm 1.75cm},clip]{figures/Mrk273_SpecWV_P0.png} 
\includegraphics[width=1.\columnwidth,trim={1.3cm 0.cm 3.cm 1.75cm},clip]{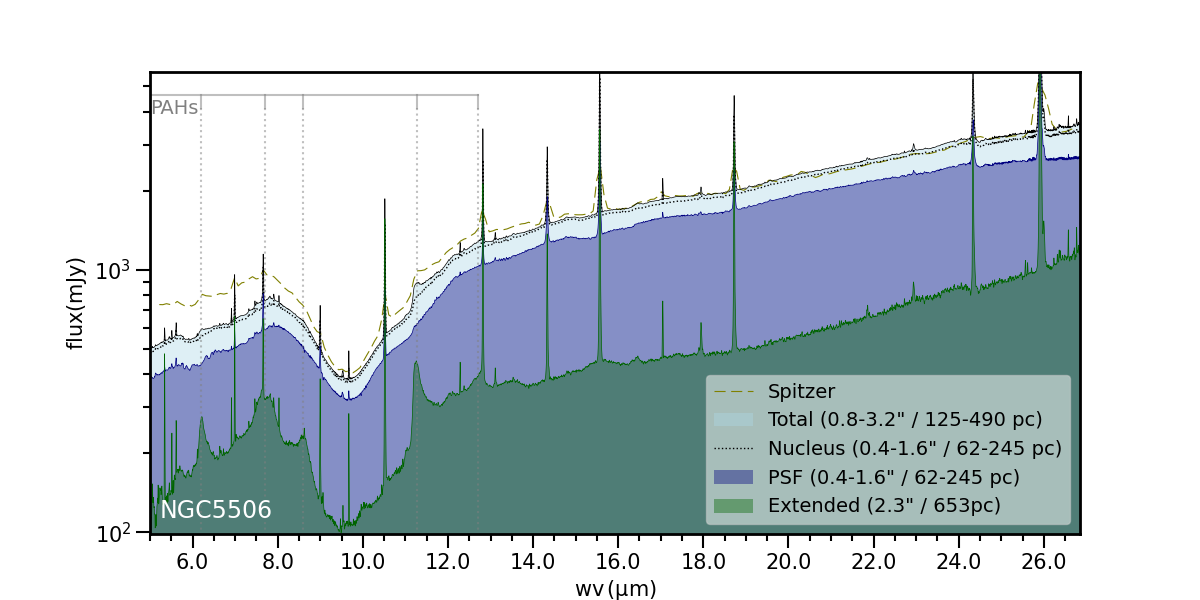} \\
\includegraphics[width=1.\columnwidth,trim={1.3cm 0.cm 3.cm 1.75cm},clip]{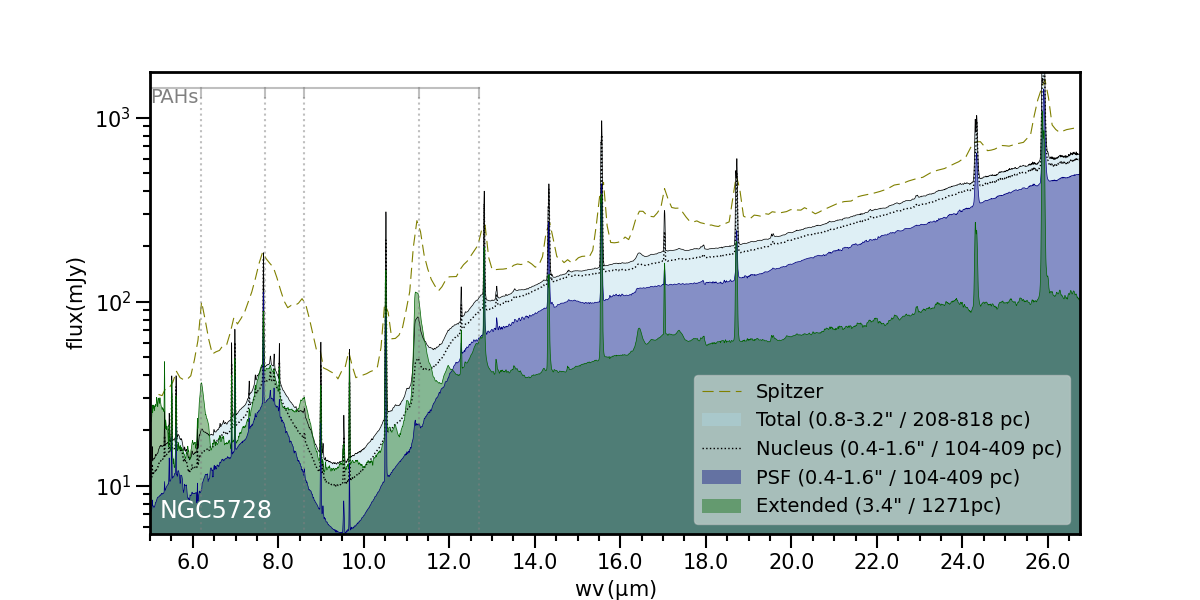}
\includegraphics[width=1.\columnwidth,trim={1.3cm 0.cm 3.cm 1.75cm},clip]{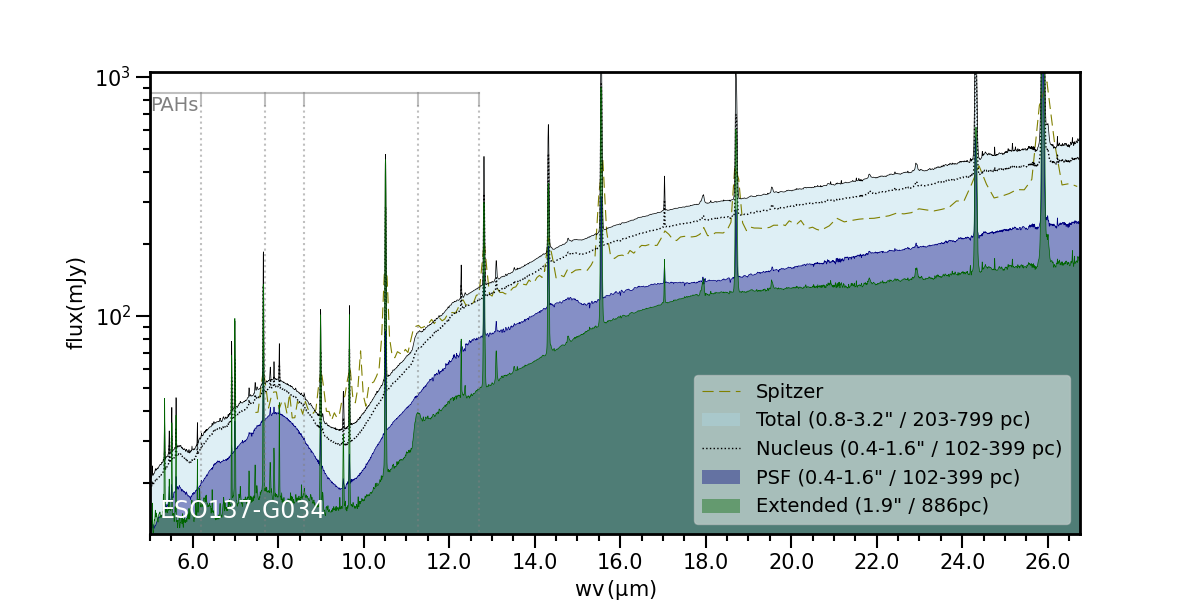} 
\label{fig:decomposedspectra2}
\caption{Each panel shows the total (solid-black line filled in light blue), nuclear (dotted-black line), PSF (solid-blue line filled in blue), and circumnuclear (solid-green line filled in green) spectra. When available, \emph{Spitzer}/IRS spectrum is shown with the long-dashed green line. Legends show the spatial scales for each spectral extraction radius.}
\end{figure*}

\begin{figure*}
\includegraphics[width=1.\columnwidth,trim={1.3cm 0.cm 3.cm 1.75cm},clip]{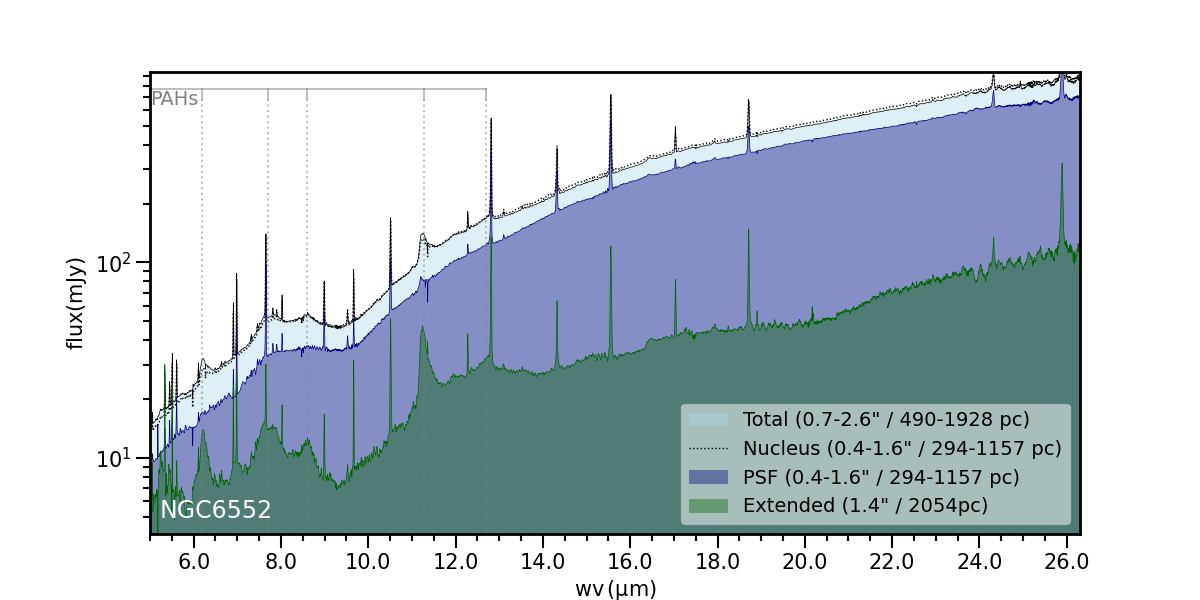} 
\includegraphics[width=1.\columnwidth,trim={1.3cm 0.cm 3.cm 1.75cm},clip]{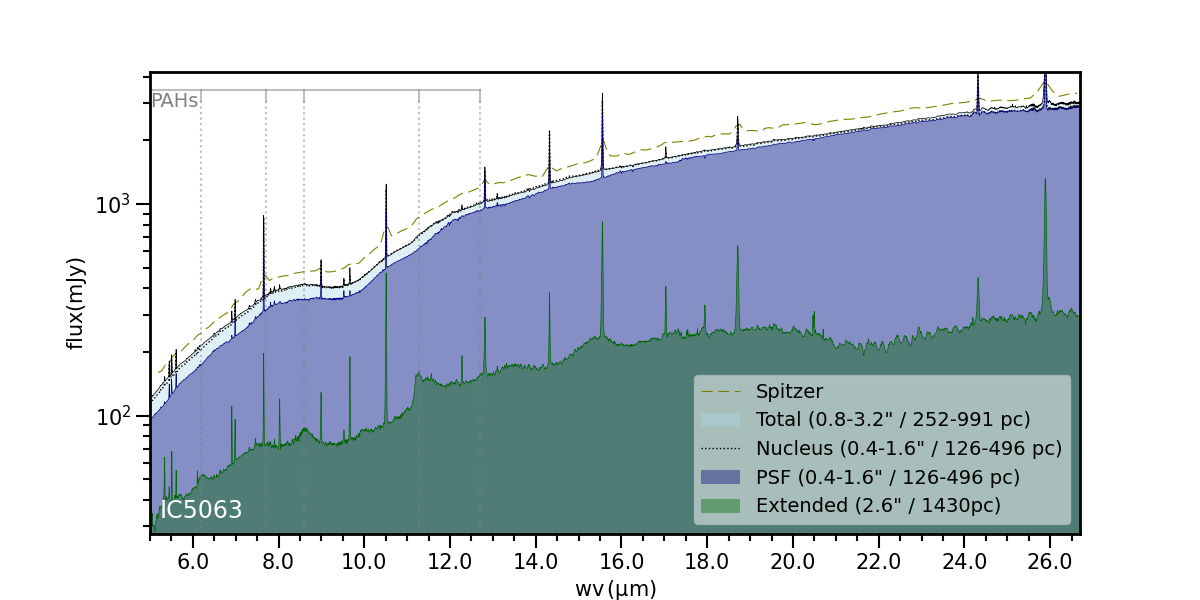} \\
\includegraphics[width=1.\columnwidth,trim={1.3cm 0.cm 3.cm 1.75cm},clip]{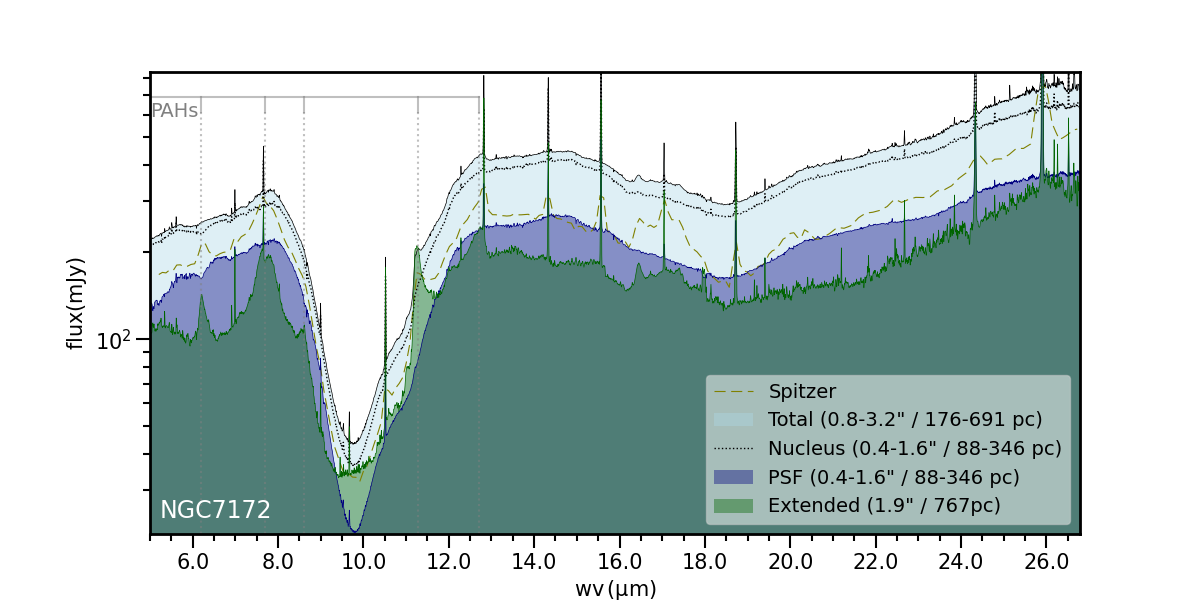} 
\includegraphics[width=1.\columnwidth,trim={1.3cm 0.cm 3.cm 1.75cm},clip]{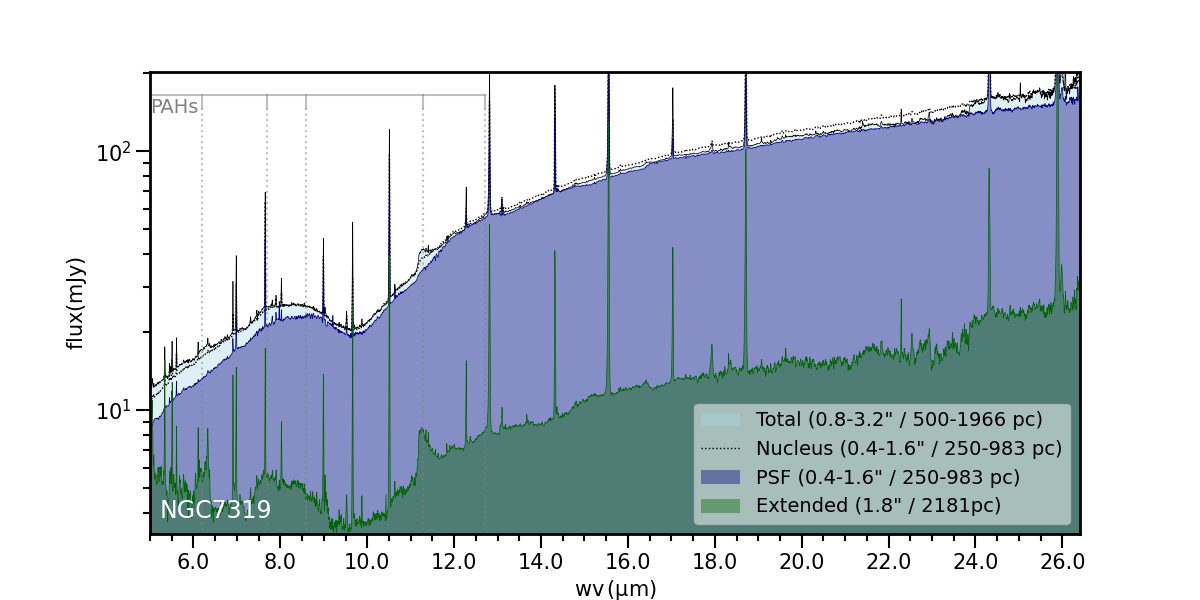} \\ 
\includegraphics[width=1.\columnwidth,trim={1.3cm 0.cm 3.cm 1.75cm},clip]{figures/NGC7469_SpecWV_P0.png}
\label{fig:decomposedspectra3}
\caption{Each panel shows the total (solid-black line filled in light blue), nuclear (dotted-black line), PSF (solid-blue line filled in blue), and circumnuclear (solid-green line filled in green) spectra. When available, \emph{Spitzer}/IRS spectrum is shown with the long-dashed green line. Legends show the spatial scales for each spectral extraction radius.}
\end{figure*}

\section{Table of emission lines}\label{app:emissionlines}

\begin{table}
	\centering
    \footnotesize
    \renewcommand{\tabcolsep}{0.07cm}
\caption{Emission lines excluded for the spectral fitting}
\label{tab:emissionlines}
\begin{tabular}{ll |ll |ll}
\hline \hline
{Line}       &   {wv.}     &  {Line}    &    {wv. }    &  {Line} &  {wv.} \\
{}       &   {($\rm{\mu m}$)}     &  {}    &    {($\rm{\mu m}$)}    &  {} &  {($\rm{\mu m}$) } \\ \hline

$\rm{[Fe\small{II}]}$      &  5.053 & $\rm{[Na\small{III}]}$      & 7.318       & H$_2$S(2)               & 12.279 \\
H$_2$S(8)                  &  5.063 & Pf$\alpha$                  & 7.465       & $\rm{[Ne\small{II}]}$   & 12.814 \\
$\rm{[He\small{II}]}$      &  5.228 & $\rm{[Ne\small{VI}]}$       & 7.652       & $\rm{[Ar\small{V}]}$    & 13.100   \\
$\rm{[Fe\small{II}]}$      &  5.340 & $\rm{[Fe\small{VII}]}$      & 7.814       & $\rm{[Mg\small{V}]}$    & 13.520  \\ 
$\rm{[Fe\small{VIII}]}$    &  5.447 & $\rm{[Ar\small{V}]}$        & 7.899       & $\rm{[Ne\small{V}]}$    & 14.322 \\
$\rm{[Mg\small{VII}]}$     &  5.503 & H$_2$S(4)                   & 8.025       & $\rm{[Ne\small{III}]}$  & 15.555 \\
H$_2$S(7)                  &  5.511 & $\rm{[Ar\small{III}]}$      & 8.991       & H$_2$S(1)               & 17.035 \\ 
$\rm{[Mg\small{V}]}$       &  5.610 & $\rm{[Mg\small{VII}]}$      & 9.040       & $\rm{[S\small{III}]}$   & 18.713 \\ 
H$_2$S(6)                  &  6.110 & $\rm{[Fe\small{VII}]}$      & 9.527       & $\rm{[Ne\small{V}]}$    & 24.318 \\ 
H$_2$S(5)                  &  6.910 & H$_2$S(3)                   & 9.665       & $\rm{[O\small{IV}]}$    & 25.890  \\
$\rm{[Ar\small{II}]}$      &  6.985 & $\rm{[S\small{IV}]}$        & 10.511      & $\rm{[Fe\small{II}]}$   & 26.000   \\
\hline \hline
	\end{tabular}
\end{table}
%%%%%%%%%%%%%%%%%%%%%%%%%%%%%%%%%%%%%%%%%%%%%%%%%%

% Don't change these lines
\bsp	% typesetting comment
\label{lastpage}
\end{document}